\documentclass[a4paper,11pt]{article}
\pdfoutput=1 

\usepackage{jheppub} 

\usepackage[T1]{fontenc} 
\usepackage{blindtext}
\usepackage{epstopdf}
\usepackage{graphicx}
\usepackage{epsfig}
\usepackage{dcolumn}  
\usepackage{bm}    
\usepackage{caption}
\usepackage{subcaption}
\usepackage{amssymb} 
\usepackage{tikz}
\usetikzlibrary{arrows}
\usepackage{nccmath}
\usepackage{xcolor}
\usepackage{epstopdf}
\usepackage{graphicx,wrapfig,lipsum}
\usepackage{epsfig}
\usepackage{amsmath,bm}
\usepackage{amsfonts}  
\usepackage{amsmath}  
\usepackage{slashed}  
\usepackage{enumitem}
\usepackage[mathscr]{euscript}
\usepackage{tabu}
\usepackage{epsfig}
\usepackage{float}
\usepackage[percent]{overpic}
\usepackage{graphicx}
\usepackage{caption}
\usepackage{fancyhdr}
\usepackage{tikz}

\makeatletter
\g@addto@macro\bfseries{\boldmath}
\makeatother

\DeclareMathOperator{\csch}{csch}

\def\l1{{{1-loop}}}

\def\n1{\Bigg|_{n=1}}

\def\n{{(n)}}

\usepackage[T1]{fontenc} 
\usepackage{tikz}
\usepackage{amsmath,amssymb}
\usepackage{relsize}
\usepackage{latexsym}
\usepackage{leftidx}
\usepackage{xcolor}
\usepackage{diagbox}
\usepackage[T1]{fontenc}
\usepackage{array}
\usepackage{makecell}
\usepackage{csquotes}
\usepackage{tikz}
\usepackage{enumitem}
\usepackage{setspace}
\usepackage{multirow}
\usepackage{amsmath,amssymb}
\usepackage{relsize}
\usepackage{latexsym}
\usepackage{leftidx}
\usepackage{xcolor}
\usepackage{csquotes}
\usepackage{tikz}
\usetikzlibrary{decorations.pathmorphing}
\usetikzlibrary{arrows.meta}
\usepackage{enumitem}
\usetikzlibrary{decorations.markings}
\usetikzlibrary{decorations.pathmorphing}
\usetikzlibrary{decorations.markings}
\usetikzlibrary{decorations.pathmorphing}
\usepackage{pifont}
\usepackage{hyperref}
\usepackage{url}

\usepackage{bookmark}

\title{\textbf{\textsf{High to low temperature: $O(N)$ model at large $N$
}}}
\author[a]{Justin R. David}
\author[a,b]{, Srijan Kumar}

\affiliation[a]{\vspace{.1cm} Centre for High Energy Physics,  Indian Institute of Science,
	C. V. Raman Avenue, Bangalore 560012, India.}
\affiliation[b]{International Centre for Theoretical Sciences (ICTS-TIFR),
	Tata Institute of Fundamental Research, Shivakote, Hesaraghatta, Bengaluru 560089, India.}

\emailAdd{justin@iisc.ac.in}
\emailAdd{srijankumar@iisc.ac.in}

\abstract{
	We study the $O(N)$ vector model for scalars with quartic interaction at large $N$ on $S^1\times S^2$ without the singlet constraint. The non-trivial fixed point of the model is described by a thermal mass satisfying the gap equation at large $N$. We obtain the free energy and the energy density for the model as a series at low temperature in units of the radius of the sphere. We show these results agree with the Borel-Pad\'{e} extrapolations of the high temperature expansions of the free energy and energy density obtained in our previous work. This agreement validates both the expansions and demonstrates that low temperature expansions obtained here correspond to the same solution of the gap equation 
	studied earlier at high temperature. We obtain the ratio of the free energy of the theory at the non-trivial fixed point to that of the Gaussian theory at all values of temperature. This ratio begins at $4/5$ when the temperature is infinity, decreases to a minimum value of $0.760753$, then increases and approaches unity as the temperature is decreased. 
	
}
\begin{document}
	\maketitle
	\flushbottom
		\section{Introduction}
		The $O(N)$ vector model is among the simplest and most studied examples of quantum field theories, important in high energy physics as well as condensed matter physics applications. The model is exactly 
		solvable  at large $N$. 
		The quartic interaction present in its action can be linearised using 
		the  Hubbard-Stratanovich transformation \cite{Moshe:2003xn}. An exact solution for the theory 
		can be obtained at large $N$ even at strong coupling. 
		The model  in $(2+1) d $ admits an infrared fixed point at strong coupling, which is often studied 
		as a prototype example of interacting conformal field theory.  
		In the singlet sector, this model  at large $N$ is conjectured to be 
		the  holographic dual of Vasiliev  higher spin gravity  in $AdS_4$ \cite{Klebanov:2002ja}.
		
		Given the importance of this model, it is natural that several works have investigated the 
		finite temperature physics of the critical $O(N)$ model in $2+1$ dimensions beginning with \cite{Rosenstein:1989sg}. 
		In  \cite{Chubukov:1993aau,Sachdev:1993pr} , it was shown that model  at large $N$, 
		admits a non-trivial fixed point at infinite coupling 
		characterized by the presence of a thermal mass. The ratio of the free energies  of the 
		model on $S^1\times R^2$ at the strong coupling 
		critical 
		point to the free theory was evaluated to be $4/5$, very similar to the famous ratio of 
		$3/4$ for ${\cal N}=4$ SYM \cite{Gubser:1998nz,Blaizot:2006tk}.


		In this paper, we would like to study the thermodynamics of the large $N$, critical $O(N)$ model at strong coupling on $S^1\times S^2$. In our earlier paper \cite{David:2024pir}, we developed a high temperature 
		expansion for the 
		free energy of this model in terms of a series in 
		powers of $\beta/r$, where $\beta$ is the length of 
		$S^1$ and $r$ is the radius of $S^2$. In this paper, we develop a low temperature expansion and 
		use Pad\'{e}-Borel methods to obtain the behavior of the model at all temperatures. 
		One particular quantity of interest in this paper is how the ratio of the free energy of this model to that 
		of the Gaussian model varies as a function of $\beta/r$. 
		
		Before we proceed, it is important to mention that there has been a recent spurt in the study of 
		conformal field theories at finite temperature both on geometries of the kind $S^1\times R^{d-1}$ as 
		well as $S^1 \times S^{d-1}$. 
		These studies have been motivated by the need to push bootstrap methods to finite temperature 
		for which there are lesser symmetries and therefore fewer constraints as well as the 
		need to understand results  on black holes physics from holography. Early studies in the direction 
		of developing CFT methods for finite temperature begin with the work of \cite{Dolan:2005wy}
		which expresses the CFT partition functions on spheres  in terms of conformal characters. 
		Thermal  conformal blocks  on
		$S^1\times S^2$ were obtain in \cite{Dolan:2003hv,Dolan:2005wy,Gobeil:2018fzy}. Very 
		recently one point functions have been written as partial wave decompositions in terms of 
		conformal blacks on $S^1\times S^2$ \cite{Buric:2024kxo,Buric:2025uqt} with very explicit
		tests performed for free CFT's. Conformal blocks for higher point functions have been 
		found in  \cite{Ammon:2025cdz}.   Bootstrap methods on $S^1\times R^2$ were initiated 
		by the work of  \cite{Iliesiu:2018fao}, which provided a way of obtaining
		writing one point functions which occur in the OPE expansions of a given thermal 
		2 point function provided it satisfied certain analytical as well as boundedness properties.  
		These were developed systematically for large $N$ vector models 
		and generalized to  fermions,  situations
		with chemical potentials as well as supersymmetry in \cite{Petkou:2018ynm,Iliesiu:2018zlz,David:2023uya,Karydas:2023ufs,David:2024naf,Kumar:2025txh}. 
		The $1/N$ corrections to the thermal one point functions were computed in \cite{Diatlyk:2023msc} for the large $N$ vector models on $S^1\times R^2$ and the critical  long range $O(N)$ model has been analyzed in \cite{Benedetti:2023pbt}.
		Bootstrap techniques based on broken symmetries, sum rules and Tauberian theorem \cite{Marchetto:2023xap,Barrat:2025wbi,Barrat:2025nvu}, 
		allow for the evaluation of  thermal OPE coefficients  in general and more specifically for the $O(N)$ model at finite $N$.

		Recent studies of holographic computations of thermal correlators include
		one point functions in \cite{Grinberg:2020fdj,David:2022nfn,Berenstein:2022nlj,Singhi:2024sdr}, 
		and other correlators in  \cite{Alday:2020eua,Karlsson:2021duj,Dodelson:2022yvn,Bhatta:2022wga,Karlsson:2022osn,Esper:2023jeq,Ceplak:2024bja,Buric:2025anb,Buric:2025fye} .  
		Another direction of development is the ambient space formalism introduced in 
		\cite{Parisini:2022wkb,Parisini:2023nbd} for studying  thermal CFTs on curved geometries. A recent development introduces a thermal bootstrap program using neural networks \cite{Niarchos:2025cdg}.
		
		%
		%
		%
		
		Though most of the studies of the $O(N)$ model at large $N$ has been for the unconstrained model, 
		as mentioned earlier it is the singlet model which is holographically dual to higher spin gravity 
		in $AdS_4$.  In this context too, it is important to study this model 
		on the geometry $S^1\times S^2$. This model with 
		with the $U(N)$ singlet constraint  has been studied on this geometry in \cite{Shenker:2011zf,Giombi:2014yra}. The theory with $U(N)$ singlet constraint undergoes the Gross-Witten-Wadia phase transition at a temperature $T\sim \sqrt{N}$ found in \cite{Shenker:2011zf}. 
		%
		\\

		In the rest of the introduction, we present  a summary of this work and the main results obtained, followed by the organization of the paper.
		\subsection*{Summary of the results}
		We consider the $O(N)$ model at large $N$ on $S^1\times S^2$ without the singlet constraint. The length of the compact direction $S^1$ is identified as the inverse temperature $\beta=T^{-1}$.  The action describing the  model for $N$ real scalar fields interacting through a quartic coupling term has the following form
		\begin{align}\label{action intro}
			S[\phi]=\frac{1}{2} \int d^3x \sqrt{g} [\partial^\mu\phi_i \partial_\mu\phi_i +\frac{R}{8}\phi_i\phi_i+\frac{\lambda}{N}(\phi_i\phi_i)^2],
		\end{align}
		where $i=1,\cdots, N$;
		$g$ is the determinant of the metric in this geometry.  The conformal coupling involves the Ricci scalar $R=\frac{2}{r^2}$ for a 2-sphere,
		$\lambda$ characterizes the coupling strength of the quartic interaction among the scalars. At large $N$, the model exhibits a non-trivial fixed point at infinite coupling,  $\lambda\to\infty$. Substituting $\lambda=0$,  we recover the Gaussian fixed point of the model or the free fixed point.\\
		
		In \cite{David:2024pir}, we evaluated the free energy, thermal expectation of energy and the thermal one point functions of conserved currents as an expansion in small $\frac{\beta}{r}$, where $\beta=\frac{1}{T}$ and $r$ is the radius of $S^2$.   This high temperature expansion was obtained based on the development of a new technique to perform an infinite sum over angular modes on the sphere $S^2$ for a massive scalar\footnote{The method has  been subsequently used by \cite{Gupta:2025ala}, which discussed generalizations to hemisphere and squashed sphere.}.  In terms of the thermal effective field theory  \cite{Kang:2022orq,Benjamin:2023qsc,Benjamin:2024kdg}, 
		the coefficient of the leading finite size correction to the free energy saturates the bound conjectured in \cite{Allameh:2024qqp}. In this work, we evaluate the low temperature expansions of the free energy and thermal expectation of energy. This involves solving the gap equation as a low temperature expansion for the thermal mass.  The free energy as an expansion in $e^{-\frac{\beta}{2r}}$ takes the following form\footnote{In this paper we often refer to $\log Z$ as the free energy, it is understood that it is up to the factor of $-\frac{1}{\beta}$.}
		\begin{align}\label{low temp intro}
			&	\log Z=\log (Z_{\rm free})-\frac{4 \beta  e^{-\frac{\beta }{r}}}{\pi ^2 r}+e^{-\frac{3 \beta }{2 r}} \Big(\frac{32 \beta ^2}{\pi ^4 r^2}+\frac{8 \left(24-5 \pi ^2\right) \beta }{3 \pi ^4 r}\Big)
			+O(e^{-\frac{2\beta}{r}}).
		\end{align}
		where $Z_{\rm free}$ is the partition function for the free theory on $S^1\times S^2$.
		\begin{align} \label{exponentials}
			\log (Z_{\rm free})=\sum_{l=0,n=1}^\infty(2l+1)\frac{e^{-\frac{n\beta}{r}(l+\frac{1}{2})}}{n}.
		\end{align}  
		The above expansion with a few more higher order terms in orders of $e^{-\frac{\beta}{2 r}}$ can be found in \eqref{log Z interacting}\footnote{We provide the low temperature expansions for the free energy and thermal expectation of the energy till the order of $O(e^{-\frac{6\beta}{r}}) $ in the ancillary file  \texttt{low\_temp\_expansions.txt} attached to the arXiv version of this paper. }.
		And the thermal expectation of energy admits the low temperature expansion as given by the differentiation of the expression \eqref{low temp intro} with respect to $\beta$ with an overall negative sign, shown below
		\begin{align}
			&	\langle E\rangle=		-\partial_\beta \log Z\nonumber\\
			&=E_{\rm free}+e^{-\frac{\beta }{r}} \left(\frac{4}{\pi ^2 r}-\frac{4 \beta }{\pi ^2 r^2}\right)+e^{-\frac{3 \beta }{2 r}} \left(\frac{48 \beta ^2}{\pi ^4 r^3}-\frac{20 \beta }{\pi ^2 r^2}+\frac{32 \beta }{\pi ^4 r^2}+\frac{40}{3 \pi ^2 r}-\frac{64}{\pi ^4 r}\right)
			+O(e^{-\frac{2\beta}{r}}).
		\end{align}
		where	$E_{\rm free}$ denotes the thermal expectation of energy of the free CFT obtained in \eqref{E free}
		\begin{align}
			E_{\rm free }=\sum_{l=0,n=1}^\infty\frac{(2 l+1)^2 }{2 r}e^{-\frac{\beta  (l+\frac{1}{2}) n}{r}}.
		\end{align}
		Again, the low temperature expansion of the thermal expectation of energy with a few more higher order terms can be found in \eqref{en den interacting}.\\
		
		We show that the low temperature expansions given above and the high temperature expansions \cite{David:2024pir} correspond to the same solution of the gap equation or  fixed point of the model
		\footnote{The solutions of the gap equation result in thermal CFT's and therefore we call them fixed points.}
		At first glance, it seems that the expansions 
		should obviously correspond to the same fixed point. However as we will see, 
		in deriving these expansions, 
		we would need to solve the gap equation  which is a transcendental equation (\ref{saddle pt cond}). 
		The solution necessarily involves expanding this equation at high or low temperatures and it is not completely guaranteed that these expansions
		naturally choose the same branch in these limits. 
		Furthermore in a free CFT,  one can explicitly show that
		a re-summation of an infinite number of terms from the low temperature expansion reproduces its high temperature expansion \cite{Cardy:1991kr}. This involves rewriting the exponentials   in the low temperature expansion \eqref{exponentials} as an integral representation followed by a contour deformation. 
		In contrast, for the theory obtained by solving the gap equation  from the low temperature fixed point, the free energy  can be evaluated only order by order in  $e^{-\frac{\beta}{2r}}$ as a low temperature expansion. 
		At present we do not have a general formula for each term in the low temperature expansion
		(or the high temperature expansion) for the 
		free energy, so 
		the technique of re-summation of an infinite set of terms  introduced in  \cite{Cardy:1991kr} cannot be applied.

		Due to the above reasons we turn to numerical techniques to study the behaviour of the free energy at all temperatures. We apply a Borel-Pad\'e re-summation \cite{Ellis:1995jv,Florio:2019hzn,Costin:2019xql,Andersen:2021bgw,Muller:2025qkw} to the high temperature expansion of the free energy to extrapolate it to  lower values of the temperature (in units of $r$) and demonstrate its agreement with the low temperature expansion. The high temperature expansion for the free energy turns out to be an asymptotic series in orders of small $\frac{\beta}{r}$. The technique of Borel sum provides  a powerful tool to re-sum an infinite number of terms of an asymptotic series to give a closed-form expression that is valid for all values of the expansion parameter. Most importantly, the singularities of the Borel transform can encode non-perturbative effects accessible from the perturbative series \cite{Dunne:ResurgenceIntro,Marino:2021Resurgence}.  But often in practice, the asymptotic series of interest, including the series for our free energy, can only be computed till  a finite order of terms. In such cases, the method of Borel-Pad\'e resum provides an effective framework to extrapolate  the series beyond the perturbative regime just from the knowledge of a finite number of terms. This relies on the approximation of the Borel transform of the series till a finite number of terms as a rational function using the technique of Pad\'e. The final step involves the Laplace transform of this Pad\'e approximant.  In our case we will avoid all the poles arising in the Pad\'e approximant of the Borel transform by following the principal value prescription in the integral for the Laplace transform. This prescription is consistent with the real-valuedness of the free energy and also avoids ambiguities in the final result, as also been used in \cite{Dondi:2021buw}. The agreement of this Borel-Pad\'e extrapolated free energy with the low temperature expansion computed in the current work further justifies the consistency  of this prescription. \\
		
\begin{figure}[ht]
			\begin{subfigure}{.46\linewidth}
				\includegraphics[width=1\linewidth]{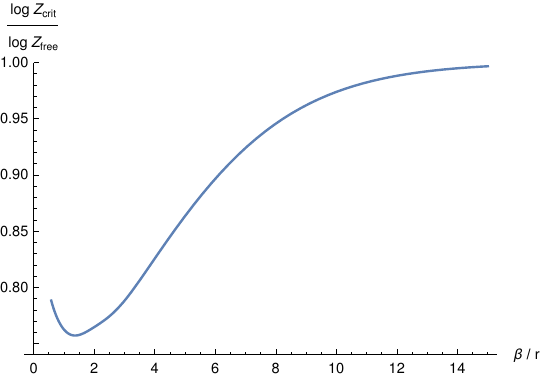}
				\caption{From the low temperature expansions }
				\label{intro fig a}
			\end{subfigure}\hfill
			\begin{subfigure}{.46\linewidth}
				\includegraphics[width=1\linewidth]{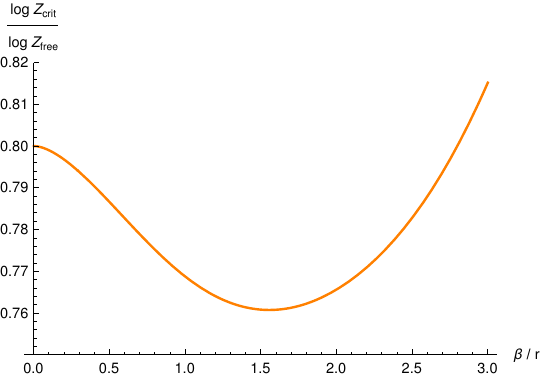}
				\caption{Borel-Pad\'e  of high temp. expansions}
				\label{intro fig b}
			\end{subfigure}
			\caption{Plots for the ratio of free energies at the non-trivial fixed point to the Gaussian fixed point. Here $Z_{\rm crit}$ and $Z_{\rm free} $ denote the partition functions at the non-trivial fixed point and the Gaussian fixed point respectively. Figure \ref{intro fig a} plots this ratio computed from the low temperature expansions of $\log Z$, truncated till $O(e^{-\frac{6\beta}{r}})$ for both the fixed points.  Figure \ref{intro fig b} is obtained using the Borel-Pad\'e re-summation of  the high temperature expansions of $\log Z$ at both the fixed points, using the Pad\'e of order $[14,14]$. 
				Figures \ref{intro fig a} and \ref{intro fig b} plot the same function.
				The low temperature expansion, Figure \ref{intro fig a} describes the function correctly when $\beta/r$ is large, as the effect of truncation of the higher order terms is negligible. In contrast, the Borel-Pad\'e extrapolations of the high temperature expansions, Figure \ref{intro fig b}  remain sufficiently accurate for small values of $\beta/r$.  Both the expansions consistently exhibit the minimum and the value at the minimum.}
			\label{intro fig}
		\end{figure}

		As an illustration and a highlight of our results, 
		we present the graph for the ratio of the free energy at the non-trivial fixed point to that at the Gaussian fixed point plotted as a function of $\frac{\beta}{r}$ in figure \ref{intro fig}. With increasing $\frac{\beta}{r}$, this ratio initially decreases from $\frac{4}{5}$, the result on $S^1\times R^2$ \cite{Sachdev:1993pr}, to attain the 
		minimum value of $0.760753$ at around $\frac{\beta}{r} \approx 1.55649$  and then it increases with increasing $\frac{\beta}{r}$, finally saturating to $1$.
		We have evaluated this ratio using the Borel-Pad\'e sum of the high temperature. 
		This is consistently seen both from the low temperature expansion, 
		\ref{intro fig a} and 
		the Borel-Pad\'{e} resummation of the high temperature expansion, \ref{intro fig b}.		
		On evaluating the ratio from the low temperature expansion, the location of the minimum and the value at the minimum differ by 
		$12.2\%$ and $.46 \%$ respectively, with respect to the values obtained from the Borel-Pad\'e approximation at high temperature. 
		This is because the low temperature expansion is affected by errors due to truncating higher-order terms when $\beta/r$ is not sufficiently large.
		\\
		The organization of the paper is as follows, Section \ref{sec 2} reviews
		the partition function and  thermal expectation of energy in
		the free CFT on $S^1\times S^2$, both at low and high temperature. 
		Section \ref{sec 3} studies the model at the non-trivial fixed point
		at an infinite coupling;
		Subsection \ref{sec 3.1} presents our previous results \cite{David:2024pir} for the high temperature expansions of the free energy and thermal expectation of energy; Subsection \ref{sec 3.2} contains the computations of their low-temperature expansions.
		Section \ref{sec 4} contains the Borel-Pad\'e re-summation of the high temperature expansions and its agreement with the low temperature expansion. Section \ref{Discussions} has our discussions. 
		Appendix \ref{app A} shows the equivalence between the 
		the low and high  
		temperature expansion of the $\log Z_1(-1/2)$ defined in \eqref{logZ after Mat sum} for a massive scalar.
		Appendix \ref{appn B} has an alternative derivation for the high temperature expansion of $\log Z_2$ defined in \eqref{logZ after Mat sum}, generalizing the method by Cardy \cite{Cardy:1991kr} for the massive scalar.

		\section{The free theory on $S^1\times S^2$}\label{sec 2}
		The free $O(N)$ model is a simple and well-studied example of  conformal field theory on $S^1\times S^2$, constructed just by combining $N$ massless conformally coupled scalars $\phi_i$'s on this geometry. We obtain the action describing the Gaussian fixed point of the $O(N)$ model by substituting $\lambda=0$ in the action given in \eqref{action intro}, and it is given by
		\begin{align}\label{free action}
			S{[\phi]}=\frac{1}{2}\int d^3x \sqrt{g}  \Big[\partial^\mu\phi_i \partial_\mu\phi_i+\frac{R}{8}\phi_i\phi_i\Big],
		\end{align}
		where $i=1,\cdots,N$, and $g$ is the determinant of the metric in this geometry.  The conformal coupling involves the Ricci scalar $R=\frac{2}{r^2}$ for a 2-sphere. \\
		
		The partition function for the free CFT given by the action \eqref{free action} can be computed as a low temperature expansion  by counting the number of scaling operators in the theory \cite{Cardy:1991kr}. We will reproduce this calculation for the partition function by evaluating the Euclidean path integral on the geometry of $S^1\times S^2$. The method of path integral gets naturally adapted for the model at the non-trivial fixed point at an infinite coupling discussed in the next section. A technique developed in \cite{Cardy:1991kr} can resum the low temperature expansion  for the free energy in free theory to organize it as a high temperature expansion(also see \cite{Kutasov:2000td,Melia:2020pzd,Lei:2024oij}). We will also review this method in this section and evaluate the high temperature expansion for the free energy.
		
		The partition function for this action can be given by the following Euclidean path integral representation
		\begin{align}
			Z=\int_{S^1\times S^2} {\cal D} \phi_i e^{-S[\phi]}.
		\end{align}
		The metric for the geometry of $S^1\times S^2$ is given by
		\begin{align}
			ds^2=d\tau^2+r^2(d\theta^2+\sin^2\theta d\phi^2), \qquad 
			{\rm where} \ \tau\sim \tau+\beta.
		\end{align}
		$\tau$ denotes the Euclidean time direction with periodicity of length $\beta$; and $\theta,\phi$ are the angular directions on the sphere $S^2$ of radius $r$.
		We can perform  this path integral by a suitable mode expansion of the field $\phi$ in terms of  Fourier modes along the compact $\tau$ direction and spherical harmonics on the sphere $S^2$ as given by
		\begin{align}
			\phi=\sum_{n=-\infty}^{\infty}\sum_{l=0}^\infty \sum_{m=-l}^{l} a_{n,l,m} e^{\frac{i2\pi n\tau}{\beta}} Y_{l,m}(\theta,\varphi),
		\end{align}
		where $Y_{l,m}(\theta,\varphi)$'s are spherical harmonics and $n$ denotes the Matsubara frequencies due to the periodicity imposed on the $\tau$ direction. Integrating over $a_{n,l,m}$, we obtain
		\begin{align}
			\log Z=-\frac{N}{2}\sum_{n=-\infty}^\infty \sum_{l=0}^\infty (2l+1) \log\Big((\frac{2n\pi}{\beta})^2+\frac{(l+\frac{1}{2})^2}{r^2}\Big)\equiv N \log Z_{\rm free}.
		\end{align}
		The sum over $n$ due to the Matsubara frequencies is performed using the following formula for the  regulated sum \cite{Klebanov:2011uf}
		\begin{align}\label{regulated sum}
			\sum_{n=0}^{\infty} \log (\frac{n^2}{q^2}+a^2)=2 \log (2\sinh (\pi q|a|)).
		\end{align}
		Thus, we obtain the following expression 
		\begin{align}
			\log Z_{\rm free}=-\sum_{l=0}^\infty (l+\frac{1}{2}) \Big(\frac{\beta}{r} (l+\frac{1}{2})+2\log (1-e^{-\frac{\beta}{r}(l+\frac{1}{2})})\Big).
		\end{align}
		Now the sum in the first term in the above equation is diverging and it is regularized using the scheme of the Zeta function regularization as given below
		\begin{align}
			\sum _{l=0}^{\infty } \left(l+\frac{1}{2}\right)^{-s}=\left(2^s-1\right) \zeta (s).
		\end{align}
		And using the fact that $\zeta(-2)=0$, the first term vanishes in this scheme of regularization, thus we have the following expression for the partition function
		\begin{align}\label{log Z free at low temp}
			\log Z_{\rm free}=-\sum_{l=0}^\infty ( 2l+1)\log (1-e^{-\frac{\beta}{r}(l+\frac{1}{2})}).
		\end{align}
		Now such an expression \eqref{log Z free at low temp} admits the following series expansion in terms of exponentially suppressed terms at $\frac{\beta}{r}\to \infty$ given by
		\begin{align}\label{free small b/r} 
			\log (Z_{\rm free})=\sum_{l=0,n=1}^\infty(2l+1)\frac{e^{-\frac{n\beta}{r}(l+\frac{1}{2})}}{n}.
		\end{align} 
		The thermal expectation of energy can be given by the negative differentiation of $\log Z$ with respect to $\beta$
		\begin{align}\label{E free}
			E_{\rm free }=-\partial_\beta \log Z=\sum_{l=0,n=1}^\infty\frac{(2 l+1)^2 }{2 r}e^{-\frac{\beta }{r} (l+\frac{1}{2}) n}.
		\end{align}
		We can also derive the small $\frac{\beta}{r}$  expansion of the free energy from the large $\frac{\beta}{r}$ expansion \eqref{free small b/r} following the method given in \cite{Cardy:1991kr}. We  use the following representation of the $e^{-\tau}$ as the inverse Mellin transform of the Gamma function 
		\begin{align}
			e^{-\tau} =\frac{1}{2\pi i}\int_{-i\infty+a}^{i\infty+a} \tau^{-s} \Gamma(s) ds, \qquad {\rm where} \ \ a>2.
		\end{align}
		in the equation \eqref{free small b/r} to obtain
		\begin{align}\label{Mellin pat}
			\log Z_{\rm free }& =\frac{1}{2\pi i} \sum_{l=0}^\infty\sum_{n=1}^\infty \int_{-i\infty+a}^{i\infty+a} ds\frac{\Gamma(s)}{(2l+1)^{s-1}n^{s+1}}  \Big(\frac{\beta}{2r}\Big)^{-s} \nonumber, \\
			&=\frac{1}{2\pi i} \int_{-i\infty+a}^{i\infty+a} ds (2^s-2) \zeta (s+1) \zeta (s-1) \Gamma (s) \Big(\frac{\beta }{r}\Big)^{-s}.
		\end{align}
		The infinite sums over $l$ and $n$ are performed in terms of zeta functions. The integrand has simple poles at negative integer values of $s$; at $s=0$ it has a simple pole as well as a double pole and at $s=2$ again it has only a simple pole. Thus, the integral in the above equation can be evaluated by adding the residues of the integrand at all these poles.\\
		
		The residue at $s=2$
		\begin{align}\label{res at 2}
			(2^s-2) \zeta (s+1) \zeta (s-1) \Gamma (s) \Big(\frac{\beta }{r}\Big)^{-s}\Big|_{{\rm  Res\ at}\ s=2}=\frac{2 r^2 \zeta (3)}{\beta ^2}.
		\end{align}
		
		The residue at $s=0$
		\begin{align}\label{res at 0}
			(2^s-2) \zeta (s+1) \zeta (s-1) \Gamma (s) \Big(\frac{\beta }{r}\Big)^{-s}\Big|_{{\rm  Res\ at}\ s=0}=-\frac{1}{12} \log \frac{\beta}{r} -\frac{\log 2}{12}-\zeta'(-1).
		\end{align}
		
		And the residue at the pole $s=n$ where $n$ is a negative integer, 
		\begin{align}\label{res at n}
			(2^s-2) \zeta (s+1) \zeta (s-1) \Gamma (s) \Big(\frac{\beta }{r}\Big)^{-s}\Big|_{{\rm  Res\ at}\ s=n}=\frac{(-1)^n \left(2^n-2\right) \zeta (n-1) \zeta (n+1) \big(\frac{\beta }{r}\big)^{-n}}{(-n)!}.
		\end{align}
		Finally combining \eqref{res at 0}, \eqref{res at 2} and \eqref{res at n} we obtain the small $\frac{\beta}{r}$ expansion of the free energy from \eqref{Mellin pat}  to be
		\begin{align}\label{free pat fn using Cardy}
			\log Z_{\rm free} =&\frac{2 r^2 \zeta (3)}{\beta ^2}-\frac{1}{12} \log \frac{\beta}{r} -\frac{\log 2}{12}-\zeta'(-1)\nonumber\\
			&	+\sum_{n\in 2\mathbb Z} \frac{(-1)^{n+1}(2-2^{-n}) \zeta (-n-1) \zeta (1-n) \big(\frac{\beta }{r}\big)^n}{n!}.
		\end{align}
		An identical result has been reproduced in a completely different approach from the small $\frac{\beta}{r}$ expansion of the free energy of  a massive scalar in the massless limit \cite{David:2024naf}\footnote{See Appendix B of \cite{David:2024naf}.}. One can easily obtain the low temperature or large $\frac{\beta}{r}$ expansion of the thermal expectation of energy of the free theory by differentiating the above expression with respect to $\beta$, with an overall negative sign as shown below
		\begin{align}\label{en den free th}
			\langle E\rangle_{\rm free}=\frac{4 r^2 \zeta (3)}{\beta ^3}+\frac{1}{12 \beta }+\sum_{n\in 2 \mathbb{Z}} \frac{(-1)^{n} \left(2-2^{-n}\right) \zeta (-n-1) \zeta (1-n) (\frac{\beta }{r})^{n-1}}{r (n-1)!}.
		\end{align}

		\section{The interacting theory on $S^1\times S^2$}
		\label{sec 3}
		The action describing the  $O(N)$ model for $N$ real scalar fields interacting through a quartic coupling term was given by \eqref{action intro},  as written below
		\begin{align}\label{quartic action}
			S[\phi]=\frac{1}{2}\int d^3x \sqrt{g} [\partial^\mu\phi_i \partial_\mu\phi_i +\frac{1}{4r^2}\phi_i\phi_i+\frac{\lambda}{N}(\phi_i\phi_i)^2],
		\end{align}
		where $i=1,\cdots, N$; $\lambda$ characterizes the coupling strength of the quartic interaction. The model  admits a solvable limit at $N\to\infty$. Similar to the case of free theory, the partition function of the model on $S^1\times S^2$ is given by the Euclidean path integral of the above action as follows
		\begin{align}
			\tilde Z=\int_{S^1\times S^2} {\cal D}\phi e^{-S[\phi]}.
		\end{align}
		The path integral for the partition function in the above expression  is over all the $N$ scalar  fields on the geometry of $S^1\times S^2$. The use of the standard trick of Hubbard-Stratanovich transformation  linearises the quartic interaction term in the action by introducing an auxiliary field $\zeta$ in the path integral as presented in the following equation 
		\begin{align}\label{intro of zeta}
			\tilde{Z}=\int_{S^1\times S^2}^{} {\cal D}\phi {\cal D}\zeta e^{-\frac{1}{2}\int_0^\beta\int d\tau d^2x [\partial_\mu\phi_i\partial_\mu\phi_i+\frac{1}{4r^2}\phi_i\phi_i+\frac{\zeta^2N}{4\lambda}+i\zeta\phi_i\phi_i]}.
		\end{align}
		In this representation the action turns out to be quadratic in the field $\phi$.
		We isolate the zero mode of the auxiliary field $\zeta$ from the non-zero modes of it by using the definition
		\begin{align}\label{zero+nz}
			\zeta=\zeta_0+\tilde \zeta,
		\end{align}
		where $\zeta_0$ stands for the zero mode of the field $\zeta$, and $\tilde \zeta$ denotes non-zero modes of $\zeta$. Now at the leading order in large $N$ only the zero mode of the auxiliary field survives as shown in appendix \ref{appen C} and it plays the role of a mass in the action. The contribution due to the non-zero modes arises in the systematic corrections to the partition function which are suppressed compared to the leading term at large $N$. In our work we will focus only on the leading order in large $N$, thus non-zero modes of $\zeta$ will not occur in our study. As a result of this, we have the partition function at the leading order in large $N$ given by the following
		\begin{align}\label{path int to int}
			\tilde Z=\int d\zeta_0 \exp\Big[-4\pi r^2 \beta N\Big(\frac{\zeta_0^2}{8\lambda}-\frac{1}{4 \pi r^2\beta}\log Z(\zeta_0,\frac{\beta}{r})\Big)\Big],
		\end{align}
		with
		\begin{align}\label{log Z free}
			\log Z(\zeta_0,\frac{\beta}{r})=-\frac{1}{2} \sum_{n=-\infty}^\infty \sum_{l=0}^\infty (2 l+1) \log\bigg(\Big(\frac{2\pi  n}{\beta}\Big)^2+\frac{{(l+\frac{1}{2})^2}}{r^2}+\tilde{m}^2\bigg),
		\end{align}
		and here
		\begin{align}
			\tilde m^2=i\zeta_0.
		\end{align}
		Note that performing the path integral over field $\phi$  led to the term $\log Z(\zeta_0,\frac{\beta}{r})$. 
		The integral over $\zeta_0$ 
		in \eqref{path int to int} will be done 
		using the saddle point approximation. The  thermal mass $\tilde m$ at the saddle point satisfies the following condition known as the gap equation at $\lambda\to \infty$ \cite{Sachdev:1993pr},
		\begin{align}\label{saddle pt cond}
			\partial_{\tilde m}\log \tilde Z(\tilde m)=0.
		\end{align}
		First, we need to simplify the expression for the $\log Z$ given in \eqref{log Z free} and make it suitable to solve for $\tilde m$. This can be done expanding  it either as a  high temperature or low temperature expansion and solving (\ref{saddle pt cond}) perturbatively in either of these expansions.  We perform the sum over the Matsubara frequencies in equation \eqref{log Z free} using the formula for the regulated sum given in  \eqref{regulated sum} in the similar manner as was done for the free theory, we have
		\begin{align}\label{logZ after Mat sum}
			\log Z&=-\frac{1}{2} \sum_{l=0}^\infty (2l+1) \Big(\beta\sqrt{\frac{(l+\frac{1}{2})^2}{r^2}+\tilde m^2}+2 \log \Big[1-e^{-\beta\sqrt{\frac{(l+\frac{1}{2})^2}{r^2}+\tilde m^2}}\Big]\Big)\nonumber,\\
			&\equiv \log Z_1(-\frac{1}{2})+\log Z_2.
		\end{align}
		It is important to emphasize, 
		the  above expression is not known to admit a closed-form representation in general.
		But it can be expressed as a systematic order by order expansion at small $\frac{\beta}{r}$ using the techniques developed for evaluating the sum over the angular modes $l$ in \cite{David:2024naf}. We will review the free energy for the $O(N)$ model on $S^1\times S^2 $ with the systematic finite size corrections in powers of $\frac{\beta}{r}$ at the strong coupling $\lambda\to \infty$, obtained in \cite{David:2024naf},  in the following subsection.

		\subsection{High temperature expansion}\label{sec 3.1}
		In \cite{David:2024naf}, we have introduced a technique  to carry out the sum over angular momentum modes $l$ by implementing a set of mathematical manipulations. This allowed us to express the right-hand side of \eqref{logZ after Mat sum} as a series expansion in powers of $\frac{\beta}{r}$, as given below
		\begin{align}\label{small beta log Z}
			\log Z=&\frac{4\pi r^2}{\beta^2} \sum_{p=0}^\infty (\frac{\beta}{r})^{2p} (-1)^{p+1} \frac{(2^{2p-1}-1)}{4\pi} B_{2p} \nonumber\\
			&	\times \Big(\frac{\Gamma(p-\frac{3}{2})\Gamma(p+\frac{1}{2})}{2\pi (2p)!(\tilde m\beta)^{2p-3}}+\sum_{j=0}^{|p-\frac{3}{2}|-\frac{1}{2}}\frac{(|p-\frac{3}{2}|-j+\frac{1}{2})_{2j}{\rm Li}_{j-p+2}(e^{-\tilde m\beta})}{2^{j+3p-2}j!\Gamma(p+1)(\tilde m\beta)^{j+p-1}}\Big).
		\end{align}
		Where $B_{p}$ denotes the Bernoulli numbers. 
		An alternative derivation of the 2nd term in the above expression is presented in Appendix \ref{appn B}, using a similar technique used for the free theory in Section \ref{sec 2}, combined with the method to sum over angular modes on $S^2$ \cite{David:2024pir}.\\
		
		Now using this result, the saddle point condition \eqref{saddle pt cond} at $\lambda\to \infty$ can be given by the following equation, called the gap equation\footnote{Note that there were typos in the gap equation given in equation (3.11) of \cite{David:2024naf}. It is corrected in this paper.}
		\begin{align}\label{gap eq high temp}
			&	\sum_{p=0}^\infty(-1)^p(4^p-2)B_{2p} \Big(\frac{\beta}{r}\Big)^{2p-2} \Big[\frac{\Gamma(p-\frac{1}{2})\Gamma(p+\frac{1}{2})}{2\pi (2p)!}(\tilde m\beta)^{2-2p}\nonumber\\
			&+\sum_{j=0}^{|p-\frac{3}{2}|-\frac{1}{2}}\frac{(|p-\frac{3}{2}|-j+\frac{1}{2})_{2j}}{j!p! 2^{j+3p-1} (\tilde m\beta)^{j+p}}
			\big(\beta  \tilde m \text{Li}_{j-p+1}(e^{-m \beta })+(j+p-1) \text{Li}_{j-p+2}(e^{-\tilde m \beta })\big)\Big]=0.
		\end{align}
		Now, one can easily verify that $\tilde m$ satisfying the above equation has the  Taylor series expansion  in $\frac{\beta}{r}$ as given below
		\begin{align}\label{thermal mass expns}
			\tilde m=\frac{1}{\beta} \Big(2\log \frac{1+\sqrt{5}}{2}+\frac{\beta^2}{r^2}\frac{1}{48 \csch^{-1}2}+\frac{\beta^4}{r^4} \frac{55+64\sqrt{5}\csch^{-1}2}{230400(\csch^{-1}2)^3}+O(\frac{\beta^6}{r^6})\Big).
		\end{align}
		Note that, for illustration, here we present the solution for $\tilde m$ up to a few orders in powers of $\frac{\beta}{r}$, but we have solved it till a very high orders of $O(\frac{\beta^{34}}{r^{34}})$ given later in \eqref{thermal mass numeric}.
		
		Now substituting the saddle point value of the thermal mass $\tilde m$ as given in \eqref{thermal mass expns}, in the equation \eqref{small beta log Z} we obtain the free energy at the non-trivial fixed point as a Taylor series expansion in $\frac{\beta}{r}$
		\begin{align}\label{high temp expns logZ 1}
			\log Z(\frac{\beta}{r})=\frac{4\pi r^2}{\beta^2} \Big(\frac{2}{5\pi}\zeta(3)-\frac{\beta^4}{r^4}\frac{1}{576\sqrt{5}\csch^{-1}2}+O(\frac{\beta^6}{r^6})\Big).
		\end{align}
		Thus one also has  the thermal expectation of energy given by a similar expansion as follows
		\begin{align}
			\langle E\rangle_\beta=-\partial_\beta \log Z(\tilde m,\frac{\beta}{r})=\frac{4\pi r^2}{\beta^3}\Big(\frac{4}{5\pi} \zeta(3)-\frac{\beta^4}{r^4 288 \sqrt{5}\csch^{-1}2}+O(\frac{\beta^6}{r^6})\Big).
		\end{align}
		We have also  given high temperature expansion of the $\log Z$ and $\langle E\rangle_\beta$ to very high orders of $O(\frac{\beta^{32}}{r^{32}})$ in \eqref{log Z numeric} and \eqref{E numeric} respectively.	We have computed the pressure from $\log Z$ by differentiating it with respect to the volume of the sphere $S^2$ in \cite{David:2024pir}.
		\begin{align}
			P=\frac{1}{4\pi \beta}\frac{\partial \log Z}{\partial r^2}.
		\end{align}
		And it can be easily checked that the trace of the stress tensor vanishes in each perturbative order.

		\subsection{Low temperature expansion}\label{sec 3.2}
		In this subsection, we will examine the model on $S^1\times S^2$ in the limit $\frac{\beta}{r}\to\infty$. First, we will evaluate $\log Z$ given in \eqref{logZ after Mat sum} as an expansion at large $\frac{\beta}{r}$. Then we will find the thermal mass satisfying the gap equation \eqref{saddle pt cond} at this limit. Similar to the case at small $\frac{\beta}{r}$, we can evaluate the free energy for the critical $O(N)$ model on $S^1\times S^2$ at $\frac{\beta}{r}\to \infty$, by substituting the saddle point value of the thermal mass in the integrand of the equation \eqref{path int to int}.
		
		Let us now consider the first term from the equation \eqref{logZ after Mat sum}, it is diverging in general. But we use a prescription \cite{Giombi:2019upv} to define this sum as an analytic continuation of the sum $\log Z_1(\alpha)$ where $\alpha>1$, as given in the following. 
		\begin{eqnarray}
			\log Z_1 (\alpha) &=& 
			- \frac{1}{2}\sum_{l=0}^{\infty} (2l+1) 
			\Big({\frac{(l+\frac{1}{2})^2\beta^2}{r^2}+{\tilde m}^2\beta^2}\Big)^{-\alpha}, \\ \nonumber
			&=&-\frac{1}{2\Gamma(\alpha)} \sum_{l=0}^\infty (2l+1) \int_0^\infty {d\tau \tau^{\alpha-1}}e^{-[\frac{(l+\frac{1}{2})^2\beta^2}{r^2}+{\tilde  m}^2\beta^2]\tau}.
		\end{eqnarray}
		In the second line of the above equation, this is expressed as  an integral representation which is straightforward to verify.
		At large $\frac{\beta}{r}\to \infty$, if $\tilde m\beta$ is finite and $\tilde m^2<\frac{1}{4r^2}$, one can expand the exponential in the above expression in the following manner
		\begin{align}\label{Z1 result}
			\log Z_1 (\alpha) &=-\frac{1}{2\Gamma(\alpha)} \sum_{l,n=0}^\infty (2l+1)\frac{(-1)^n}{n!} (\tilde m^2\beta^2)^n \int_0^\infty {d\tau \tau^{\alpha+n-1}} e^{-\frac{\beta^2}{r^2}(l+\frac{1}{2})^2\tau},\\
			&=-\sum_{l,n=0}^\infty \frac{(-1)^n2^{2\alpha+2n-1}\Gamma(n+\alpha)}{\Gamma(\alpha) n!(2l+1)^{2\alpha+2n-1}} (\beta^2\tilde m^2)^n (\frac{r}{\beta})^{2\alpha+n}.
		\end{align}
		Finally with $\alpha=-\frac{1}{2}$, and performing the sum over $l$ from $0$ to $\infty$, we obtain
		\begin{align}\label{sum from n=2}
			\log Z_1(-\frac{1}{2})
			&=\sum_{n=2}^\infty\frac{\Gamma(n-\frac{1}{2})}{2\sqrt{\pi}n!} (-\beta^2\tilde m^2)^n (\frac{r}{\beta})^{2n-1} (2^{2n-2}-1) \zeta(2n-2).
		\end{align}
		Note that after summing over $l$, the terms due to $n=0,1$ turn out to be vanishing using the formula as given below
		\begin{align}
			\sum _{l=0}^{\infty } (2 l+1)^{2-2 n}=	4^{-n} \left(4^n-4\right) \zeta (2 n-2).
		\end{align} 
		Thus, the sum over $n$ in the equation \eqref{sum from n=2} starts from $n=2$. 
		We show that this low temperature expansion of $\log Z_1(-1/2)$  reproduces the high temperature expansion given in the first term of equation \eqref{small beta log Z}  with the use of techniques closely related to the technique of Borel re-summation in Appendix \ref{app A}. This justifies the validity of analytic continuations used in evaluating \eqref{sum from n=2}.\\
		
		The second term in the equation \eqref{logZ after Mat sum} is convergent and can be organized as a series expansion at large $\frac{\beta}{r}$ following the steps as described below. One can get rid of the square root in the exponent by recasting the exponential in terms of an integral representation, as demonstrated below.
		\begin{align}
			\log Z_2 &= 
			2 \sum_{l=0}^\infty \Big(l+\frac{1}{2}\Big) \sum_{n=1}^\infty \frac{e^{-n\beta\sqrt{\frac{(l+\frac{1}{2})^2+}{r^2}  \tilde m^2}}}{n}, \\
			\nonumber
			&=\frac{1}{\sqrt{\pi} } \sum_{l=0}^\infty \sum_{n=1}^\infty 
			\frac{1}{n}\Big(l+\frac   {1}{2}\Big) \int_0^\infty \frac{d\tau}{\tau^{3/2}}e^{-\tau n^2\beta^2[\frac{(l+\frac{1}{2})^2}{r^2}+ \tilde m^2]-\frac{1}{4\tau}}. \\ \nonumber
		\end{align}
		Expanding the exponential $e^{-\tau n^2\beta^2\tilde m^2}$ for a small value of the argument such that $\tilde m^2<\frac{1}{4r^2}$ and performing the integral over $\tau$ in each term order by order
		\begin{align}\label{log Z2 sum}
			\log Z_2
			&=\frac{1}{\sqrt{\pi}}\sum_{l,p=0,n=1}^\infty \frac{(-\tilde m^2\beta^2)^p}{(2l+1)^{p-\frac{3}{2}}p!} \big(\frac{n r}{\beta}\big)^{p-\frac{1}{2}} K_{p-\frac{1}{2}}\big(\frac{\left(l+\frac{1}{2}\right) n \beta }{r}\big) 
			\nonumber,\\
			&=\sum_{l,p=0,n=1}^\infty \frac{(-\tilde m^2\beta^2)^p e^{-\frac{\beta n (l+\frac{1}{2}) }{r}}}{(2l+1)^{p-\frac{3}{2}}p!} \big(\frac{n r}{\beta}\big)^{p-\frac{1}{2}}
			\sum _{k=0}^{\left| \frac{1}{2}-p\right| +\frac{1}{2}} \frac{ (\left| \frac{1}{2}-p\right| +\frac{1}{2}-k)_{2 k}  }{ k! n^{k+\frac{1}{2}} (2l+1)^{k+\frac{1}{2}}} \big(\frac{r}{\beta}\big)^{k+\frac{1}{2}}.
		\end{align}
		Now by combining \eqref{sum from n=2} and \eqref{log Z2 sum} we can obtain $\log Z$ as given  according to \eqref{logZ after Mat sum}
		\begin{align}\label{pat fn at low temp}
			\log Z(\tilde m,\frac{\beta}{r})&=\sum_{n=2}^\infty\frac{\Gamma(n-\frac{1}{2})}{2\sqrt{\pi}n!} (-\beta^2\tilde m^2)^n (\frac{r}{\beta})^{2n-1} (2^{2n-2}-1) \zeta(2n-2)\nonumber\\
			&	+\sum_{l,p=0,n=1}^\infty \frac{(-\tilde m^2\beta^2)^p e^{-\frac{\beta n (l+\frac{1}{2}) }{r}}}{(2l+1)^{p-\frac{3}{2}}p!} \big(\frac{n r}{\beta}\big)^{p-\frac{1}{2}}
			\sum _{k=0}^{\left| \frac{1}{2}-p\right| +\frac{1}{2}} \frac{ (\left| \frac{1}{2}-p\right| +\frac{1}{2}-k)_{2 k}  }{ k! n^{k+\frac{1}{2}} (2l+1)^{k+\frac{1}{2}}} \big(\frac{r}{\beta}\big)^{k+\frac{1}{2}}.
		\end{align}
		Note that the first term in the above equation is a series expansion in powers of $\frac{r}{\beta} $, while the second term admits an expansion in terms of decaying exponentials, each multiplied by algebraic powers of $\frac{r}{\beta}$.		We have evaluated the free energy for a scalar of mass $\tilde m$ on $S^1\times S^2$ as a series expansion at large $\frac{\beta}{r}$. Now one can find the critical point of the theory of $O(N) $ model on this geometry by evaluating the saddle point condition of the integral \eqref{path int to int} at large $N$  and at  $\lambda\to\infty$. The thermal mass $\tilde m$, characterizing the critical point, satisfies the following gap equation. 
		\begin{align}
			\partial_{\tilde m}\log Z(\tilde m,\frac{\beta}{r})=0.
		\end{align}
		Using \eqref{pat fn at low temp} in the above equation, we have the gap equation organized as a systematic series expansion valid at large $\frac{\beta}{r}$ to be 
		\begin{align}
			&	\sum_{n=2}^\infty\frac{\Gamma(n-\frac{1}{2})}{\sqrt{\pi}(n-1)!} (-\beta^2)^n \tilde m^{2n-1} (\frac{r}{\beta})^{2n-1} (2^{2n-2}-1) \zeta(2n-2)
			\nonumber\\
			&	+\sum_{l,p=0,n=1}^\infty \frac{2(-\beta^2)^p \tilde m^{2p-1} e^{-\frac{\beta n (l+\frac{1}{2}) }{r}}}{(2l+1)^{p-\frac{3}{2}}(p-1)!} \big(\frac{n r}{\beta}\big)^{p-\frac{1}{2}}
			\sum _{k=0}^{\left| \frac{1}{2}-p\right| +\frac{1}{2}} \frac{ (\left| \frac{1}{2}-p\right| +\frac{1}{2}-k)_{2 k}  }{ k! n^{k+\frac{1}{2}} (2l+1)^{k+\frac{1}{2}}} \big(\frac{r}{\beta}\big)^{k+\frac{1}{2}}=0.
		\end{align}
		We can solve the thermal mass $\tilde m$ as a large $\frac{\beta}{r}$ expansion satisfying this gap equation in each systematic order of the expansion, has the following form 
		\footnote{ The solution for the $\tilde m$ \eqref{m at low temp} is given till a very high order of $O(e^{-\frac{21 \beta }{4 r}})$  in the ancillary file  \texttt{low\_temp\_expansions.txt} attached to the arXiv version of this paper. In this file, $\log Z$ \eqref{log Z interacting} and the thermal expectation of energy \eqref{en den interacting} are also provided till $O(e^{-\frac{6 \beta }{r}})$. }
		{	\begin{align}\label{m at low temp}
				\tilde m&=\frac{2 \sqrt{2} e^{-\frac{\beta }{4 r}}}{\pi  r}+	{e^{-\frac{3 \beta }{4 r}} \Big(-\frac{8 \sqrt{2} \beta }{\pi ^3 r^2}+\frac{\sqrt{2} \left(3 \pi ^2-16\right)  }{\pi ^3 r}\Big)}\nonumber\\
				& 	+e^{-\frac{5 \beta }{4 r}} \Big(\frac{80 \sqrt{2} \beta ^2}{\pi ^5 r^3}-\frac{12 \sqrt{2} \left(5 \pi ^2-32\right) \beta }{\pi ^5 r^2}+\frac{6912-1248 \pi ^2+77 \pi ^4}{6 \sqrt{2} \pi ^5 r}\Big)\nonumber\\
				& +e^{-\frac{7\beta}{4r}} \Big(-\frac{3136 \sqrt{2} \beta ^3}{3 \pi ^7 r^4}+\frac{168 \left(7 \pi ^2-48\right) \sqrt{2} \beta ^2}{\pi ^7 r^3}-\frac{\sqrt{2} \left(72960-14784 \pi ^2+853 \pi ^4\right) \beta }{3 \pi ^7 r^2}\nonumber\\
				&\qquad+\frac{-2027520+458496 \pi ^2-32048 \pi ^4+803 \pi ^6}{36 \sqrt{2} \pi ^7 r}\Big)\nonumber\\
				&+e^{-\frac{9\beta}{4r} }\Big(\frac{15552 \sqrt{2} \beta ^4}{\pi ^9 r^5}-\frac{2592 \sqrt{2} (9 \pi ^2-64) \beta ^3}{\pi ^9 r^4}+\frac{2 \sqrt{2} \beta ^2 (3379968-723168 \pi ^2+42523 \pi ^4)}{9 \pi ^9 r^3}\nonumber\\
				&\qquad-\frac{\left(-30449664+7734528 \pi ^2-647072 \pi ^4+19047 \pi ^6\right) \beta }{9 \sqrt{2} \pi ^9 r^2}\nonumber\\
				&\qquad+\frac{13686865920-3784458240 \pi ^2+360317440 \pi ^4-13248960 \pi ^6+143997 \pi ^8}{4320 \sqrt{2} \pi ^9 r}\Big)\nonumber\\
				&+O(e^{-\frac{11\beta}{4r}}).
		\end{align}}
		Let us provide a heuristic reasoning to justify the above expansion. 
		At low temperatures, but say fixed $r$, it is natural the thermal mass is measured in units of $r$, then the  solution one may expect is 
		\begin{eqnarray} \label{formalmsol}
			\tilde m r = f_0(r T)  + \exp ( -\beta E_1/r) f_1(rT)  + \exp ( -\beta E_2/r) f_2(rT) + \cdots 
		\end{eqnarray}
		where $f_i(rT)$ are functions which admit a Taylor series expansion, $E_1,E_2,\cdots$ are energy levels.  We have organized the solution for the thermal mass as a general expansion of perturbative and non-perturbative  terms which go as 
		$\big( \exp ( -\frac{\beta E_i}{r} ) \big)$ for $i=1,2,\cdots$.  The non-perturbative terms have to arise since at low temperatures we can expand partition functions  of systems in a compact geometry such as a sphere 
		in terms of Boltzmann weights $\exp(-\beta E) $, with 
		$E$ being energy levels.
		One point to note is that it must be the case, the leading perturbative solution has  to be of the form
		\begin{eqnarray}
			f_0(0) = 0
		\end{eqnarray}
		If this were not so, then there would be a mass gap in the theory characterized by $\tilde m = f(0)/r$
		at zero temperature. The theory then  would not be conformal. 
		The explicit result in (\ref{m at low temp}) indeed bears out this expectation. 
		The thermal mass therefore must vanish at zero temperature. 
		As we will see subsequently, this leads to the fact that strictly at 
		zero temperature, the free energies  at the IR fixed point and the free Gaussian fixed point must coincide since the Gaussian theory is massless. 
		
		Now to obtain the expansions \eqref{Z1 result} and \eqref{log Z2 sum},  we have used $\tilde m>\frac{1}{4r^2}$. Thus by substituting the leading order term for $\tilde m$ from the above expression we can re-express this inequality as
		\begin{align}
			\frac{\beta}{r}>2\log (\frac{32}{\pi^2}).
		\end{align}
		This gives the region where the solution \eqref{m at low temp} is consistently valid.
		Substituting the expansion for thermal mass \eqref{m at low temp} in equation \eqref{pat fn at low temp} yields  the free energy as an expansion valid at large $\frac{\beta}{r}$ 
		{	\begin{align}\label{log Z interacting}
				&	\log Z=\log (Z_{\rm free})-\frac{4 \beta  e^{-\frac{\beta }{r}}}{\pi ^2 r}+e^{-\frac{3 \beta }{2 r}} \Big(\frac{32 \beta ^2}{\pi ^4 r^2}+\frac{8 \left(24-5 \pi ^2\right) \beta }{3 \pi ^4 r}\Big)\nonumber\\
				&+e^{-\frac{2 \beta }{r}} \Big(-\frac{1024 \beta ^3}{3 \pi ^6 r^3}+\frac{256 \left(\pi ^2-6\right) \beta ^2}{\pi ^6 r^2}\nonumber
				-\frac{4 \left(1536-336 \pi ^2+31 \pi ^4\right) \beta }{3 \pi ^6 r}\Big)\\
				&      +e^{-\frac{5 \beta }{2 r}} \Big(\frac{12800 \beta ^4}{3 \pi ^8 r^4}-\frac{1280 \left(11 \pi ^2-72\right) \beta ^3}{3 \pi ^8 r^3}
				+\frac{32 \left(7872-1776 \pi ^2+125 \pi ^4\right) \beta ^2}{3 \pi ^8 r^2}\nonumber\\
				&+\frac{8 \left(483840-126720 \pi ^2+12040 \pi ^4-541 \pi ^6\right) \beta }{45 \pi ^8 r}\Big) +O(e^{-\frac{3\beta}{r}}),
		\end{align}}
		where $\log (Z_{\rm free})$ is the logarithm of the free partition function obtained in \eqref{log Z free at low temp}
		\begin{align} \label{log Z free at low temp 1}
			\log (Z_{\rm free})=\sum_{l=0,n=1}^\infty(2l+1)\frac{e^{-\frac{n\beta}{r}(l+\frac{1}{2})}}{n}.
		\end{align}  
		Note the presence of the polynomials in powers of $\frac{\beta}{r}$ multiplied with each of the exponential terms in the expansion for the free energy for $O(N) $ model at large $\frac{\beta}{r}$ given in \eqref{log Z interacting}, in contrast to the structure of the expansion obtained for the free CFT \eqref{log Z free at low temp 1}. Such a structure of the expansion in terms of exponentials enveloped with polynomials is inherited from the structure of the expansion of the thermal mass $\tilde m$ satisfying the gap equation as given in \eqref{m at low temp}.
		Similarly, we can evaluate the thermal expectation of energy as a   derivative of the $\log Z$ with respect to $\beta$ and it is given by
		\begin{align}\label{en den interacting}
			&	\langle E\rangle=		-\partial_\beta \log Z\\
			&=E_{\rm free}+e^{-\frac{\beta }{r}} \left(\frac{4}{\pi ^2 r}-\frac{4 \beta }{\pi ^2 r^2}\right)+e^{-\frac{3 \beta }{2 r}} \left(\frac{48 \beta ^2}{\pi ^4 r^3}-\frac{20 \beta }{\pi ^2 r^2}+\frac{32 \beta }{\pi ^4 r^2}+\frac{40}{3 \pi ^2 r}-\frac{64}{\pi ^4 r}\right)\nonumber\\&
			+e^{-\frac{2 \beta }{r}} \left(-\frac{2048 \beta ^3}{3 \pi ^6 r^4}+\frac{512 \beta ^2}{\pi ^4 r^3}-\frac{2048 \beta ^2}{\pi ^6 r^3}-\frac{248 \beta }{3 \pi ^2 r^2}+\frac{384 \beta }{\pi ^4 r^2}-\frac{1024 \beta }{\pi ^6 r^2}+\frac{124}{3 \pi ^2 r}-\frac{448}{\pi ^4 r}+\frac{2048}{\pi ^6 r}\right)\nonumber\\&
			+e^{-\frac{5 \beta }{2 r}} \Big(\frac{32000 \beta ^4}{3 \pi ^8 r^5}-\frac{35200 \beta ^3}{3 \pi ^6 r^4}+\frac{179200 \beta ^3}{3 \pi ^8 r^4}+\frac{10000 \beta ^2}{3 \pi ^4 r^3}-\frac{33280 \beta ^2}{\pi ^6 r^3}+\frac{117760 \beta ^2}{\pi ^8 r^3}\nonumber\\
			&-\frac{2164 \beta }{9 \pi ^2 r^2}+\frac{24160 \beta }{9 \pi ^4 r^2}-\frac{18432 \beta }{\pi ^6 r^2}+\frac{47104 \beta }{\pi ^8 r^2}+\frac{4328}{45 \pi ^2 r}-\frac{19264}{9 \pi ^4 r}+\frac{22528}{\pi ^6 r}-\frac{86016}{\pi ^8 r}\Big)+O(e^{-\frac{3\beta}{r}}),
		\end{align}
		Again	$E_{\rm free}$ denotes the thermal expectation of energy of the free CFT obtained in \eqref{E free}
		\begin{align}
			E_{\rm free }=\sum_{l=0,n=1}^\infty\frac{(2 l+1)^2 }{2 r}e^{-\frac{\beta  (l+\frac{1}{2}) n}{r}}.
		\end{align}
		These results are organized as a systematic series expansion in $e^{-\frac{\beta}{2r}}$. It  will be interesting to express the partition function and thermal expectation of energy at the non-trivial fixed point of the model, obtained at an infinite coupling, in terms of
		of conformal characters \cite{Buric:2024kxo}.
		The pressure can be evaluated as 
		{\small	\begin{align}
				P=P_{\rm free}+e^{-\frac{\beta }{r}} \Big(\frac{1}{2 \pi ^3 r^3}-\frac{\beta }{2 \pi ^3 r^4}\Big)+e^{-\frac{3 \beta }{2 r}} \Big(\frac{6 \beta ^2}{\pi ^5 r^5}-\frac{\left(5 \pi ^2-8\right) \beta }{2 \pi ^5 r^4}+\frac{5 \pi ^2-24}{3 \pi ^5 r^3}\Big)\nonumber\\
				+e^{-\frac{2 \beta }{r}} \Big(-\frac{256 \beta ^3}{3 \pi ^7 r^6}+\frac{64 \left(\pi ^2-4\right) \beta ^2}{\pi ^7 r^5}-\frac{\left(384-144 \pi ^2+31 \pi ^4\right)
					\beta }{3 \pi ^7 r^4}+\frac{1536-336 \pi ^2+31 \pi ^4}{6 \pi ^7 r^3}\Big)\nonumber\\
				+e^{-\frac{5 \beta }{2 r}} \Big(\frac{4000 \beta ^4}{3 \pi ^9 r^7}-\frac{400 \left(11 \pi ^2-56\right) \beta ^3}{3 \pi ^9 r^6}+\frac{10 \left(4416-1248 \pi ^2+125
					\pi ^4\right) \beta ^2}{3 \pi ^9 r^5}\nonumber\\
				-\frac{\left(-105984+41472 \pi ^2-6040 \pi ^4+541 \pi ^6\right) \beta }{18 \pi ^9 r^4}+\frac{-483840+126720 \pi ^2-12040 \pi
					^4+541 \pi ^6}{45 \pi ^9 r^3}\Big)\nonumber\\
				+O(e^{-\frac{3\beta}{r}})
		\end{align}}
		where $P_{\rm free}$ is the pressure for the free CFT on $S^1\times S^2$ as given below
		\begin{align}
			P_{\rm free}=\sum_{l=0,n=1}^\infty	\frac{(2 l+1)^2 }{16 \pi  r^3}e^{-\frac{\beta  \left(l+\frac{1}{2}\right) n}{r}}
		\end{align}
		Using the above expression for pressure and the thermal expectation of energy \eqref{en den interacting}, one can easily verify the tracelessness of the stress tensor at each order in the exponentials. 
		
		%

		\section{High to low temperature expansion: Borel-Pad\'{e} sum}\label{sec 4}
		
		In section \ref{sec 2} we have reproduced the small $\frac{\beta}{r}$ expansion of the free energy for the free CFT directly from the re-summation of its large $\frac{\beta}{r}$ expansion, confirming the equality of the two expansions.
		In section \ref{sec 3} we have evaluated the free energy for the critical $O(N) $ model on $S^1\times S^2$ as a small $\frac{\beta}{r} $ expansion \eqref{high temp expns logZ 1} and large $\frac{\beta}{r}$ expansion \eqref{log Z interacting} separately.  The expansions \eqref{high temp expns logZ 1} and \eqref{log Z interacting} are obtained by solving the gap equation order by order in either of the  expansions.  At present we do not have a general formula 
		for each term of these expansions, unlike the case 
		in the  free theory.  Due to the absence of the general formula we cannot re-sum one of the expansions say the low temperature expansion and re-cast it as a high temperature expansion which can  be done for the free theory. 
		Since such a re-summation cannot be done analytically,  this implies that we cannot be sure that 
		these expansions have picked up the same solution branch of the gap equation (\ref{saddle pt cond}). 
		Furthermore, both the expansions were obtained using several mathematical manipulations and analytical continuations. There were  tests of these methods 
		before the use of the solution of the gap equation that the low temperature expansion can be re-summed and cast as a high temperature expansion in appendix \ref{app A} and \ref{appn B}. However as we have emphasized 
		we do not at present have such methods to  re-sum the expansions  after using the solution of the gap equation. 
		
		Therefore to demonstrate the consistency of both these expansions as well as the fact that we are in the 
		same branch of the solution of the gap equation we will examine these expansions numerically. 
		We will use a method of Borel-Pad\'{e} sum to extrapolate the small $\frac{\beta}{r}$
		expansion to finite values of $\frac{\beta}{r}$.
		We then show that the extrapolated function overlaps the large $\frac{\beta}{r}$ expansion over a certain range of $\frac{\beta}{r}$. This subsection is organised as follows. 
		First, we will illustrate our method of Borel-Pad\'{e} sum in general. Then we will apply this method in the free theory where the equality between the expansions at small and large $\frac{\beta}{r}$ has been analytically established. This will serve as a test of out method of Borel-Pad\'{e} re-summation. Finally,
		we will use this method to demonstrate the equality of the small and large $\frac{\beta}{r}$ expansions in the $O(N)$ model.
		\subsection{The method}
		
		Let us discuss the general method of re-summing asymptotic series using the Borel-Pad\'{e} method. 
		Consider a function $f(x)$ given by the following expansion which is asymptotic in nature.
		\begin{align}
			f(x)=\sum_{n=0}^\infty a_n x^n,
		\end{align}
		We assume that the asymptotic series has its coefficients growing factorially with $n $ as $\lim_{n\to\infty}a_n\sim n!$. Thus, the series has zero radius of convergence. The series can  approximate the function only very close to $x\to 0$. 
		
		Evaluating such a sum appears to be a challenging task.
		The technique of Borel re-summation is the most used tool to compute the sum of such a series expansion. The Borel transform of the series  defines a new series where the factorial growth of the coefficients in the original series  is removed by dividing each $n$-th   coefficient $a_n$ by $n!$ as shown below
		\begin{align}\label{borel trans}
			{\cal B} f(x)=\sum_{n=0}^\infty \frac{a_n}{n!} x^n,
		\end{align}
		where ${\cal B}f(x)$ denotes the Borel transform of $f(x)$. The Borel transformed series should be a converging series, as the factorial growth of the coefficients is canceled.  Now, the original asymptotic series is given by the Laplace transform of its Borel transform, as demonstrated below
		\begin{align}\label{laplace}
			f(x)=       {\cal L}[ {\cal B} f(x)]=\int_0^\infty e^{-t} {\cal B}f(xt) dt.
		\end{align}
		Now, if the Borel transformed series has a known closed form expression, one can perform the Laplace transform of ${\cal B}f(x) $ to obtain the sum of the original asymptotic series $f(x) $.
		A very familiar example in this context is the series $\sum_{n=0}^\infty n! x^n$, having coefficients increasing factorially with $n$. One can evaluate the sum of this series by the use of the formula \eqref{laplace} and obtain a closed form expression as $ e^{-1/x}{\rm Ei}(1/x)$, where ${\rm Ei}(x)$ is the exponential integral function. We will elaborate on this example, given the conceptual relevance of this simple example to our problem. So, let us begin with 
		\begin{align}
			f(x)=\sum_{n=0}^\infty n! x^n.
		\end{align}
		The Borel transform of the series 
		\begin{align}
			{\cal B}f(x)=\sum_{n=0}^\infty  x^n=\frac{1}{1-x}.
		\end{align}
		Finally in  the Laplace transform of $f(x t)$, the integrand admits simple pole at $t=\frac{1}{x}$  which obstructs the contour of  the integration along the real line from $0$ to $\infty$. In this case we follow the principal value prescription to perform the integral of the Laplace transform avoiding the pole at $t=\frac{1}{x}$
		\begin{align}\label{expint}
			f(x)= {\rm P.V}  \int_0^\infty e^{-t} \frac{1}{1-x t} dt=e^{-1/x}{\rm Ei}(\frac{1}{x}).
		\end{align}
		Note that any other choice of the contour will lead to an answer to this integral with a non-zero imaginary part.
		
		But in many of the examples  we encounter  in Physics, 
		the asymptotic series is  known only till a finite number of terms and still we have to find its behavior away from $x=0$. The series representation miserably fails even slightly away from $x=0$. In this case, we need a method that can approximate the function just from the knowledge of a finite number of terms of an infinite series. The method we are going to use is known as Borel-Pad\'{e} sum can extrapolate the value of the function away from $x=0$.
		Let us consider the following series expansion of a function $ f(x)$ where only a  finite number of terms are known to us
		\begin{align}
			f(x)=\sum_{n=0}^N a_n x^n+O(x^{N+1}).
		\end{align}
		where the higher order terms $O(x^{N+1})$ are not known.
		Again, the coefficients of $x^n$ in this series grow factorially as $\lim_{n\to N} a_n\sim n!$. The first step just resembles the Borel transform used in \eqref{borel trans}
		for this series with a finite number of terms
		\begin{align}\label{Borel trans}
			{\cal B}f(x) =\sum_{n=0}^N \frac{a_n}{n!} x^n+O(x^{N+1}).
		\end{align}
		Unlike the example of the exponential integral shown before,  we cannot determine a closed-form expression for this series as only  a finite number of terms are known of this infinite series.
		At this stage, the Borel transform ${\cal B}f(x)$ can be approximated by the series \eqref{Borel trans} truncated  till $O(x^{N})$, as factorial growth of $a_n$ is removed.
		Now, we approximate this series \eqref{Borel trans} truncated till $O(x^N)$ by the use of Pad\'{e} approximation. We denote the Pad\'{e} approximation of the Borel transformed series ${\cal B}f(x)$ as  $[p,q]{\cal B}f(x)$ which approximates the series as a ratio of a polynomial of degree $p$ to a polynomial of degree $q$ as demonstrated below
		\begin{align}
			&   {[p,q]}{\cal B}f(x)=\frac{c_0+c_1 x+c_2 x^2+\cdots+c_px^p}{1+d_1 x+d_2 x^2+\cdots+d_q x^q},\qquad{\rm where}\ \ p+q=N,\ \ c_i,d_i\in R,\\
			& {\rm such\ that}\qquad {[p,q]}{\cal B}f(x)={\cal B}f(x)+O(x^{N+1}).
		\end{align}
		We will use symmetric Pad\'{e} approximations, i.e., choose $p=q$. We can write $[p,p]{\cal B}f(x)$ as a partial fractions of sum over poles of the Pad\'{e} approximant.
		\begin{align}\label{partial frac}
			[p,p]{\cal B}f(x)=c'_0+{\rm Re}\sum_{i=1}^p\frac{c'_i}{x-x_i}.
		\end{align}
		where $x_i,c_i'$  can be any complex number in general, $c_0'$ is a real number.
		
		Finally, one has to perform the Laplace transform of the Pad\'{e} approximant of the  Borel transform of the asymptotic series. The Laplace transform of the terms in \eqref{partial frac} can be performed easily if the poles do not lie on the positive real axis in the complex $t$ plane. And it is given by
		\begin{align}\label{not on R}
			\int_0^{\infty } \frac{e^{-t}}{t x-x_i} \, dt=-\frac{e^{-x_i/x} {\rm Ei} 
				(\frac{x_i}{x})}{x}, \qquad {\rm for}\ \ x_i\notin R_+.
		\end{align}
		But for the terms with poles along the positive real axis in the complex $t$ plane one has to perform the integral for the Laplace transform carefully avoiding the pole at $t=x_i/x$. 
		In this work, we will always follow the principal value prescription to avoid the pole lying on the integration contour. 
		Thus, the Laplace transform of such a term from equation \eqref{partial frac} with poles on the real positive axis in the $t$-plane can be performed using the following formula for the principal value of the integral as was used in \eqref{expint}\footnote{The correct prescription to avoid poles on the positive real axis in the $t$-plane should be chosen based on the specific asymptotic series one starts with. For our cases, the principal value prescription is the correct choice, as it keeps the free energies real-valued and results in an unambiguous   answer in Borel-Pad\'e resum, as also discussed in \cite{Dondi:2021buw}.}
		\begin{align}\label{On R}
			{\rm P.V}   \int_0^{\infty} \frac{e^{-t}}{t x-x_i} \, dt=-\frac{e^{-\frac{x_i}{x}} \text{Ei}\left(\frac{x_i}{x}\right)}{x},\qquad {\rm for}\ \ x_i\in R_+.
		\end{align}
		Finally performing the Laplace transform of the Pad\'{e} approximant of the Borel transform \eqref{partial frac} by combining \eqref{not on R} and \eqref{On R}, we obtain
		\begin{align}\label{laplace trans approx}
			f(x)\approx	\int_0^\infty dt e^{-t}\times [p,p] {\cal B} f(x)=c_0'-{\rm Re}\Big[\sum_{i=1}^p\frac{e^{-\frac{x_i}{x}} \text{Ei}\left(\frac{x_i}{x}\right)}{x}\Big].
		\end{align}
		We will proceed to apply this   method of the Borel-Pad\'{e} re-summation first on the free energy for the 
		free theory and then on the free energy for the  critical $O(N)$ model.
		\subsection{Free theory}
		We consider the free theory and 
		apply the method, described above, to the asymptotic series for the high temperature expansion of the free energy. Using  the Borel-Pad\'e resum of the free energy we also study the thermal expectation of energy. Finally, we will demonstrate that the method  successfully extrapolates the high temperature expansions to the region where these agree with the low temperature expansions.  
		\subsection*{Free energy}
		Consider the high temperature expansion of the free energy for  the free theory \eqref{free pat fn using Cardy} written as 
		\begin{align}
			\log Z_{\rm free} =&\frac{2 r^2 \zeta (3)}{\beta ^2}-\frac{1}{12} \log \frac{\beta}{r} -\frac{\log 2}{12}-\zeta'(-1)\nonumber\\
			&	+\sum_{n\in 2\mathbb Z} \frac{(-1)^{n+1}(2-2^{-n}) \zeta (-n-1) \zeta (1-n) \big(\frac{\beta }{r}\big)^n}{n!}.
		\end{align}
		Now we isolate the term $-\frac{1}{12}\log \frac{\beta}{r}$ from the series expansion in powers of $\frac{\beta}{r}$ in the r.h.s, and then we multiply the series by a factor of $\frac{\beta^2}{r^2}$ so that series contains the terms with non-negative powers of $\frac{\beta}{r}$ only, in the manner as given below\footnote{We always apply the Borel-Pad\'e resummation on the series expansion with only non-negative powers.}
		\begin{align}\label{log Z free+1/12..}
			f(\frac{\beta}{r})\equiv&	\frac{\beta^2}{r^2}(\log Z_{\rm free} +\frac{1}{12}\log \frac{\beta}{r}) =2  \zeta (3)-\frac{\beta^2}{r^2}(\frac{\log 2}{12}+\zeta'(-1))\nonumber\\
			&	+\sum_{n\in 2\mathbb Z} \frac{(-1)^{n+1}(2-2^{-n}) \zeta (-n-1) \zeta (1-n) \big(\frac{\beta }{r}\big)^{n+2}}{n!}.
		\end{align}
		One can easily compute the numerical coefficients in the series expansion in $\frac{\beta}{r}$ given in the r.h.s of the equation above and observe the asymptotic nature of the series expansion. 
		
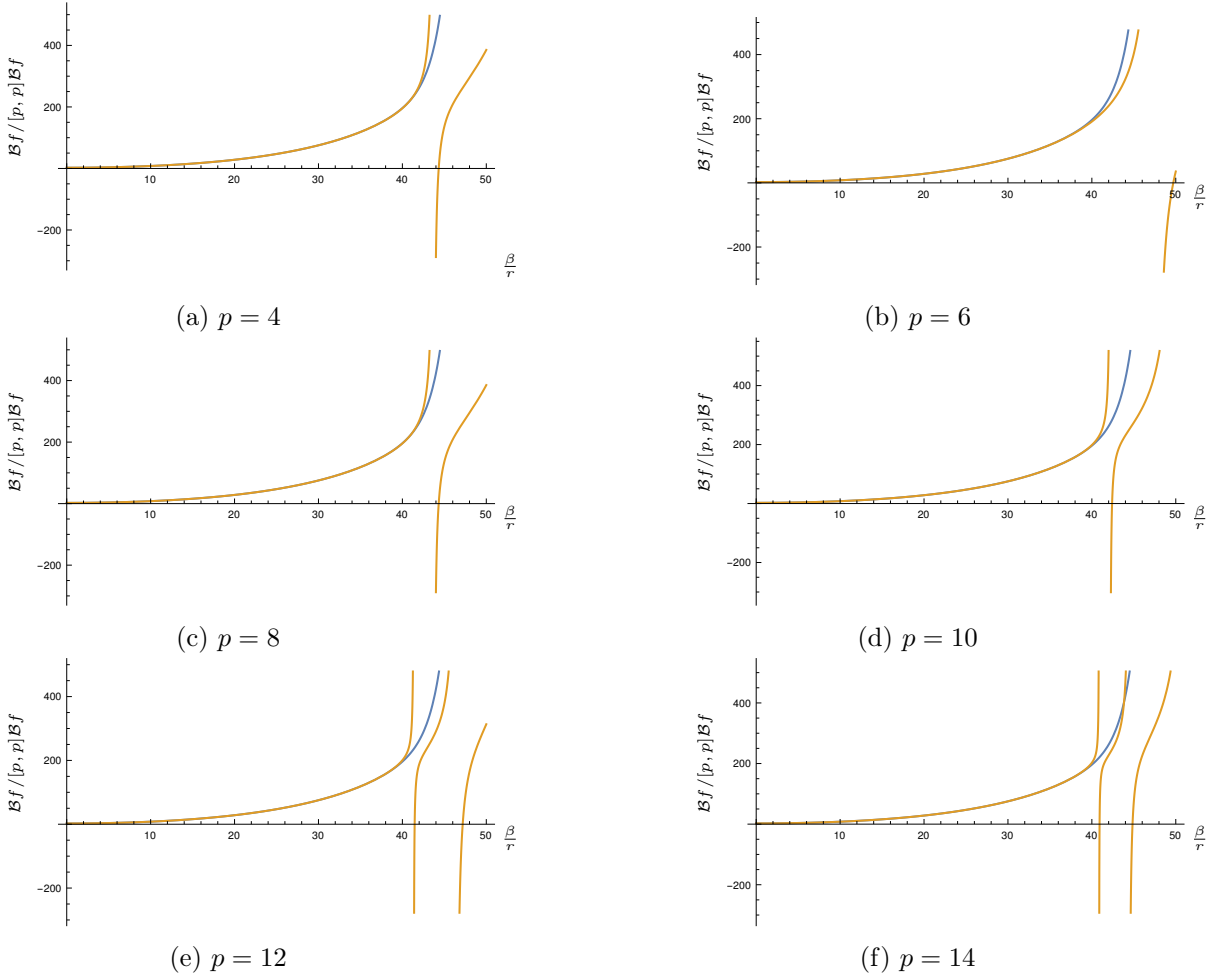
\begin{figure}[h]
			\begin{subfigure}{.4\linewidth}
				\begin{tikzpicture}
					\node[inner sep=0] (img) at (-1,0) {\includegraphics[width=1\linewidth]{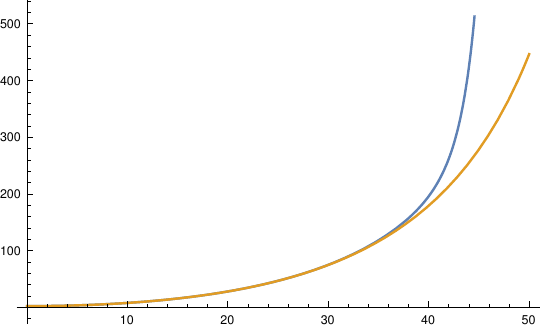}};
					\node at (2.2, -1.7) {\tiny $\frac{\beta}{r}$};
					\node[rotate=90] at (-4.3, .4) {\tiny $\mathcal{B}f/[p,p]{\cal B}f$};
				\end{tikzpicture}
				\caption{$p=4$}
			\end{subfigure}\hfill 
			\begin{subfigure}{.4\linewidth}
				\begin{tikzpicture}
					\node[inner sep=0] (img) at (-1,0) {\includegraphics[width=1\linewidth]{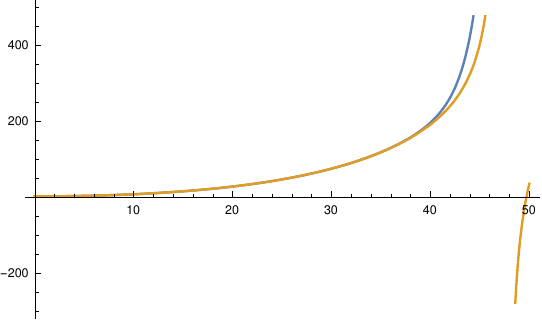}};
					\node at (2.2, -.6) {\tiny $\frac{\beta}{r}$};
					\node[rotate=90] at (-4.3, .4) {\tiny $\mathcal{B}f/[p,p]{\cal B}f$};
				\end{tikzpicture}
				\caption{$p=6$}
			\end{subfigure}
			\begin{subfigure}{.4\linewidth}
				\begin{tikzpicture}
					\node[inner sep=0] (img) at (-1,0) {\includegraphics[width=1\linewidth]{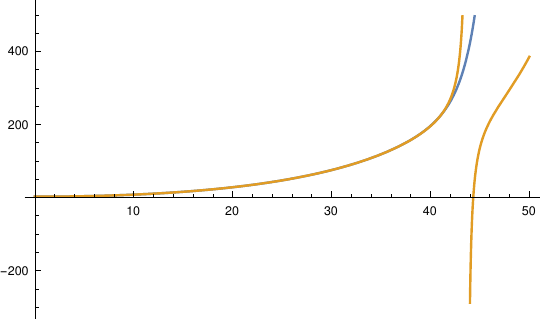}};
					\node at (2.2, -.6) {\tiny $\frac{\beta}{r}$};
					\node[rotate=90] at (-4.3, .4) {\tiny $\mathcal{B}f/[p,p]{\cal B}f$};
				\end{tikzpicture}
				\caption{$p=8$}
			\end{subfigure}
			\begin{subfigure}{.4\linewidth}
				\begin{tikzpicture}
					\node[inner sep=0] (img) at (-1,0) {\includegraphics[width=1\linewidth]{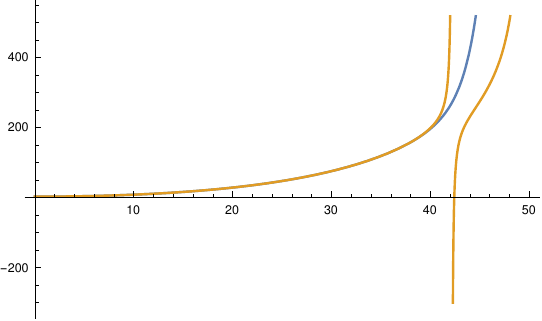}};
					\node at (2.2, -.6) {\tiny $\frac{\beta}{r}$};
					\node[rotate=90] at (-4.3, .4) {\tiny $\mathcal{B}f/[p,p]{\cal B}f$};
				\end{tikzpicture}
				\caption{$p=10$}
			\end{subfigure}
			\begin{subfigure}{.4\linewidth}
				\begin{tikzpicture}
					\node[inner sep=0] (img) at (-1,0) {\includegraphics[width=1\linewidth]{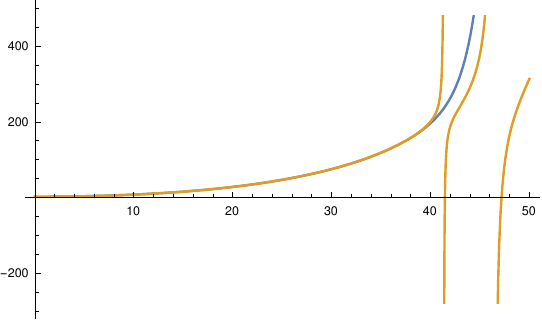}};
					\node at (2.2, -.6) {\tiny $\frac{\beta}{r}$};
					\node[rotate=90] at (-4.3, .4) {\tiny $\mathcal{B}f/[p,p]{\cal B}f$};
				\end{tikzpicture}
				\caption{$p=12$}
			\end{subfigure}\hfill
			\begin{subfigure}{.4\linewidth}
				\begin{tikzpicture}
					\node[inner sep=0] (img) at (-1,0) {\includegraphics[width=1\linewidth]{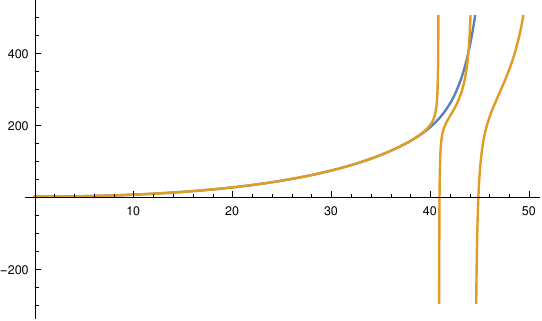}};
					\node at (2.2, -.6) {\tiny $\frac{\beta}{r}$};
					\node[rotate=90] at (-4.3, .4) {\tiny $\mathcal{B}f/[p,p]{\cal B}f$};
				\end{tikzpicture}
				\caption{$p=14$}
			\end{subfigure}
			\caption{The figures compare the Pad\'e approximant of the Borel transform of $f(\frac{\beta}{r})$ given in \eqref{log Z free+1/12..} described by the orange curve with the Borel transform ${\cal B}f(\frac{\beta}{r})$ itself given by the blue curve, for different orders $p$ of the Pad\'e approximation. The agreement between the two curves occurs till the Pad\'e approximant of the Borel transform $[p,p]{\cal B}f(\frac{\beta}{r})$ admits its first pole on the positive real axis. The agreement provides an internal consistency check to the numerical implementation of the Borel-Pad\'e re-summation technique for the free theory. Note the occurrence of closely located poles and zeros for higher order Pad\'e approximations. In general, such poles usually turn out to be  spurious poles arising from Pad\'e approximation and may not be present in the actual Borel transform. }
			\label{fig 1}
		\end{figure}

\begin{figure}[h]
			
			\begin{subfigure}{.4\linewidth}
				\begin{overpic}[width=1\textwidth]{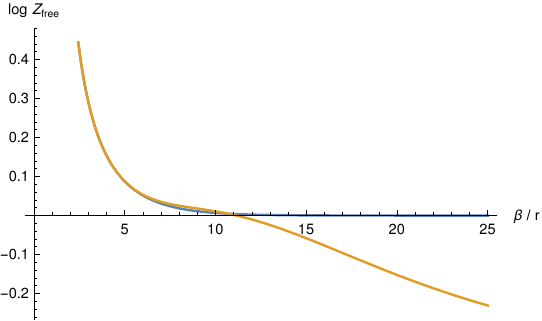}
					\put(40,25){%
						\setlength{\fboxsep}{1pt}
						\setlength{\fboxrule}{0.5pt}
						\fcolorbox{black}{white}{\includegraphics[width=0.56\textwidth]{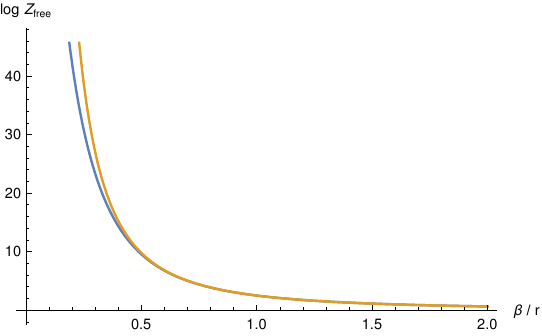}}%
					}
				\end{overpic}
				\caption{Using Pad\'e $[4,4]$}
			\end{subfigure}\hfill 
			\begin{subfigure}{.4\linewidth}
				\begin{overpic}[width=1\textwidth]{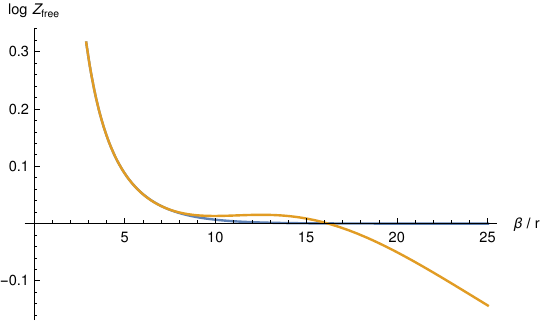}
					\put(40,25){%
						\setlength{\fboxsep}{1pt}
						\setlength{\fboxrule}{0.5pt}
						\fcolorbox{black}{white}{\includegraphics[width=0.56\textwidth]{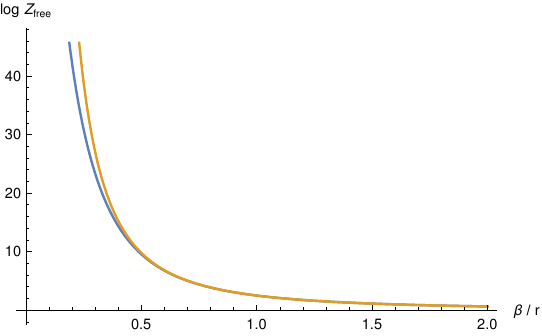}}%
					}
				\end{overpic}
				\caption{Using Pad\'e $[6,6]$}
			\end{subfigure}
			\par \bigskip\bigskip
			\begin{subfigure}{.4\linewidth}
				\begin{overpic}[width=1\textwidth]{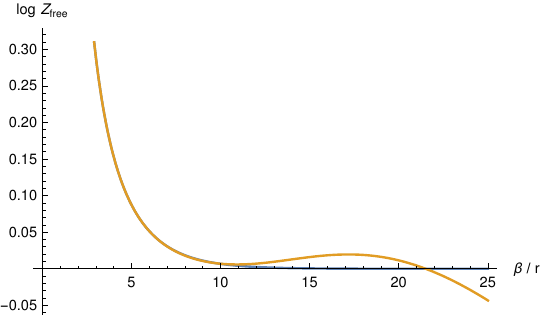}
					\put(40,25){%
						\setlength{\fboxsep}{1pt}
						\setlength{\fboxrule}{0.5pt}
						\fcolorbox{black}{white}{\includegraphics[width=0.56\textwidth]{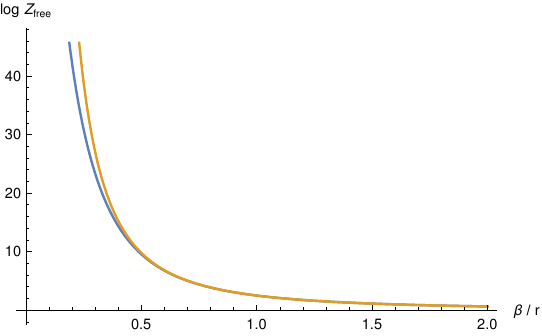}}%
					}
				\end{overpic}
				\caption{Using Pad\'e $[8,8]$}
			\end{subfigure}\hfill
			\begin{subfigure}{.4\linewidth}
				\begin{overpic}[width=1\textwidth]{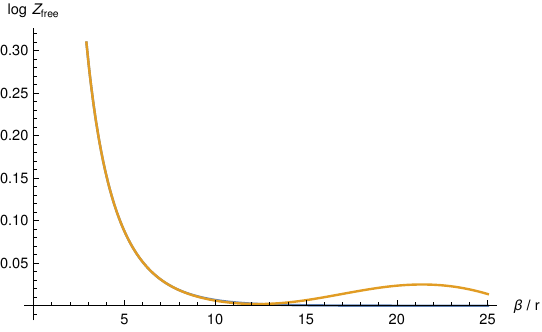}
					\put(40,25){%
						\setlength{\fboxsep}{1pt}
						\setlength{\fboxrule}{0.5pt}
						\fcolorbox{black}{white}{\includegraphics[width=0.56\textwidth]{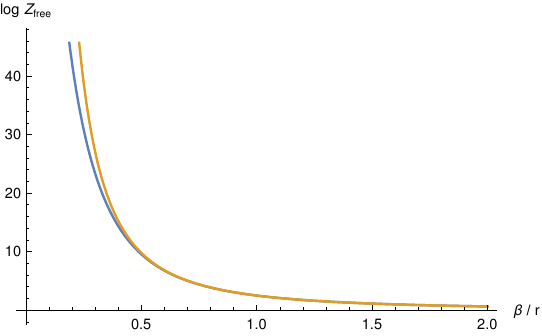}}%
					}
				\end{overpic}
				\caption{Using Pad\'e $[10,10]$}
			\end{subfigure}
			\par \bigskip\bigskip
			\begin{subfigure}{.4\linewidth}
				\begin{overpic}[width=1\textwidth]{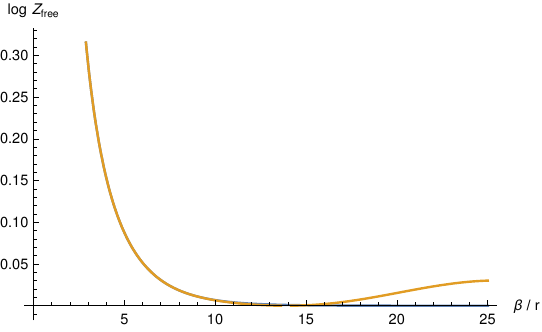}
					\put(40,25){%
						\setlength{\fboxsep}{1pt}
						\setlength{\fboxrule}{0.5pt}
						\fcolorbox{black}{white}{\includegraphics[width=0.56\textwidth]{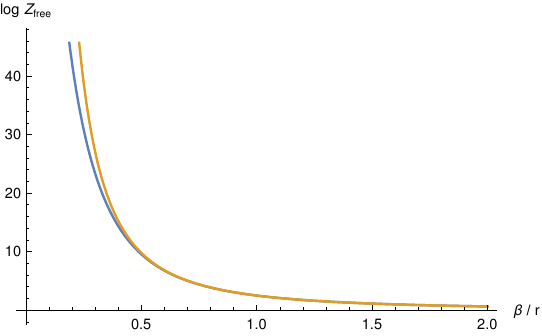}}%
					}
				\end{overpic}
				\caption{Using Pad\'e $[12,12]$}
			\end{subfigure}\hfill
			\begin{subfigure}{.4\linewidth}
				\begin{overpic}[width=1\textwidth]{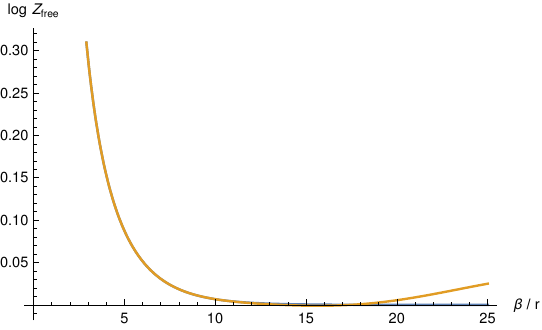}
					\put(40,25){%
						\setlength{\fboxsep}{1pt}
						\setlength{\fboxrule}{0.5pt}
						\fcolorbox{black}{white}{\includegraphics[width=0.56\textwidth]{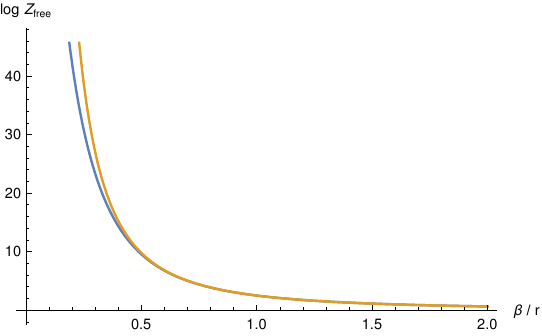}}%
					}
				\end{overpic}
				\caption{Using Pad\'e $[14,14]$}
			\end{subfigure}
			\caption{$\log Z_{\rm free}$ in the free theory; the orange curve denotes the Borel-Pad\'e re-sum of the high temperature expansion \eqref{free pat fn using Cardy} for $\log Z_{\rm free}$  and the blue curve stands for the low temperature expansion \eqref{free small b/r}(truncated till $l=n=10$). These two curves overlap on each other for a finite range of $\frac{\beta}{r}$ using Pad\'e approximants of different orders. The figures inside the boxes focus on small values of $\frac{\beta}{r}$ corresponding to each  plots.  For very large values of $\frac{\beta}{r}$ the Borel-Pad\'e sum fails to approximate the function correctly.  From the plots within the boxes, for very small values of $\frac{\beta}{r}$ two curves diverge slowly from each other. It is due to the fact that the low temperature expansion is plotted till finite orders in $e^{-\frac{\beta}{2r}}$, but as $\frac{\beta}{r}$ decreases, an increasing number of subleading contributions become significant.
			}
			\label{fig 2}
		\end{figure}

\begin{figure}[h!]
			
			\begin{subfigure}{.4\linewidth}
				\begin{tikzpicture}
					\node[inner sep=0] (img) at (-1.5,0) {\includegraphics[width=1\linewidth]{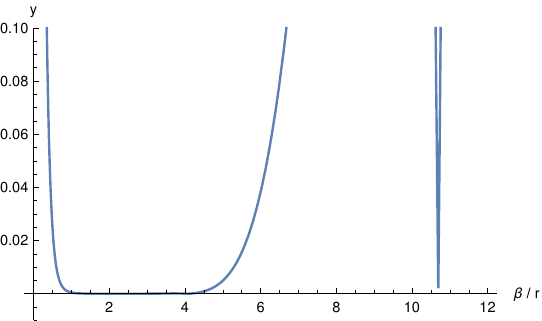}};
					\node[fill=white, draw=black, font=\tiny, anchor=north east] at (7,3) {\small $y=\Big|\frac{\text{ Borel-Pad\'{e}\ of\ high\ temp}-\text{ low\ temp\ expansion}}{\text{ low\ temp\ expansion}}\Big|$\quad {\rm for }$\log Z_{\rm free}$ Vs. $\frac{\beta}{r}$};
				\end{tikzpicture}
				\caption{Using Pad\'e $[4,4]$}
			\end{subfigure}\hfill 
			\begin{subfigure}{.4\linewidth}
				\par \bigskip\bigskip\bigskip
				\includegraphics[width=1\linewidth]{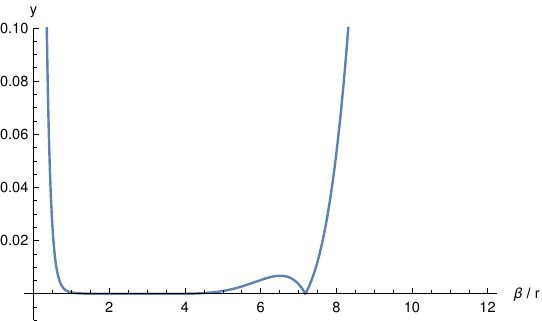}
				\caption{Using Pad\'e $[6,6]$}
			\end{subfigure}
			\begin{subfigure}{.4\linewidth}
				\includegraphics[width=1\linewidth]{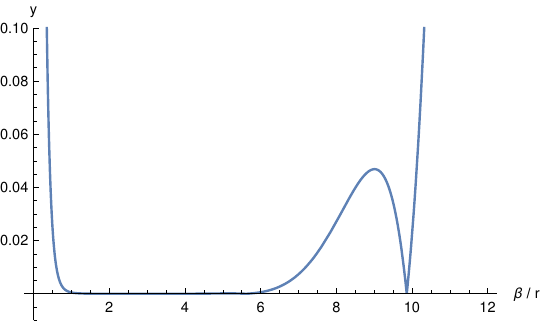}
				\caption{Using Pad\'e $[8,8]$}
			\end{subfigure}
			\begin{subfigure}{.4\linewidth}
				\includegraphics[width=1\linewidth]{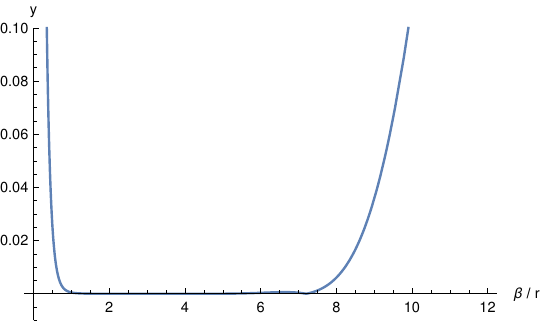}
				\caption{Using Pad\'e $[10,10]$}
			\end{subfigure}
			\begin{subfigure}{.4\linewidth}
				\includegraphics[width=1\linewidth]{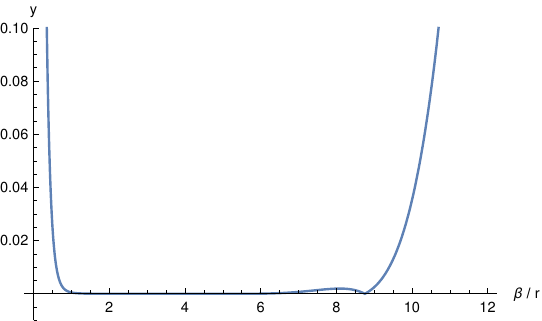}
				\caption{Using Pad\'e $[12,12]$}
				\label{fig4e}
			\end{subfigure}\hfill
			\begin{subfigure}{.4\linewidth}
				\includegraphics[width=1\linewidth]{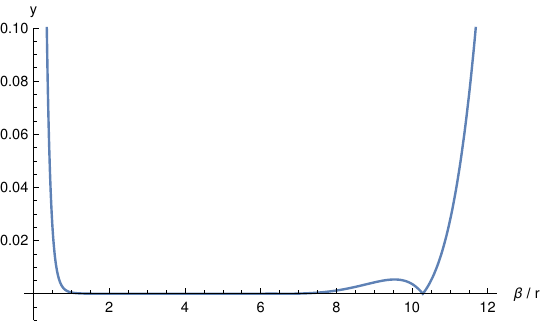}
				\caption{Using Pad\'e $[14,14]$}
				\label{fig4f}
			\end{subfigure}
			\caption{Free theory; here we plot  the absolute value of the difference between  the low temperature expansion \eqref{free small b/r} and the Borel-Pad\'e re-sum of high temp expansion  for the $\log Z_{\rm free}$ divided by its low temperature expansion against $\frac{\beta}{r}$. This demonstrates the numerical accuracy of the agreement between the Borel-Pad\'e resum of high temp expansion and low temperature expansion \eqref{free small b/r} plotted in figure \ref{fig 2}.}
			\label{fig 3}
		\end{figure}

		We apply the method of the Borel-Pad\'{e} re-summation introduced at the beginning of this section to the asymptotic series given on the r.h.s of \eqref{log Z free+1/12..} by truncating it till a finite number of terms in the expansion. In the first step we find the Borel transform \eqref{Borel trans} of this truncated series. We then evaluate the Pad\'{e} approximant of the truncated Borel transform. As an internal consistency check we show the agreement of this Pad\'{e} approximant with the truncated Borel transformed series in figure \ref{fig 1} for different symmetric orders of Pad\'e denoted earlier by $p$. This agreement holds till the first pole of the Pad\'e approximant of the Borel transform on the positive real axis. At the final step, we perform the 
		the Laplace transform \eqref{laplace trans approx}  on the 
		the Pad\'{e} approximant of the Borel transformed series. 
		This results in the re-summation for the r.h.s of \eqref{log Z free+1/12..}. 
		The Borel-Pad\'{e} re-summed $\log Z_{\rm free}$ is obtained by adding $-\frac{1}{12} \log \frac{\beta}{r}$ to the answer to this Laplace transform, followed by an overall multiplication by $\frac{r^2}{\beta^2}$.
		The Borel-Pad\'{e} resummed $\log Z_{\rm free}$  is plotted in figure \ref{fig 2} in orange and it coincides with the low temperature expansion \eqref{free small b/r} plotted in blue, for a certain range of values of $\frac{\beta}{r}$. The difference between these two curves is analyzed in the figure \ref{fig 3}. For the free theory, the agreement in these expansions can be established by analytic methods as described in Section \ref{sec 2}. The high temperature expansion is obtained just by performing a set of mathematical manipulations on the low temperature expansion. Now in figure \ref{fig 2} the agreement between the Borel-Pad\'{e} re-summation of the high temperature expansion and the low temperature expansion \eqref{free small b/r} is limited in a range of $\frac{\beta}{r}$. The reason for this is the Borel-Pad\'e re-summation can correctly extrapolate the high temperature expansion for lower values of temperatures but it fails after a certain value of $\frac{\beta}{r}$. Thus in figure \ref{fig 2} the two curves do not match when $\frac{\beta}{r}$ is very large. Again they do not agree when $\frac{\beta}{r}$ is very small which is clear from the graphs shown in the figures given in insets of each of the plots in figure \ref{fig 2}. Here, the Borel-Pad\'{e} re-sum should work very well when $\frac{\beta}{r}$ is very small. But one has to include  more higher orders terms  from the low temperature expansion \eqref{free small b/r} in this limit, for this to agree with the Borel-Pad\'{e} re-sum of the high temperature expansion.  
		
		An important observation from examining the difference between the approximations
		in figure \ref{fig 3} is the following. 
		Observe  that the graph is minimum and flat at intermediate ranges of $\beta/r$. This is consistent with the 
		fact the one of the expansions is not good  at either at low temperature or high temperatures. 
		Note also that as we go to higher orders in the Borel-Pad\'{e} approximation, we find the error graphs flattens more and more indicating that the consistency of the approximations over increased domains of temperatures. 
		Consider for example figures  \ref{fig4e} and \ref{fig4f}  which plots the differences in the $[12, 12]$ and $[14, 14]$ approximants from the low temperature expansion. 
		The range over which the curve stays flat has increased in the $[14, 14]$ case. 

		\subsection*{Thermal expectation of energy}		
\begin{figure}[h]
			
			\begin{subfigure}{.4\linewidth}
				\begin{overpic}[width=1\textwidth]{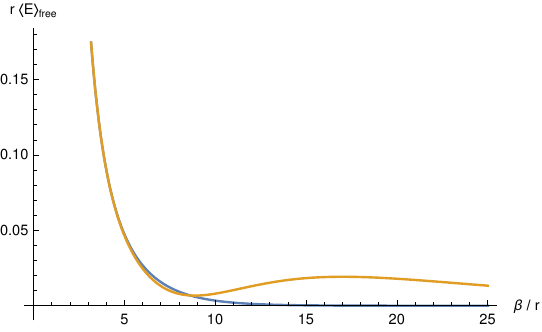}
					\put(40,26){%
						\setlength{\fboxsep}{1pt}
						\setlength{\fboxrule}{0.5pt}
						\fcolorbox{black}{white}{\includegraphics[width=0.56\textwidth]{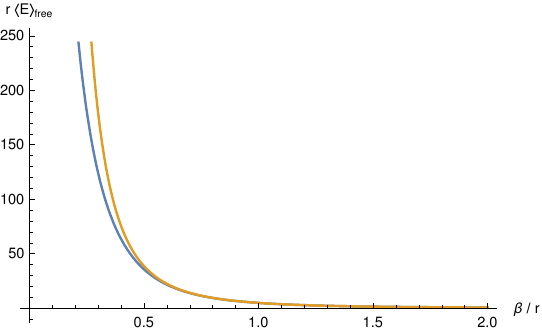}}%
					}
				\end{overpic}
				\caption{Using Pad\'e $[4,4]$}
			\end{subfigure}\hfill 
			\begin{subfigure}{.4\linewidth}
				\begin{overpic}[width=1\textwidth]{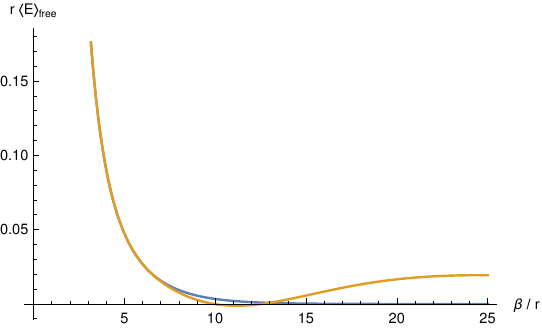}
					\put(40,26){%
						\setlength{\fboxsep}{1pt}
						\setlength{\fboxrule}{0.5pt}
						\fcolorbox{black}{white}{\includegraphics[width=0.56\textwidth]{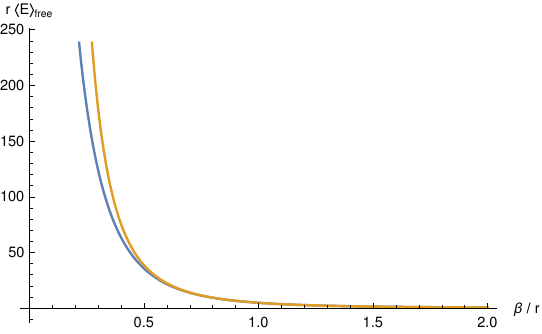}}%
					}
				\end{overpic}
				\caption{Using Pad\'e $[6,6]$}
			\end{subfigure}
			\par \bigskip\bigskip
			\begin{subfigure}{.4\linewidth}
				\begin{overpic}[width=1\textwidth]{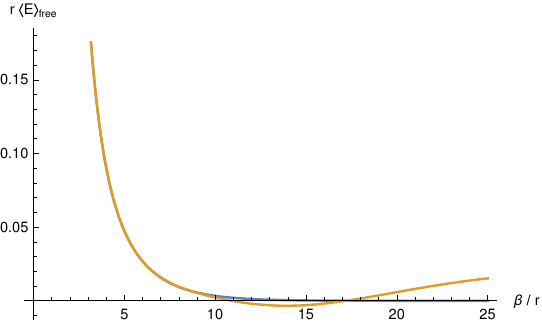}
					\put(40,26){%
						\setlength{\fboxsep}{1pt}
						\setlength{\fboxrule}{0.5pt}
						\fcolorbox{black}{white}{\includegraphics[width=0.56\textwidth]{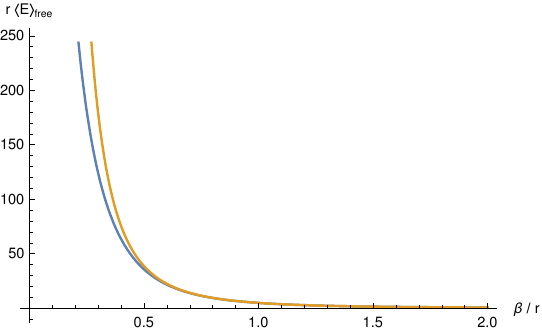}}%
					}
				\end{overpic}
				\caption{Using Pad\'e $[8,8]$}
			\end{subfigure}\hfill
			\begin{subfigure}{.4\linewidth}
				\begin{overpic}[width=1\textwidth]{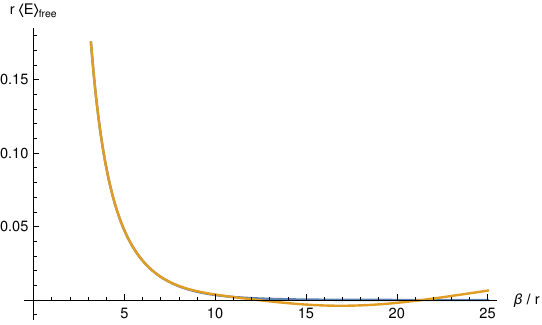}
					\put(40,26){%
						\setlength{\fboxsep}{1pt}
						\setlength{\fboxrule}{0.5pt}
						\fcolorbox{black}{white}{\includegraphics[width=0.56\textwidth]{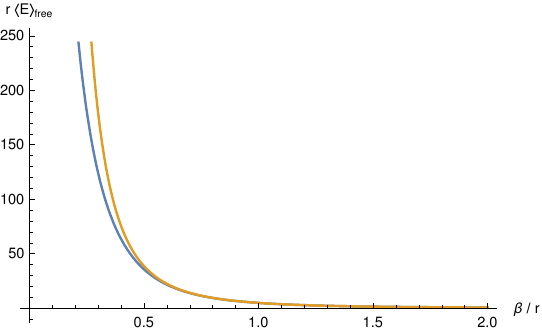}}%
					}
				\end{overpic}
				\caption{Using Pad\'e $[10,10]$}
			\end{subfigure}
			\par \bigskip\bigskip
			\begin{subfigure}{.4\linewidth}
				\begin{overpic}[width=1\textwidth]{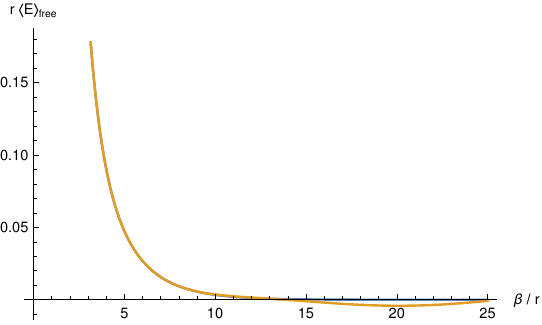}
					\put(40,26){%
						\setlength{\fboxsep}{1pt}
						\setlength{\fboxrule}{0.5pt}
						\fcolorbox{black}{white}{\includegraphics[width=0.56\textwidth]{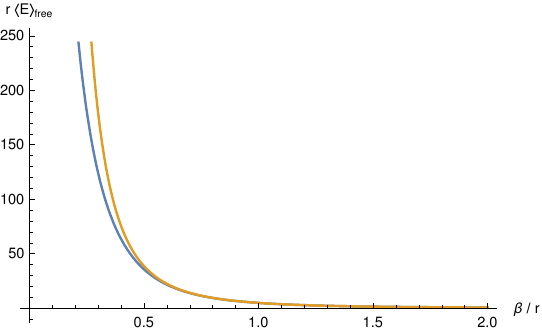}}%
					}
				\end{overpic}
				\caption{Using Pad\'e $[12,12]$}
			\end{subfigure}\hfill
			\begin{subfigure}{.4\linewidth}
				\begin{overpic}[width=1\textwidth]{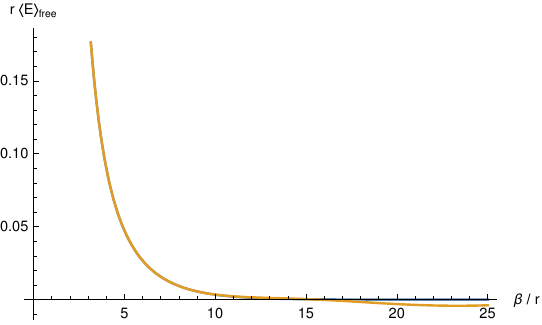}
					\put(40,26){%
						\setlength{\fboxsep}{1pt}
						\setlength{\fboxrule}{0.5pt}
						\fcolorbox{black}{white}{\includegraphics[width=0.56\textwidth]{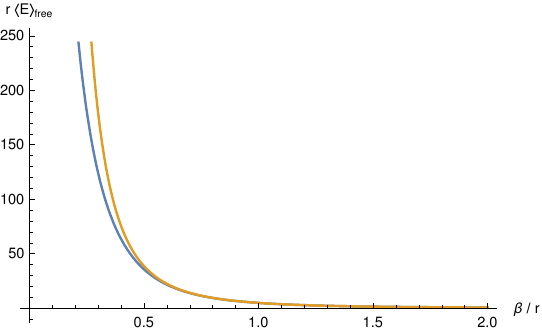}}%
					}
				\end{overpic}
				\caption{Using Pad\'e $[14,14]$}
			\end{subfigure}
			\caption{thermal expectation of energy for free theory in units of radius i.e., $r\langle E\rangle$; The Borel-Pad\'e re-sum \eqref{P B rsum E} of the high temperature expansion of the thermal expectation of energy, plotted as the orange curve, is compared against the low temperature expansion \eqref{E free} (truncated till $l=n=10$) given by the blue curve. The figures inside the boxes magnify the small $\frac{\beta}{r}$ regions for the corresponding graphs. The agreement between the orange and blue curves is observed over a finite range of $\frac{\beta}{r}$. The limited domain of validity of the Borel-Padé resummation at large 
				$\frac{\beta}{r} $ causes the two curves to deviate significantly from each other. At small $\frac{\beta}{r}$(see the figure inside boxes),   the truncated low temperature expansion starts accumulating error, as the higher order terms become significant, though Borel-Pad\'e resum for the high temperature expansion works very well in this regime.
			}
			\label{fig stress 2}
		\end{figure}

\begin{figure}[h!]
			
			\begin{subfigure}{.4\linewidth}
				\begin{tikzpicture}
					\node[inner sep=0] (img) at (-1.5,0) {\includegraphics[width=1\linewidth]{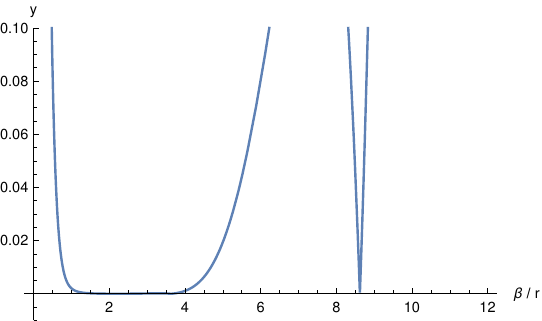}};
					\node[fill=white, draw=black, font=\tiny, anchor=north east] at (7.5,3) {\small $y=\Big|\frac{\text{ Borel-Pad\'{e}\ of\ high\ temp}-\text{ low\ temp\ expansion}}{\text{ low\ temp\ expansion}}\Big|$\ {\rm for }$\langle E\rangle_{\rm free}$ {\rm Vs.} $\frac{\beta}{r}$};
				\end{tikzpicture}
				\caption{Using Pad\'e [4,4]}
			\end{subfigure}\hfill 
			\begin{subfigure}{.4\linewidth}
				\par \bigskip\bigskip\bigskip
				\includegraphics[width=1\linewidth]{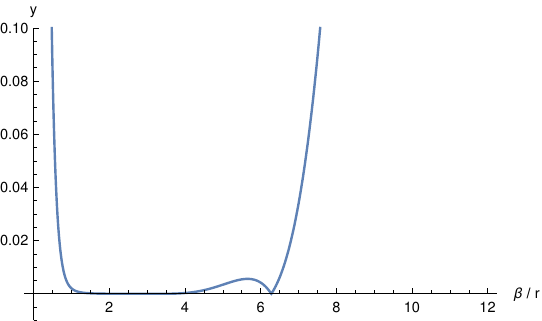}
				\caption{Using Pad\'e [6,6]}
			\end{subfigure}
			\begin{subfigure}{.4\linewidth}
				\includegraphics[width=1\linewidth]{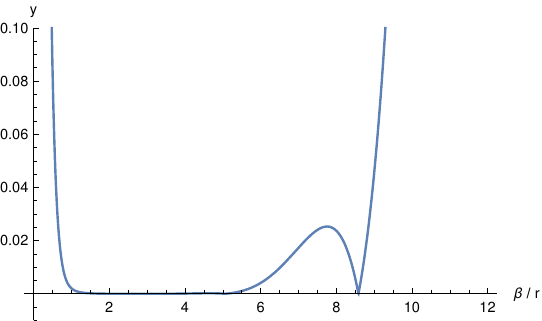}
				\caption{Using Pad\'e [8,8]}
			\end{subfigure}
			\begin{subfigure}{.4\linewidth}
				\includegraphics[width=1\linewidth]{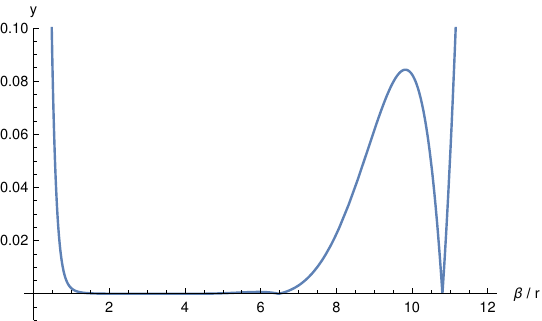}
				\caption{Using Pad\'e [10,10]}
			\end{subfigure}
			\begin{subfigure}{.4\linewidth}
				\includegraphics[width=1\linewidth]{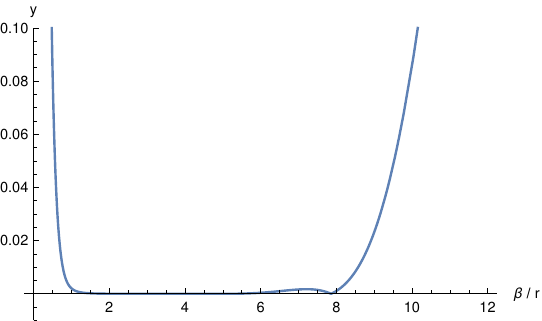}
				\caption{Using Pad\'e [12,12]}
			\end{subfigure}\hfill
			\begin{subfigure}{.4\linewidth}
				\includegraphics[width=1\linewidth]{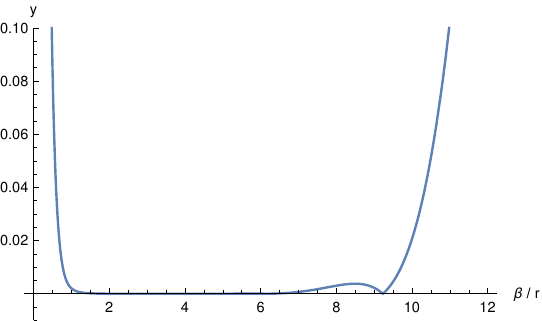}
				\caption{Using Pad\'e [14,14]}
			\end{subfigure}
			\caption{ We plot  the absolute value of the difference between  the low temperature expansion \eqref{en den free th} and the Borel-Pad\'e re-sum \eqref{P B rsum E}  for $\langle E\rangle_{\rm free}$  from the high temp expansion divided by its low temperature expansion against $\frac{\beta}{r}$. This shows the numerical accuracy of the agreement between the Borel-Pad\'e resum of high temp expansion and low temperature expansion plotted in figure \ref{fig stress 2}.}
			\label{fig stress 3}
		\end{figure}

		To extrapolate the high temperature expansion for the thermal expectation of energy to lower values of the temperature in units of the radius of the 2-sphere, we   differentiate the Borel-Pade resum of the $\log Z_{\rm free}$ with respect to $\beta$ with an overall negative sign, as given below
		\begin{align}\label{P B rsum E}
			E_{\rm free}=-\partial_\beta [\log Z_{\rm free}^{\text{Borel-Pad\'e\ resumed}}].
		\end{align}
		%
		The Borel-Pad\'{e} re-summed  thermal expectation of energy obtained in this way, agrees with the low temperature expansion \eqref{E free} for a certain range of values for $\frac{\beta}{r}$, as shown in figure \ref{fig stress 2}. The numerical accuracy of this agreement is demonstrated in figure \ref{fig stress 3}. Again the differences 
		in the approximation is small at intermediate ranges of 
		of $\frac{\beta}{r}$.   This is to be expected since  the two expansions are valid in either 
		the high temperatures are low temperatures. 
		The validity of the approximation by the Borel-Pad\'e sum is limited to when  $\frac{\beta}{r}$ is less than a finite value and  since the  low temperature expansion   \eqref{E free} is truncated 
		to finite orders in $e^{-\frac{\beta}{2r}}$ in the plot.   Note again that the curves flatten out as the order of 
		Bore-Pad\'{e} approximation is increased which indicates  consistency of the approximations over larger 
		domains of temperature.

		\subsection{Interacting theory}	
		
\begin{figure}[h]
			
			\begin{subfigure}{.4\linewidth}
				\begin{tikzpicture}
					\node[inner sep=0] (img) at (-1.2,0) {\includegraphics[width=1\linewidth]{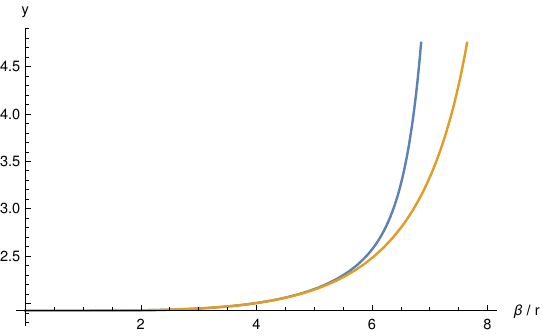}};
					\node[fill=white, draw=black, font=\tiny, anchor=north east] at (7.5,3) {\small $y=[p,p]{\cal B}(\frac{\beta^2}{r^2}\log Z)$ {\rm or} ${\cal B}(\frac{\beta^2}{r^2}\log Z)$ \ {\rm Vs.} $\frac{\beta}{r}$};
				\end{tikzpicture}
				\caption{Using Pad\'e $[4,4]$}
			\end{subfigure}\hfill 
			\begin{subfigure}{.4\linewidth}
				\par\bigskip\bigskip\bigskip
				\includegraphics[width=1\linewidth]{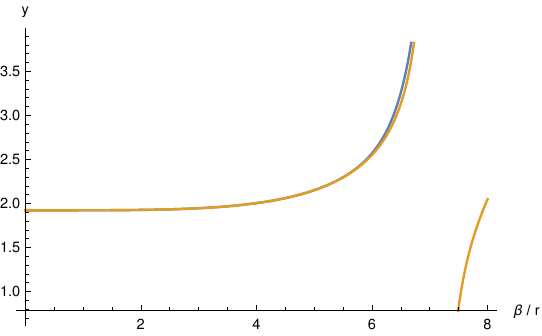}
				\caption{Using Pad\'e $[6,6]$}
			\end{subfigure}
			\begin{subfigure}{.4\linewidth}
				\includegraphics[width=1\linewidth]{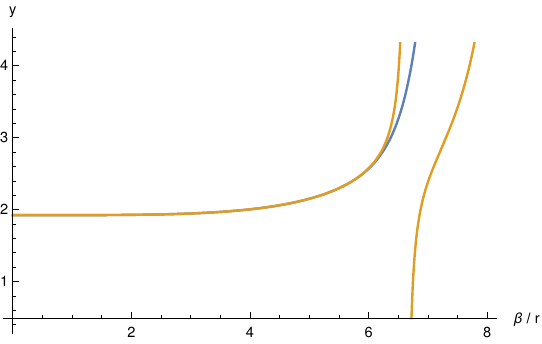}
				\caption{Using Pad\'e $[8,8]$}
			\end{subfigure}
			\begin{subfigure}{.4\linewidth}
				\includegraphics[width=1\linewidth]{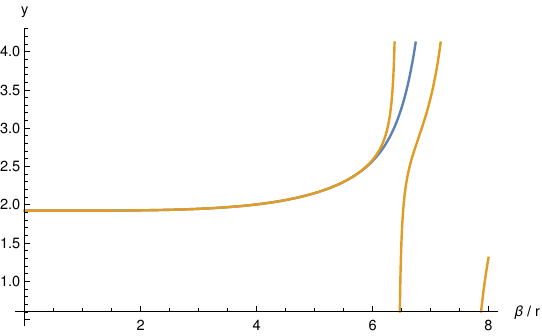}
				\caption{Using Pad\'e $[10,10]$}
			\end{subfigure}
			\begin{subfigure}{.4\linewidth}
				\includegraphics[width=1\linewidth]{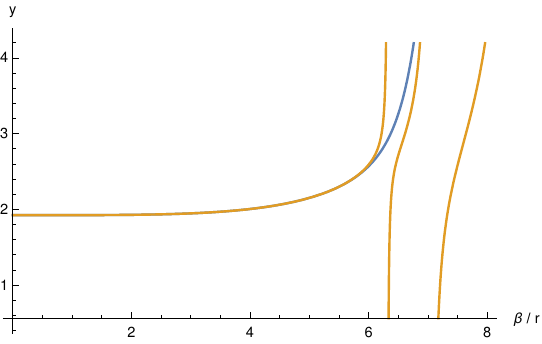}
				\caption{Using Pad\'e $[12,12]$}
			\end{subfigure}\hfill
			\begin{subfigure}{.4\linewidth}
				\includegraphics[width=1\linewidth]{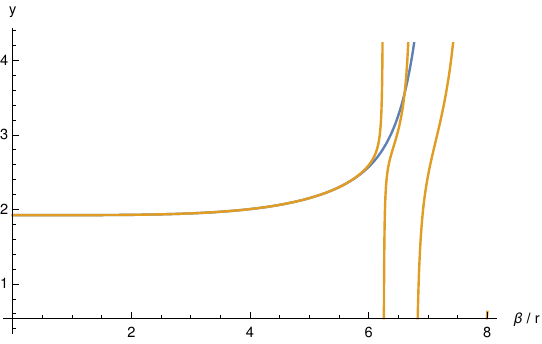}
				\caption{Using Pad\'e $[14,14]$}
			\end{subfigure}
			\caption{Interacting theory; we compare the Pad\'e approximant of the Borel transform of the asymptotic series \eqref{log Z numeric} denoted by $[p,p]{\cal B}(\frac{\beta^2}{r^2}\log Z)$ with the Borel transform ${\cal B}(\frac{\beta^2}{r^2}\log Z) $ itself for different orders of the Pad\'e approximations. The Pad\'e approximation works well till it encounters the first pole on the positive real axis. Again this agreement serves as an internal consistency check to the numerical implementation of the Borel-Pad\'e resummation for the thermal expectation of energy in the theory at infinite coupling.}
			\label{fig int 1}
		\end{figure}

\begin{figure}[h]
			
			\begin{subfigure}{.4\linewidth}
				\begin{overpic}[width=1\textwidth]{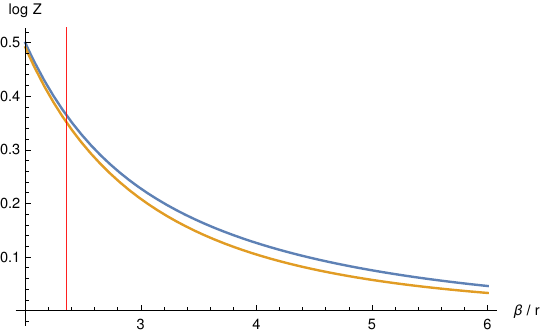}
					\put(40,25){%
						\setlength{\fboxsep}{1pt}
						\setlength{\fboxrule}{0.5pt}
						\fcolorbox{black}{white}{\includegraphics[width=0.56\textwidth]{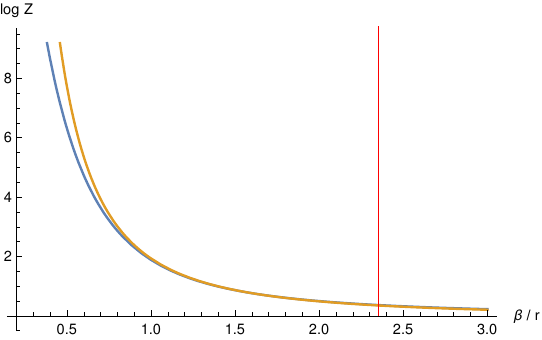}}%
					}
				\end{overpic}
				\caption{Using Pad\'e $[4,4]$}
			\end{subfigure}\hfill 
			\begin{subfigure}{.4\linewidth}
				\begin{overpic}[width=1\textwidth]{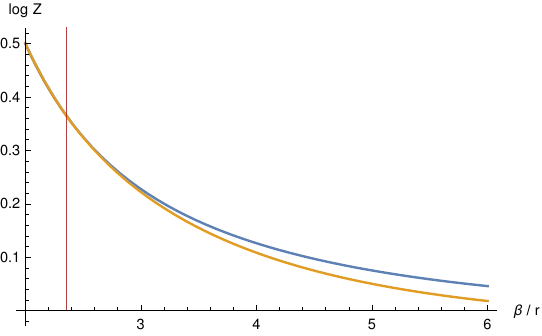}
					\put(40,25){%
						\setlength{\fboxsep}{1pt}
						\setlength{\fboxrule}{0.5pt}
						\fcolorbox{black}{white}{\includegraphics[width=0.56\textwidth]{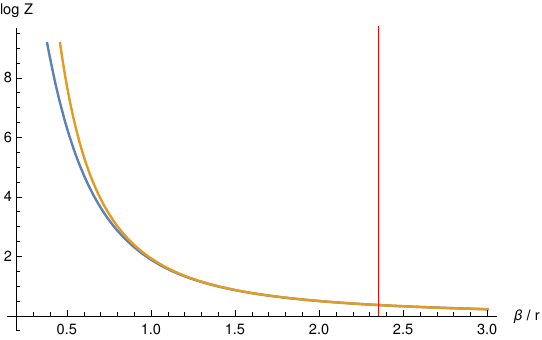}}%
					}
				\end{overpic}
				\caption{Using Pad\'e $[6,6]$}
			\end{subfigure}
			\par \bigskip\bigskip
			\begin{subfigure}{.4\linewidth}
				\begin{overpic}[width=1\textwidth]{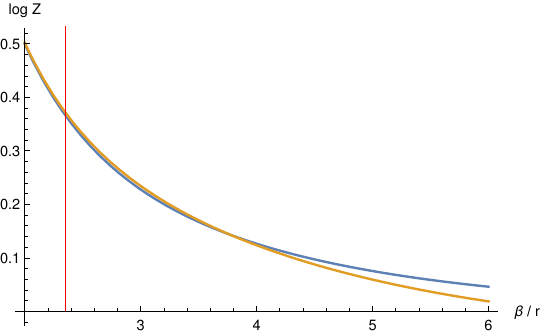}
					\put(40,25){%
						\setlength{\fboxsep}{1pt}
						\setlength{\fboxrule}{0.5pt}
						\fcolorbox{black}{white}{\includegraphics[width=0.56\textwidth]{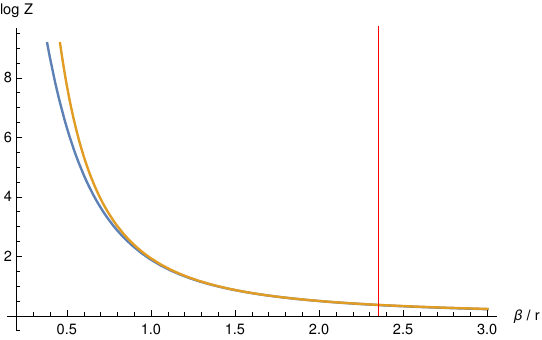}}%
					}
				\end{overpic}
				\caption{Using Pad\'e $[8,8]$}
			\end{subfigure}\hfill
			\begin{subfigure}{.4\linewidth}
				\begin{overpic}[width=1\textwidth]{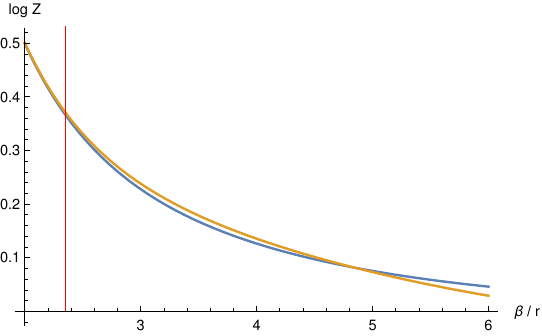}
					\put(40,25){%
						\setlength{\fboxsep}{1pt}
						\setlength{\fboxrule}{0.5pt}
						\fcolorbox{black}{white}{\includegraphics[width=0.56\textwidth]{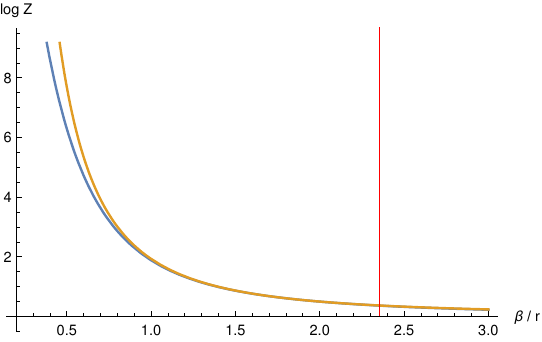}}%
					}
				\end{overpic}
				\caption{Using Pad\'e $[10,10]$}
			\end{subfigure}
			\par \bigskip\bigskip
			\begin{subfigure}{.4\linewidth}
				\begin{overpic}[width=1\textwidth]{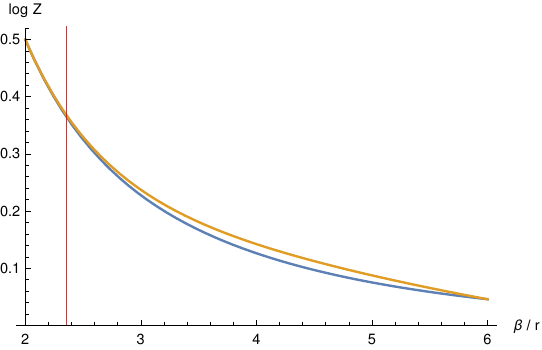}
					\put(40,25){%
						\setlength{\fboxsep}{1pt}
						\setlength{\fboxrule}{0.5pt}
						\fcolorbox{black}{white}{\includegraphics[width=0.56\textwidth]{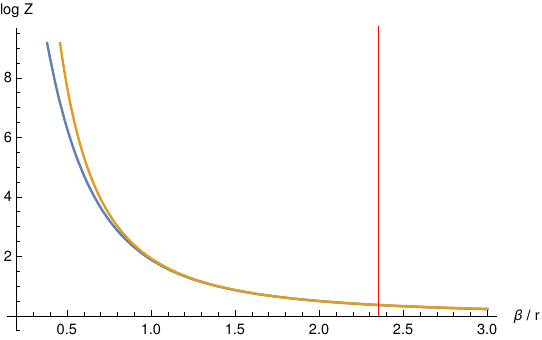}}%
					}
				\end{overpic}
				\caption{Using Pad\'e $[12,12]$}
			\end{subfigure}\hfill
			\begin{subfigure}{.4\linewidth}
				\begin{overpic}[width=1\textwidth]{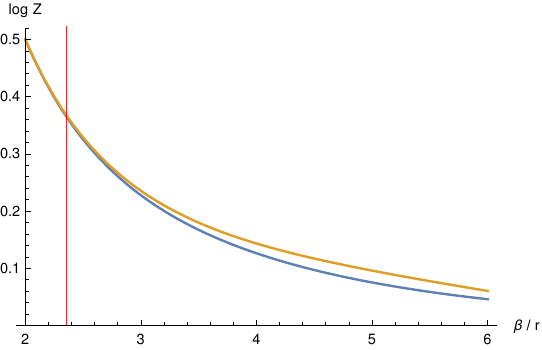}
					\put(40,25){%
						\setlength{\fboxsep}{1pt}
						\setlength{\fboxrule}{0.5pt}
						\fcolorbox{black}{white}{\includegraphics[width=0.56\textwidth]{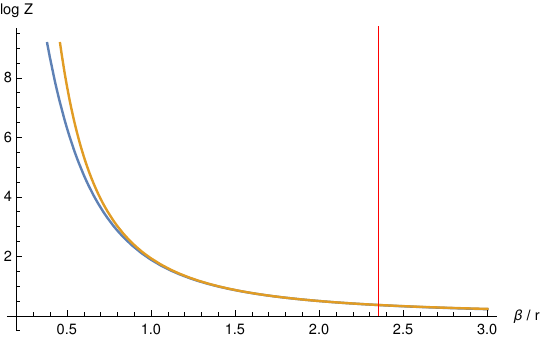}}%
					}
				\end{overpic}
				\caption{Using Pad\'e $[14,14]$}
			\end{subfigure}
			\caption{$\log Z$ for the interacting theory; The Borel-Pad\'e re-sum of the high temperature expansion \eqref{log Z numeric} for $\log Z$, plotted in orange, is compared against the low temperature expansion \eqref{log Z interacting}, including orders till $O(e^{-\frac{6\beta}{r}})$  given in  the ancillary file, \texttt{low\_temp\_expansions.txt}),  is plotted in blue. The figures inside the boxes describe the small $\frac{\beta}{r}$ regions for the corresponding graphs. 
				The vertical red line is the straight line $\frac{\beta}{r}=2\log \frac{32}{\pi^2}$, to right side of this line the low temperature expansion \eqref{log Z interacting} is self-consistent.
				The agreement between the orange and blue curves is observed over a finite range of $\frac{\beta}{r}$. The limited domain of validity of the Borel-Padé resummation at large 
				$\frac{\beta}{r} $ causes the two curves to deviate significantly from each other. At small $\frac{\beta}{r}$,   the truncated low temperature expansion  starts to accumulate error, as an increasing number of  higher order terms become significant, though Borel-Pad\'e resum for the high temperature expansion works very well in this regime.}
			\label{fig int 2}
		\end{figure}

		For the $O(N)$ model at the non-trivial fixed point obtained at $\lambda\to \infty$ we have the following expansions for the thermal mass, free energy and thermal expectation of energy.
		The thermal mass satisfying the gap equation \eqref{gap eq high temp} given as a series expansion at small $\frac{\beta}{r}$ as follows
		{\small\begin{align}\label{thermal mass numeric}
				\tilde m\beta=& 0.962424\, +0.0432935 \frac{\beta ^2}{r^2}+0.00482457\frac{ \beta ^4}{r^4}+0.00295908\frac{ \beta ^6}{r^6}+0.00376594\frac{ \beta ^8}{r^8}\nonumber\\
				&+0.00796123\frac{ \beta ^{10}}{r^{10}}
				+0.0250979\frac{ \beta
					^{12}}{r^{12}}+0.110481\frac{ \beta ^{14}}{r^{14}}+0.648485\frac{ \beta ^{16}}{r^{16}}+4.9011\frac{ \beta ^{18}}{r^{18}}+46.397\frac{ \beta ^{20}}{r^{20}}\nonumber\\
				&+538.013\frac{ \beta ^{22}}{r^{22}}+7502.54\frac{ \beta
					^{24}}{r^{24}}
				+123892.\frac{ \beta ^{26}}{r^{26}}+2.39127\times 10^6\frac{ \beta ^{28}}{r^{28}}+5.33444\times 10^7\frac{ \beta ^{30}}{r^{30}}\nonumber\\
				&+1.36209\times 10^9\frac{ \beta ^{32}}{r^{32}}+3.94716\times 10^{10}\frac{
					\beta ^{34}}{r^{34}}.
		\end{align}}
		One obtains the high temperature expansion for  the free energy by substituting the thermal mass \eqref{thermal mass numeric} in \eqref{small beta log Z} as given below
		{\small \begin{align}\label{log Z numeric}
				\frac{\beta^2}{r^2}\log Z=&1.92329\, +0.00645381\frac{ \beta ^4}{r^4}+0.00164641\frac{ \beta ^6}{r^6}+0.00135093\frac{ \beta ^8}{r^8}
				+0.00210951\frac{ \beta ^{10}}{r^{10}}\nonumber\\
				&+0.00527806\frac{ \beta ^{12}}{r^{12}}+0.0192741\frac{ \beta
					^{14}}{r^{14}}+0.0967 \frac{\beta ^{16}}{r^{16}}+0.638316\frac{ \beta ^{18}}{r^{18}}+5.36456\frac{ \beta ^{20}}{r^{20}}\nonumber\\
				&+55.9355\frac{ \beta ^{22}}{r^{22}}+708.629\frac{ \beta ^{24}}{r^{24}}+\frac{10721.2 \beta
					^{26}}{r^{26}}+190938.\frac{ \beta ^{28}}{r^{28}}+3.95394\times 10^6\frac{ \beta ^{30}}{r^{30}}\nonumber\\
				&+9.42044\times 10^7\frac{ \beta ^{32}}{r^{32}}+2.55876\times 10^9\frac{ \beta ^{34}}{r^{34}}.
		\end{align}}		
		And the thermal expectation of energy admits the following series expansion at high temperature
		{\small	\begin{align}\label{E numeric}
				\langle E\rangle=3.84658\frac{ r^2}{\beta ^3}-0.0129076\frac{ \beta }{r^2}-0.00658562\frac{ \beta ^3}{r^4}-0.00810558\frac{ \beta ^5}{r^6}-0.0168761 \frac{\beta
					^7}{r^8}\nonumber\\
				-0.0527806\frac{ \beta ^9}{r^{10}}-0.231289\frac{ \beta ^{11}}{r^{12}}-1.3538\frac{ \beta ^{13}}{r^{14}}-10.2131\frac{ \beta ^{15}}{r^{16}}-96.5621\frac{
					\beta ^{17}}{r^{18}}\nonumber\\-1118.71\frac{ \beta ^{19}}{r^{20}}
				-15589.8\frac{ \beta ^{21}}{r^{22}}-257309.\frac{ \beta ^{23}}{r^{24}}-4.96438\times 10^6\frac{ \beta
					^{25}}{r^{26}}\nonumber\\-1.1071\times 10^8 \frac{\beta ^{27}}{r^{28}}
				-2.82613\times 10^9\frac{ \beta ^{29}}{r^{30}}-8.18803\times 10^{10}\frac{ \beta ^{31}}{r^{32}}.
		\end{align}}
		Note that all the above three expansions are asymptotic series as the coefficients in powers of $\frac{\beta}{r}$  initially decrease but then keep on increasing.
		Now we implement the Borel-Pad\'e resummation technique on  the asymptotic expansion of the free energy \eqref{log Z numeric}. The Pad\'e approximant of the Borel transform of the series \eqref{log Z numeric}, denoted by $[p,p] {\cal B} (\frac{\beta^2}{r^2}\log Z(\frac{\beta}{r}))$, is computed and has been compared against the original Borel transform of the series \eqref{log Z numeric} denoted by ${\cal B} (\frac{\beta^2}{r^2}\log Z(\frac{\beta}{r}))$ in figure \ref{fig int 1}. Similar to the discussions for the free theory in the previous subsection, this serves as an internal consistency check for our method. Finally, we apply the Laplace transform \eqref{laplace trans approx} on the Pad\'e approximant of the Borel transform $[p,p] {\cal B} (\frac{\beta^2}{r^2}\log Z(\frac{\beta}{r}))$, followed by an overall multiplication by $\frac{r^2}{\beta^2}$,  to obtain the Borel-Pad\'e re-summation of $\log Z$. This is compared against the low temperature expansion of the free energy \eqref{log Z interacting} in figure \ref{fig int 2}. An agreement between the Borel-Pad\'e re-summed high temperature expansion and the low temperature expansion is observed in a finite range of $\frac{\beta}{r}$. 
		
\begin{figure}[h]
			
			\begin{subfigure}{.4\linewidth}
				\begin{tikzpicture}
					\node[inner sep=0] (img) at (-1.5,0) {				\includegraphics[width=1\linewidth]{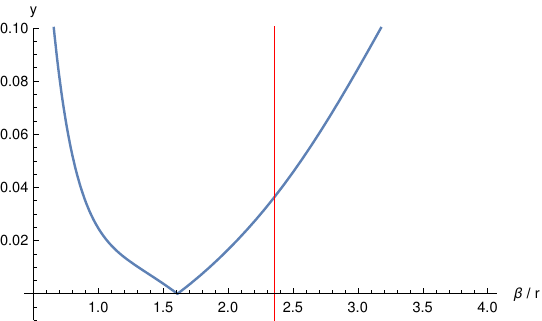}};
					\node[fill=white, draw=black, font=\tiny, anchor=north east] at (7,3) {\small $y=\Big|\frac{\text{ Borel-Pad\'{e}\ of\ high\ temp}-\text{ low\ temp\ expansion}}{\text{ low\ temp\ expansion}}\Big|$\ {\rm for }$\log Z$ vs. $\frac{\beta}{r}$};
				\end{tikzpicture}
				\caption{Using Pad\'e $[4,4]$}
			\end{subfigure}\hfill 
			\begin{subfigure}{.4\linewidth}
				\par \bigskip\bigskip\bigskip
				\includegraphics[width=1\linewidth]{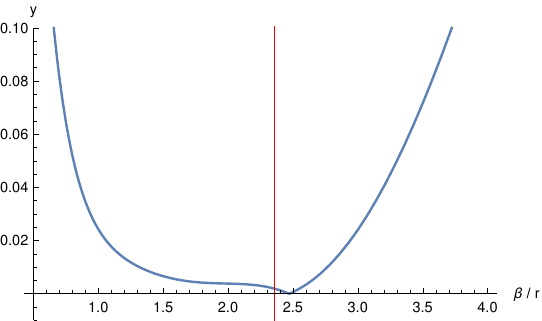}
				\caption{Using Pad\'e $[6,6]$}
			\end{subfigure}
			\begin{subfigure}{.4\linewidth}
				\includegraphics[width=1\linewidth]{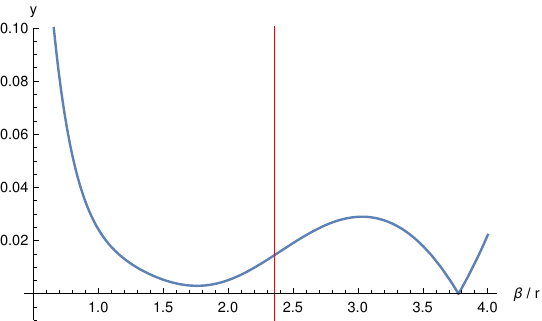}
				\caption{Using Pad\'e $[8,8]$}
			\end{subfigure}
			\begin{subfigure}{.4\linewidth}
				\includegraphics[width=1\linewidth]{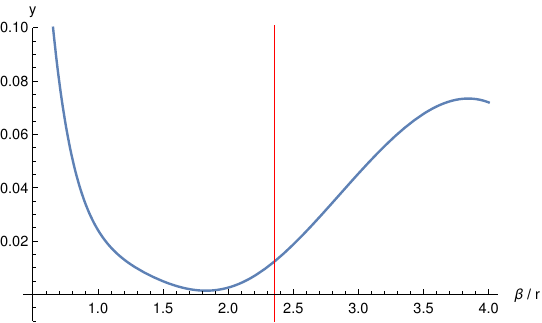}
				\caption{Using Pad\'e $[10,10]$}
			\end{subfigure}
			\begin{subfigure}{.4\linewidth}
				\includegraphics[width=1\linewidth]{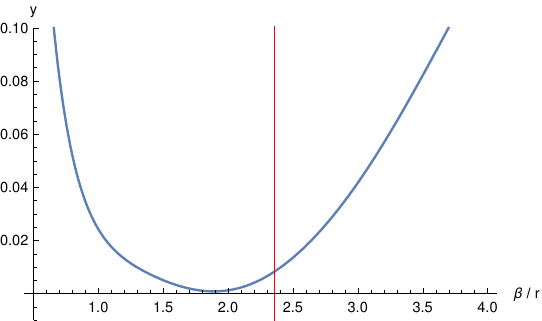}
				\caption{Using Pad\'e $[12,12]$}
			\end{subfigure}\hfill
			\begin{subfigure}{.4\linewidth}
				\includegraphics[width=1\linewidth]{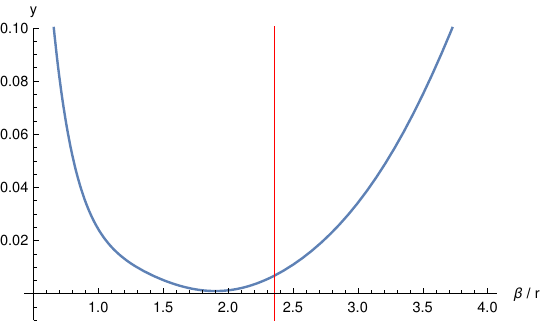}
				\caption{Using Pad\'e $[14,14]$}
			\end{subfigure}
			\caption{Interacting theory; We plot  the absolute value of the difference between  the low temperature expansion \eqref{log Z interacting} for $\log Z$ and the Borel-Pad\'e re-sum of high temp expansion \eqref{log Z numeric} for  $\log Z$ divided by its low temperature expansion against $\frac{\beta}{r}$. 
				The vertical red line is the straight line $\frac{\beta}{r}=2\log \frac{32}{\pi^2}$, to the  right of this line, the low temperature expansion \eqref{log Z interacting} is self-consistent.
				This shows the numerical accuracy of the agreement between the Borel-Pad\'e resum of high temp expansion and the low temperature expansion plotted in figure \ref{fig int 2}.}
			\label{fig int 3}
		\end{figure}

		The accuracy of this agreement is further analyzed in figure \ref{fig int 3}. Again, just as in the case of the free theory the curve flattens out at intermediate ranges of $\beta/r$. 
		Though compared to the free theory, the ranges over which the curve stays flat is not large, it is clear that 
		there is a gradual tendency to flatten out as the degree of the Pad\'{e} approximant is increased. 
		Note that the the approximant $[6,6]$ may seem to be better than $[8,8]$, but 
		we feel that it is coincidental,  it is not consistent as we increase the 
		degree to $[8, 8]$.  The errors seem to be minimised at around $\frac{\beta}{r} = 2.5$, however the position of this minima shifts at higher order of the Pad\'{e} approximation. 
		Also note that the $[6, 6]$ approximant does not remain  as flat compared to the 
		$[10, 10]$, $[12, 12]$ and $[14, 14]$ which consistently show the minima at around $\frac{\beta}{r} = 2$. 
		Furthermore, it is important to mention that the differences between the high and low temperature expansions at around the intermediate values of $\frac{\beta}{r} \sim  2$ are below $2$\%, 
		from the $[6, 6]$ approximant onwards is reasonable  given the various steps involved in 
		obtaining the numerical values of the approximations. 

		%

		\begin{figure}[h!]
			
			\begin{subfigure}{.4\linewidth}
				\begin{overpic}[width=1\textwidth]{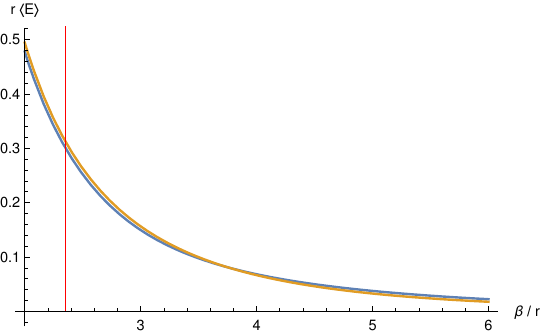}
					\put(40,25){%
						\setlength{\fboxsep}{1pt}
						\setlength{\fboxrule}{0.5pt}
						\fcolorbox{black}{white}{\includegraphics[width=0.56\textwidth]{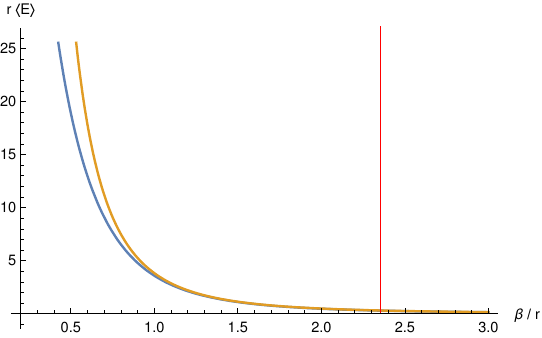}}%
					}
				\end{overpic}
				\caption{Using Pad\'e $[4,4]$}
			\end{subfigure}\hfill 
			\begin{subfigure}{.4\linewidth}
				\begin{overpic}[width=1\textwidth]{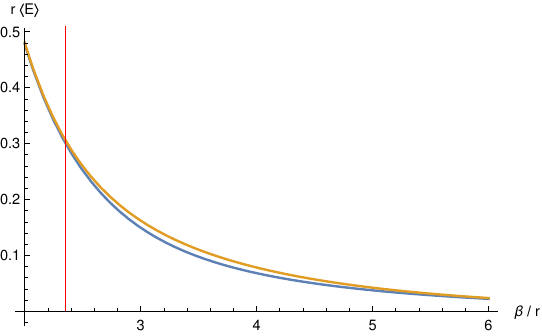}
					\put(40,25){%
						\setlength{\fboxsep}{1pt}
						\setlength{\fboxrule}{0.5pt}
						\fcolorbox{black}{white}{\includegraphics[width=0.56\textwidth]{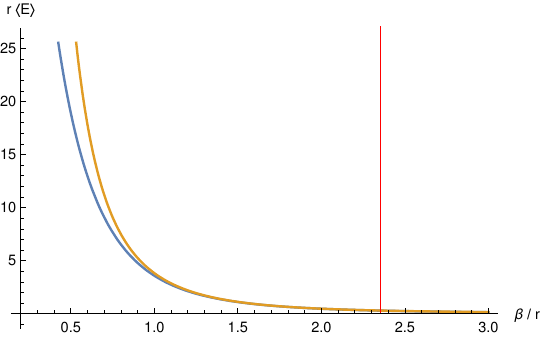}}%
					}
				\end{overpic}
				\caption{Using Pad\'e $[6,6]$}
			\end{subfigure}
			\par \bigskip\bigskip
			\begin{subfigure}{.4\linewidth}
				\begin{overpic}[width=1\textwidth]{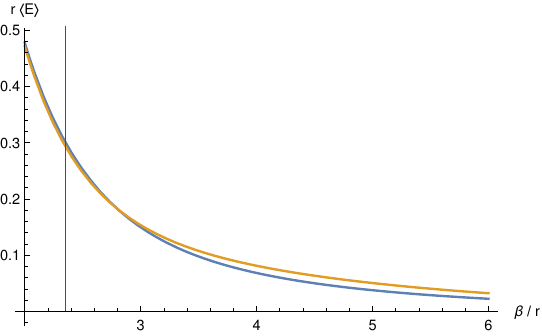}
					\put(40,25){%
						\setlength{\fboxsep}{1pt}
						\setlength{\fboxrule}{0.5pt}
						\fcolorbox{black}{white}{\includegraphics[width=0.56\textwidth]{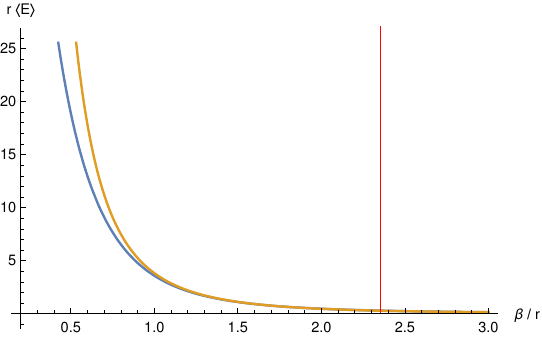}}%
					}
				\end{overpic}
				\caption{Using Pad\'e $[8,8]$}
			\end{subfigure}\hfill
			\begin{subfigure}{.4\linewidth}
				\begin{overpic}[width=1\textwidth]{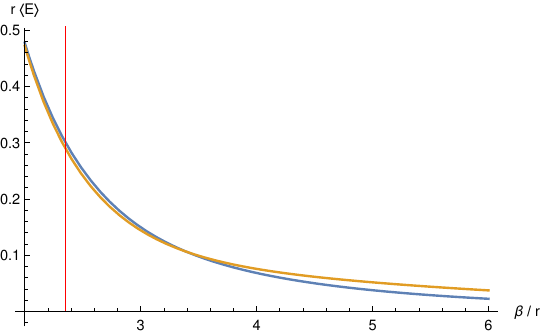}
					\put(40,25){%
						\setlength{\fboxsep}{1pt}
						\setlength{\fboxrule}{0.5pt}
						\fcolorbox{black}{white}{\includegraphics[width=0.56\textwidth]{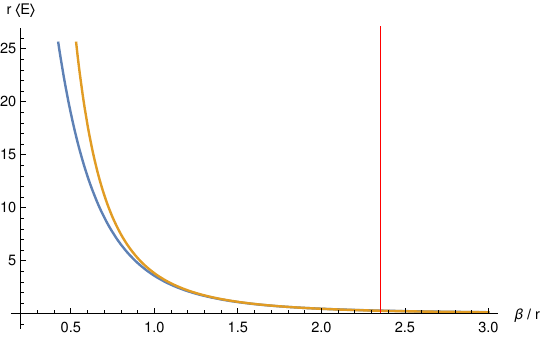}}%
					}
				\end{overpic}
				\caption{Using Pad\'e $[10,10]$}
			\end{subfigure}
			\par \bigskip\bigskip
			\begin{subfigure}{.4\linewidth}
				\begin{overpic}[width=1\textwidth]{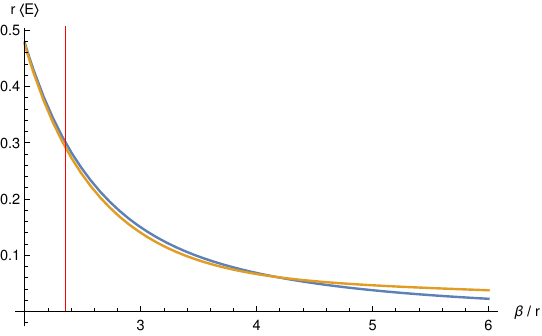}
					\put(40,25){%
						\setlength{\fboxsep}{1pt}
						\setlength{\fboxrule}{0.5pt}
						\fcolorbox{black}{white}{\includegraphics[width=0.56\textwidth]{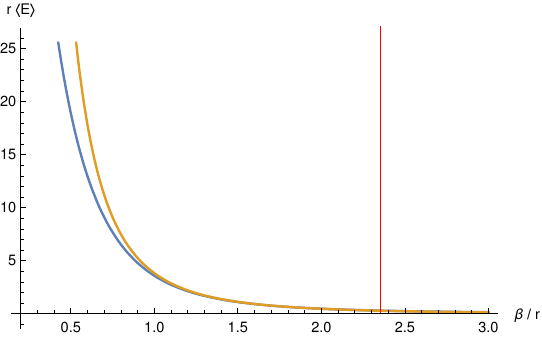}}%
					}
				\end{overpic}
				\caption{Using Pad\'e $[12,12]$}
			\end{subfigure}\hfill
			\begin{subfigure}{.4\linewidth}
				\begin{overpic}[width=1\textwidth]{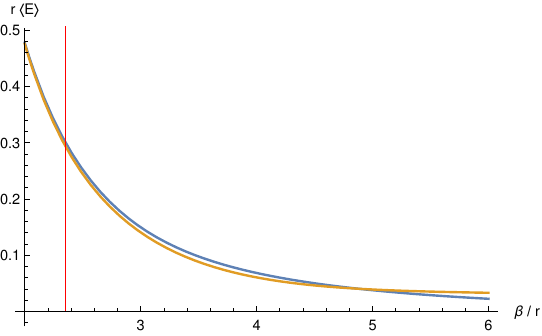}
					\put(40,25){%
						\setlength{\fboxsep}{1pt}
						\setlength{\fboxrule}{0.5pt}
						\fcolorbox{black}{white}{\includegraphics[width=0.56\textwidth]{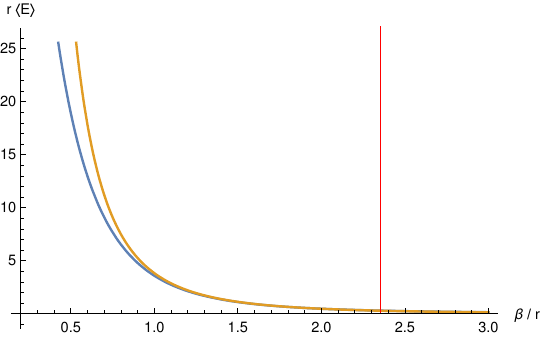}}%
					}
				\end{overpic}
				\caption{Using Pad\'e $[14,14]$}
			\end{subfigure}
			\caption{thermal expectation of energy in the interacting theory(in units of the radius i.e., $r\langle E\rangle$); the Borel-Pad\'e re-sum of the high temperature expansion \eqref{E numeric} for the thermal expectation of energy, plotted in orange, is compared against the low temperature expansion \eqref{en den interacting}, including orders till $O(e^{-\frac{6\beta}{r}}$) available in  the ancillary file  \texttt{low\_temp\_expansions.txt},  in blue. The figures inside the boxes describe the small $\frac{\beta}{r}$ regions for the corresponding graphs.
				The vertical red line is the straight line $\frac{\beta}{r}=2\log \frac{32}{\pi^2}$, to the  right  of this line,  the low temperature expansion \eqref{en den interacting} is self-consistent.
				The agreement between the orange and blue curves is observed over a finite range of $\frac{\beta}{r}$. The limited domain of validity of the Borel-Padé resummation at large 
				$\frac{\beta}{r} $ causes the two curves to deviate significantly from each other. At small $\frac{\beta}{r}$,   the truncated low temperature expansion starts to accumulate error, as the higher order terms become significant, though Borel-Pad\'e resum for the high temperature expansion works very well in this regime.}
			\label{fig int str 2}
		\end{figure}

		\begin{figure}[h]
			
			\begin{subfigure}{.4\linewidth}
				\begin{tikzpicture}
					\node[inner sep=0] (img) at (-1.5,0) {				\includegraphics[width=1\linewidth]{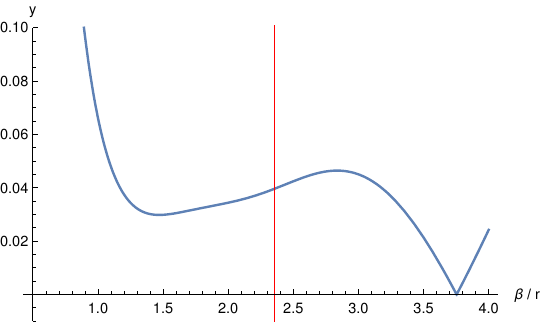}};
					\node[fill=white, draw=black, font=\tiny, anchor=north east] at (7,3) {\small $y=\Big|\frac{\text{ Borel-Pad\'{e}\ of\ high\ temp}-\text{ low\ temp\ expansion}}{\text{ low\ temp\ expansion}}\Big|$\ {\rm for }$\langle E\rangle$ vs. $\frac{\beta}{r}$};
				\end{tikzpicture}	
				\caption{Using Pad\'e $[4,4]$}
			\end{subfigure}\hfill 
			\begin{subfigure}{.4\linewidth}
				\par \bigskip\bigskip\bigskip
				\includegraphics[width=1\linewidth]{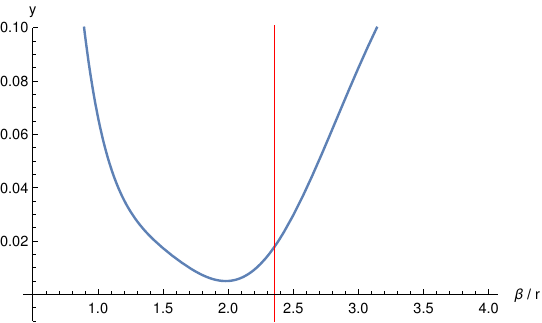}
				\caption{Using Pad\'e $[6,6]$}
			\end{subfigure}
			\begin{subfigure}{.4\linewidth}
				\includegraphics[width=1\linewidth]{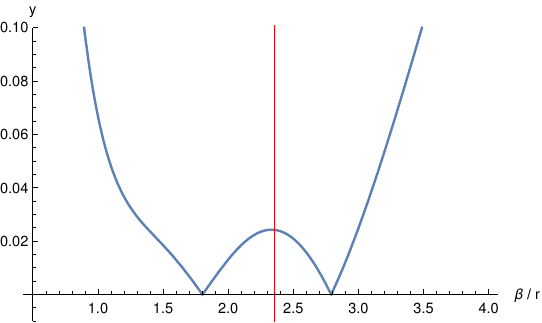}
				\caption{Using Pad\'e $[8,8]$}
			\end{subfigure}
			\begin{subfigure}{.4\linewidth}
				\includegraphics[width=1\linewidth]{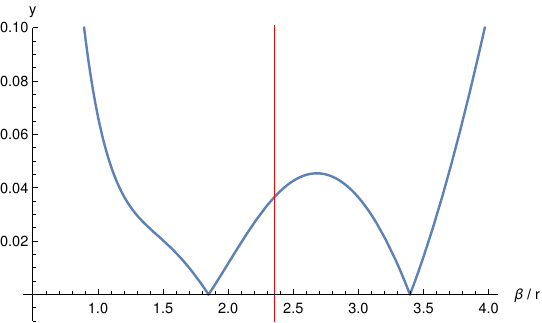}
				\caption{Using Pad\'e $[10,10]$}
			\end{subfigure}
			\begin{subfigure}{.4\linewidth}
				\includegraphics[width=1\linewidth]{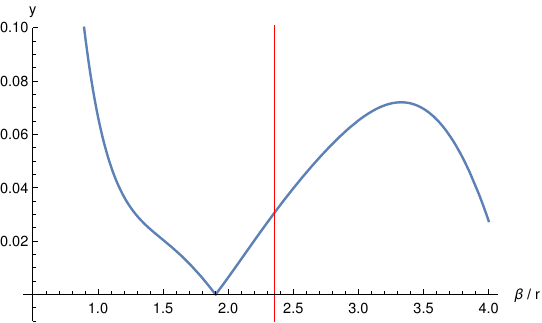}
				\caption{Using Pad\'e $[12,12]$}
			\end{subfigure}\hfill
			\begin{subfigure}{.4\linewidth}
				\includegraphics[width=1\linewidth]{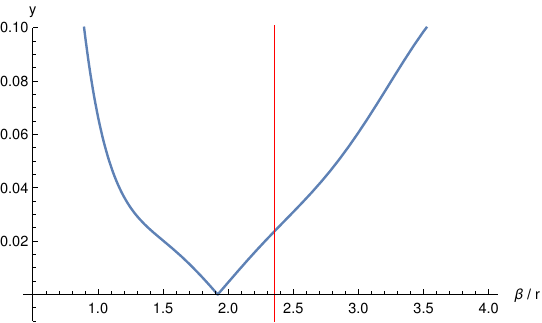}
				\caption{Using Pad\'e $[14,14]$}
			\end{subfigure}
			\caption{Interacting theory; We plot  the absolute value of the difference between  the low temperature expansion \eqref{en den interacting} for the thermal expectation of energy and the Borel-Pad\'e re-sum of high temp expansion \eqref{E numeric} for  $\langle E\rangle$ divided by its low temperature expansion against $\frac{\beta}{r}$.
				The vertical red line is the straight line $\frac{\beta}{r}=2\log \frac{32}{\pi^2}$, to the right  of this line, the low temperature expansion \eqref{en den interacting} is self-consistent.
				This demonstrates the numerical accuracy of the agreement between the Borel-Pad\'e resum of high temp expansion and low temperature expansion plotted in figure \ref{fig int str 2}.}
			\label{fig int str 3}
		\end{figure}

		The Borel-Pad\'e extrapolated thermal expectation of energy for the theory at the non-trivial fixed point can be obtained by differentiating the Borel-Pad\'e re-summed $\log Z$ with respect to $\beta$ as usual
		\begin{align}
			E=-\partial_\beta [\log Z^{\text{Borel-Pad\'e\ resumed}}].
		\end{align}
		This extrapolation of the thermal expectation of energy from the high temperature to lower values of the temperature is plotted in figure \ref{fig int str 2}  and we demonstrate its agreement with the low temperature expansion in a finite range of $\beta/r$. 
		
		The difference between the Borel-Pad\'{e} approximation and the low temperature expansion is plotted in  \ref{fig int str 3}. Note that though the curves don't flatten out as observed in the case of the free energies, 
		the error is consistently minimised at $\frac{\beta}{r} =2.0$ as seen from the $[12,12]$ and $[14, 14]$ approximants. Also note  the fact that though the  $[6, 6]$ does mimimise its error at around $\frac{\beta}{r} =2.0$, 
		the value at the minimum is higher than the $[12, 12]$ and $[14, 14]$ case. 
		The lack of consistency from $[6,6]$ to $[8,8]$ indicates the case of $[6,6]$ is coincidental. 
		Finally  the errors at around $\frac{\beta}{r} =2.0$  is below $2\%$ as observed in the case of the 
		free energies again demonstrating the consistency of our numerics.

		\clearpage
		
		\subsubsection*{Ratio of free energies}

		Finally, we evaluate the ratio of the free energy for the non-trivial fixed point at the infinite coupling $\lambda\to \infty$ to that for the Gaussian fixed point at $\lambda=0$. This ratio computed from the  low temperature expansions of free energies both at the non-trivial fixed point \eqref{log Z interacting}  and the Gaussian fixed point \eqref{free small b/r}, is plotted in \ref{intro fig a}. The ratio computed from the Borel-Pad\'e resummation(with Pad\'e of order $[14,14]$) of the high temperature expansions for the non-trivial fixed point and the Gaussian fixed point is plotted in \ref{intro fig b}.
		
		As discussed earlier around (\ref{formalmsol}), we noted that the low temperature expansion of the 
		thermal mass resulting from the gap equation vanishes  at $T=0$, else there would have been an 
		non-trivial  mass gap indicating that the fixed point is not a CFT. 
		The thermal mass in (\ref{m at low temp})  vanishes as $ \tilde m \sim \frac{1}{r } \exp (-\frac{\beta}{4r})$. 
		This implies that strictly at $T=0$, the free energy of the non-trivial fixed point has to coincide with the 
		massless Gaussian model.  The free energies of the 2 models will deviate as the temperature is increased. 
		This feature is clearly seen in the graphs and in fact a simple test of our numerics. 
		
\begin{figure}[H]
			
			\begin{subfigure}{.4\linewidth}
				\begin{tikzpicture}
					\node[inner sep=0] (img) at (-1.5,0) {				\includegraphics[width=1\linewidth]{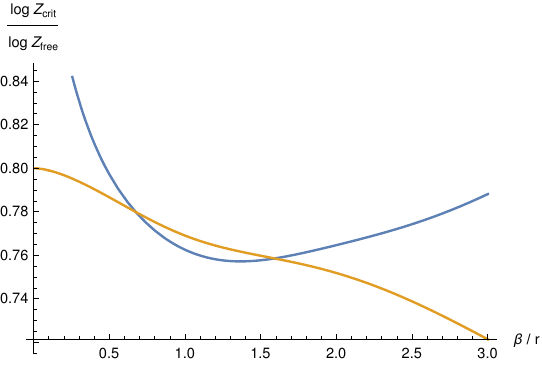}};
					\node[fill=white, draw=black, font=\tiny, anchor=north east] at (7,3) {\large Plot for $\frac{\log Z|_{\text{non-trival fixed point}}}{\log Z|_{\text{Gaussian fixed point}}}$\  vs. $\frac{\beta}{r}$};
				\end{tikzpicture}	
				\caption{Using Pad\'e $[4,4]$}
			\end{subfigure}\hfill 
			\begin{subfigure}{.4\linewidth}
				\par \bigskip\bigskip\bigskip
				\includegraphics[width=1\linewidth]{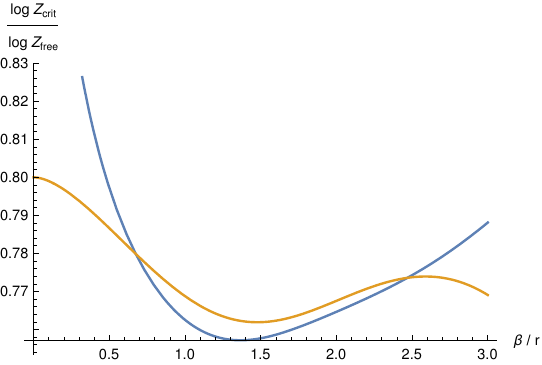}
				\caption{Using Pad\'e $[6,6]$}
			\end{subfigure}
			\begin{subfigure}{.4\linewidth}
				\includegraphics[width=1\linewidth]{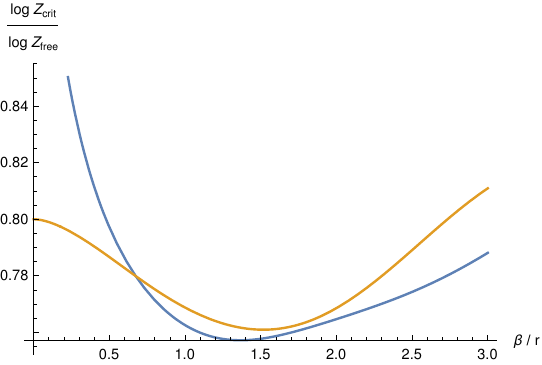}
				\caption{Using Pad\'e $[8,8]$}
			\end{subfigure}
			\begin{subfigure}{.4\linewidth}
				\includegraphics[width=1\linewidth]{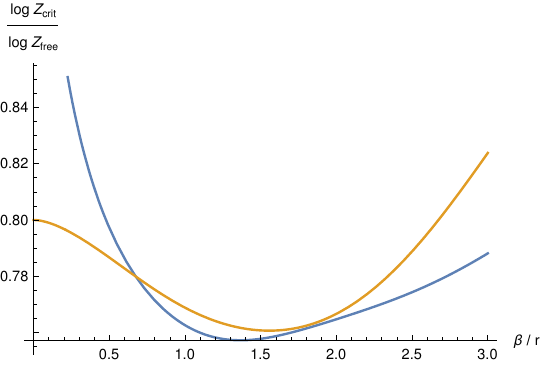}
				\caption{Using Pad\'e $[10,10]$}
			\end{subfigure}
			\begin{subfigure}{.4\linewidth}
				\includegraphics[width=1\linewidth]{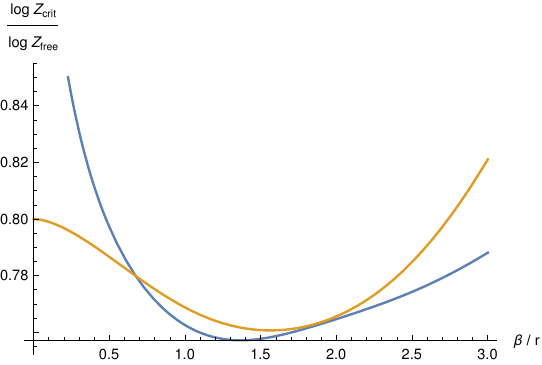}
				\caption{Using Pad\'e $[12,12]$}
			\end{subfigure}\hfill
			\begin{subfigure}{.4\linewidth}
				\includegraphics[width=1\linewidth]{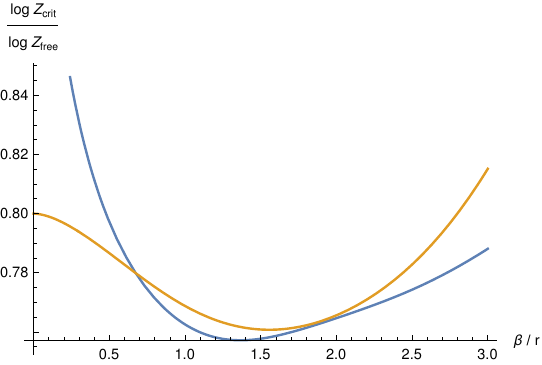}
				\caption{Using Pad\'e $[14,14]$}
			\end{subfigure}
			\caption{The figures plot the ratio  of the free energy evaluated at the non-trivial fixed point of the model to that at the Gaussian fixed point versus $\frac{\beta}{r}$, using various orders of Pad\'e approximations. The orange curves plot the ratio 
				computed from the the Borel-Pad\'e resum of the high temperature expansion of $\log Z$  at both the fixed points. The blue curves represent the ratio obtained from the low temperature expansion of $\log Z$, at  the two fixed points,  truncated till $O(e^{-\frac{6\beta}{r}})$.
				%
				Similar to the plots of $\log Z$ in figure \ref{fig int 2} and the thermal expectation of energy in figure \ref{fig int str 2}, the ratio computed from the Borel-Pad\'e resum of high temperature expansion and low temperature expansion agree well in a region around $\frac{\beta}{r}\sim2$. The accuracy of the agreement improves with the use of higher order the Pad\'e approximations. 
			}
			\label{ratioall}
		\end{figure}

		We also show the results for this ratio of the free energy at the non-trivial fixed point to that at the Gaussian fixed point evaluated using  Pad\'e approximations of different orders in figure \ref{ratioall}. The results obtained from the Borel-Pad\'e approximations of the high temperature expansions and also the ratio obtained from the series expansion at low temperature are plotted in the same graphs. Similar to the plots for $\log Z$ and the thermal expectation of energy given in figures \ref{fig int 2} and \ref{fig int str 2} respectively, the ratio evaluated from  the Borel-Pad\'e approximations of the high temperature expansions agrees with the ratio obtained from the  low temperature expansion  in a region about $\frac{\beta}{r}\sim 2$. Note in these graphs we are zooming into a smaller region of $\frac{\beta}{r}$ to clearly show the agreement between the results obtained from high temperature and low temperature expansions near $\frac{\beta}{r}\sim 2$. Thus in this region of $\frac{\beta}{r}$, the ratio obtained from both high temperature and low temperature coincides. But as one increases the value of $\frac{\beta}{r}$ further the Pad\'e-Borel resum of the high temperature  expansion fails to approximate the ratio correctly. For larger values of $\frac{\beta}{r}$ the ratio obtained from the low temperature expansion should provide a reliable approximation.
		The occurrence of the minima is consistent in each order of the Pad\'e-Borel resum of the high temperature expansion starting from the order $[6,6]$(as $[4,4]$ is not accurate enough). The position of the minima in the $\frac{\beta}{r}$ axis and the value of the ratio at these minima evaluated using different orders of Pad\'e approximation are reasonably stable as shown in the table    \ref{tab:sampledata}. 
		\begin{table}[h]
			\centering
			\begin{tabular}{ccc}
				\hline
				Pad\'e &	min attained at $\frac{\beta}{r}$	 & min value of the ratio \\
				\hline
				$[6,6]$	&	$1.47697$ & $0.761935$ \\
				$[8,8]$	&	$1.51703$ & $0.760937$ \\
				$[10,10]$ &	$1.55636$ & $0.76062$ \\
				$[12,12]$ &		$1.56313$ & $0.760685$ \\
				$[14,14]$	&	$1.55649$ & $0.760753$ \\
				\hline
			\end{tabular}
			\caption{The positions of the minima attained by the ratio of the free energy at the non-trivial fixed point to that at the Gaussian fixed point and  values of this ratio at the minima from the figures \ref{ratioall} are shown here for different orders of Pad\'e.   }
			\label{tab:sampledata}
		\end{table}
		\pagebreak
		
		\section{Discussions}\label{Discussions}
		We have evaluated   the free energy and the thermal expectation of energy for the  $O(N) $ model at large $N$ without the singlet constraint on $S^1\times S^2$ at the non-trivial fixed point as a low temperature expansion. This complements the study of the expansions at high temperature in our previous work \cite{David:2024pir}. The evaluation of the free energy at the low temperature involves solving the large $N$ gap equation in orders of $e^{-\frac{\beta}{4r}}$. Thus the free energy also organizes in such exponentials at low temperature. The free energy at the leading order in $e^{-\frac{\beta}{2r}}$ coincides with  the leading  term of a free CFT on $S^1\times S^2$.\\
		
		We have used a Borel-Pad\'e re-summation technique to demonstrate that the low temperature expansion for the free energy and thermal expectation of energy obtained here and their high temperature expansions obtained in \cite{David:2024pir} 
		correspond to the same solution  of the gap equation and therefore the same fixed point.
		We apply this technique of Borel-Pad\'e re-summation on the asymptotic series for the free energy at high temperature to extrapolate it to lower values of the temperature in units of the radius of the sphere $S^2$. We have observed the agreement between this Borel-Pad\'e extrapolation of the high temperature expansion and the low temperature expansion for a finite range of $\frac{\beta}{r}$. The reason for the agreement only in a finite range of $\frac{\beta}{r}$ is as follows. The validity of the Borel-Pad\'e resummation of the high temperature series is limited at large $\frac{\beta}{r}$. At small $\frac{\beta}{r}$, the truncated low temperature expansion accumulates error as an increasing number of sub-leading terms become important in this region. Thus the agreement persists in an  intermediate range of $\frac{\beta}{r}$ where both the Pad\'e-Borel resum of high temperature and the truncated low temperature expansion itself are sufficiently accurate.\\
		
		One of the highlights of our investigation is the evaluation of the ratio of free energies of 
		interacting theory to that of the free theory at all temperatures. From the Borel-Pad\'{e} high temperature expansion, the graph is given in \ref{intro fig b}, while the ratio from the low temperature expansion 
		is given in \ref{intro fig a}. We see that the ratio when the temperature is infinity begins with the 
		well known value of $\frac{4}{5}$, decreases to a minimum to $0.760753$ at $\frac{\beta}{r} =1.55649$ and 
		then starts to increase again. The ratio asymptotically tends to unity as the temperature is dialled to zero. 
		It is important to mention that this behaviour of the ratio was found due to our numerical methods. 
		The interesting minimum of this ratio lies at intermediate values of $\beta/r$ which was accessible 
		onlly through the Borel-Pad\'{e} extrapolation and the low temperature expansion.\\
		
		It will be interesting to generalise our study to vector models with other potentials, one such example is the model with sextic interaction studied recently in \cite{Romatschke:2019ybu} and models with fermions 
		\cite{DeWolfe:2019etx,Kumar:2025txh}. \\
		
		The study of the $O(N)$ model with the singlet constraint on $S^1\times S^2$ is of particular  interest as it is proposed to be dual to higher spin gravity in $AdS_4$ \cite{Klebanov:2002ja}. The model with singlet constraint undergoes the Gross-Witten-Wadia transition at temperature $T= b \sqrt{N}$ \cite{Shenker:2011zf}, in units of the radius of $S^2$ and the coupling constant $\lambda=\gamma N$, where $b$ and $\gamma$ are numerical constants.  These constants are determined by solving the gap equation and the condition for vanishing $U(N)$ holonomy eigenvalue density simultaneously.
		The method of summing over angular modes on the geometry of $S^1\times S^2$ developed in our previous work \cite{David:2024pir} can be implemented to find out the finite size corrections to the transition temperature. For this computation, one should carefully account for the sub-leading $\frac{1}{N}$ corrections, and higher orders as necessary,  in the partition function as well as the $U(N)$ holonomy eigenvalue density.  It is also useful to evaluate the partition function with finite size corrections in the non-trivial fixed point of the model as this should agree with the results from the calculations in higher spin $AdS_4$ gravity.\\
		
		Thermal one point functions of higher spin currents for the $O(N)$ model, without singlet constraint, on $S^1\times S^2$  had been evaluated as a high temperature expansion \cite{David:2024pir}.
		This involved applying OPE inversion formula in each order of small $\frac{\beta}{r}$ for the thermal 2-point function.
		At a large spin limit, these thermal one point functions at the non-trivial fixed point tends to the answer for the Gaussian fixed point.
		An interesting direction would be to evaluate these thermal one point functions as a low temperature expansion and test if the large spin behavior remains universal across the temperature.

		\section*{Acknowledgements}
		J.R.D  thanks the organizers of the `Black hole physics from strongly coupled thermal dynamics' 
		for a stimulating program and 
		gratefully acknowledges support from the Simons Center for Geometry and Physics, Stony Brook University  during which 
		some of  the research for this paper was performed. J.R.D thanks the organizers of the 14th Crete Regional Meeting on String Theory for the warm hospitality where a preliminary version of this work was presented. J.R.D thanks Spenta Wadia for discussions and sharing his  lectures on the vector model with 
		sextic interaction. S.K thanks the String theory group at ICTS-TIFR Bengaluru for the warm hospitality during the final stage of this work.
		
		\appendix
		\section{$\log Z_1(-\frac{1}{2})$: low temperature to high temperature}
		\label{app A}
		In this appendix we will show that the low temperature expansion of $\log Z_1{(-\frac{1}{2})}$ given in the 1st term of  \eqref{sum from n=2}  can be reorganized to reproduce its high temperature  expansion given in the first term of \eqref{small beta log Z}. We will implement a technique  closely related to the method of Borel sum as described below
		\begin{align}
			\log Z_1(-\frac{1}{2})
			&=\sum_{n=2}^\infty\frac{\Gamma(n-\frac{1}{2})}{2\sqrt{\pi}n!} (-\beta^2\tilde m^2)^n (\frac{r}{\beta})^{2n-1} (2^{2n-2}-1) \zeta(2n-2).
		\end{align}
		Now we  use the following formula for the $\zeta(2n-2)$ rewriting it in terms of Bernoulli numbers $B_{2n-2}$,
		\begin{align}
			\zeta(2n-2)=B_{2n-2}\frac{(2\pi)^{2n-2}}{(-1)^n 2 (2n-2)!},
		\end{align}
		in the above equation and obtain the following expression  with the use of the definition $x=mr \pi$,
		\begin{align}\label{StuveL BesselI}
			\log Z_1{(-\frac{1}{2})}&=\frac{\beta x^3}{4r\pi^{5/2}}\sum_{n=2}^\infty \frac{\Gamma(n-\frac{1}{2})}{(2n-2)!n!}x^{2n-3}2^{2n-2}(2^{2n-2}-1)B_{2n-2}\nonumber,\\
			&=\frac{\beta x^3}{4r\pi^{5/2}}\sum_{n=1}^\infty \frac{\Gamma(n+\frac{1}{2})}{(2n)!\Gamma(n+2)}x^{2n-1}2^{2n}(2^{2n}-1)B_{2n}\nonumber,\\
			&=\frac{\beta x^3}{2r\pi^{3}}\int_{0}^{1}f(\sqrt{t}x)\sqrt{1-t} dt,
		\end{align}
		Where
		\begin{align}
			f(x)=\sum_{n=1}^\infty \frac{B_{2n}}{(2n)!}x^{2n-1}2^{2n}(2^{2n}-1)&=\tanh x
			=1+2\sum_{n=1}^\infty (-1)^ne^{-2n x}.
		\end{align}
		We can perform the integral \eqref{StuveL BesselI} by the use of the expansion of $f(x)$ as a geometric series as given above. Each term can be integrated to result in the following expression for $\log Z_1(-1/2)$
		%
		\begin{align}
			\log Z_1(-1/2)=\frac{\beta  x^3}{3 \pi ^3 r}+\sum_{n=1}^\infty \frac{\beta  (-1)^n x^2 (3 \pi  \pmb{L}_2(2 n x)+4 n x-3 \pi  I_2(2 n x))}{6 \pi ^3 n r}.
		\end{align}
		where $\pmb{L}_2, I_2$ refer to the modified Struve and modified Bessel function of 1st kind respectively. 
		Now again expanding this at large $x$, followed by performing the sum over $n$, we reproduce the high temperature expansion of $\log Z_1(-1/2)$ as was given in the first term of \eqref{small beta log Z}.
		\begin{align}
			\log Z_1(-1/2)=&\frac{1}{3} \beta  m^3 r^2-\frac{\beta  m}{24}+\frac{7 \beta }{1920 m r^2}	+\frac{31 \beta }{64512 m^3 r^4}+\frac{127 \beta }{491520 m^5 r^6}+\frac{2555 \beta }{8650752 m^7 r^8}\nonumber\\&+\frac{1414477 \beta }{2453667840 m^9 r^{10}}
			+\frac{57337 \beta }{33554432 m^{11} r^{12}}+\frac{1303700629 \beta }{182536110080 m^{13} r^{14}}\nonumber\\
			&\qquad+\frac{822205892651 \beta }{20564303413248 m^{15} r^{16}}+\cdots.
		\end{align}
		One can explicitly write down terms from the first term of \eqref{small beta log Z} and verify the agreement with the above expression.

		\section{High temperature expansion of $\log Z_2$: an alternative method}
		\label{appn B}
		
		Here we present an alternative method to obtain the high temperature expansion for $\log Z_2$ given in the 2nd term of \eqref{small beta log Z} which reproduces the identical result found in \cite{David:2024pir}\footnote{We refer to result for the small $\frac{\beta}{r}$ expansion for $\log Z_2$ given in equation (2.15) of \cite{David:2024pir}.}. The method, we are going to present here, is based on rewriting the exponentials in equation \eqref{logZ after Mat sum} as an inverse Mellin  transform of Gamma function followed by a contour deformation in the integral involved in the Mellin transform as was done in \cite{Cardy:1991kr} for the massless free conformal scalars, also reviewed in section \ref{sec 2}. But due to the presence of the thermal mass $\tilde m$, the sum over the angular modes($l$) gets more involved compared to the case encountered for the free massless conformal scalars reviewed in \ref{sec 2}. We handle this angular sum by implementing the techniques developed in \cite{David:2024pir}. We begin with the definition for $\log Z_2$
		found in equation \eqref{logZ after Mat sum}
		\begin{align}\label{appen log Z2}
			\log Z_2 &= 
			2 \sum_{l=0}^\infty \Big(l+\frac{1}{2}\Big) \sum_{n=1}^\infty \frac{e^{-n\beta\sqrt{\frac{(l+\frac{1}{2})^2+}{r^2}  \tilde m^2}}}{n}.
		\end{align}
		We can represent  $e^{-\tau}$ as an inverse Mellin transform of the Gamma function as given below
		\begin{align}
			e^{-\tau} =\frac{1}{2\pi i}\int_{-i\infty+a}^{i\infty+a} \tau^{-s} \Gamma(s) ds, \qquad {\rm where} \ \ a>2.
		\end{align}
		Using this representation for the exponentials in \eqref{appen log Z2}, we can write
		\begin{align}\label{after Mellin}
			\log Z_2	&=\frac{1}{2\pi i} \int_{-i\infty+a}^{i\infty+a} ds \frac{\Gamma(s)}{\beta^s} \sum_{l=0}^\infty (2l+1) \Big(\frac{(l+\frac{1}{2})^2}{r^2}+\tilde m^2\Big)^{-s/2}
			\sum_{n=1}^\infty n^{-s-1}.
		\end{align}
		Now the infinite sum over $l$ in the above expression can be performed by following the techniques developed in \cite{David:2024pir}.\footnote{The method is described from equation (2.6) to (2.10) in \cite{David:2024pir}. } As a result of this, we have
		\begin{align}\label{l sum}
			\beta^{-s}  \sum_{l=0}^\infty (2l+1) \Big(\frac{(l+\frac{1}{2})^2}{r^2}+\tilde m^2\Big)^{-s/2}
			=\sum_{p=0}^\infty
			\frac{(-1)^{p+1}  \left(1-2^{1-2p}\right) (\tilde m r)^2 B_{2 p}  (\frac{\beta }{r})^{2 p}  \Gamma \left(p+\frac{s}{2}-1\right)}{(\tilde m \beta)^{2p+s}\Gamma (p+1) \Gamma \left(\frac{s}{2}\right)}.
		\end{align}
		Now combining \eqref{after Mellin} and \eqref{l sum}, we have
		\begin{align}\label{contour int}
			\log Z_2&=\frac{1}{2\pi i} \int_{-i\infty+a}^{i\infty+a} ds {\Gamma(s)} \sum_{p=0}^\infty
			\frac{(-1)^{p+1}  \left(1-2^{1-2p}\right) (\tilde m r)^2 B_{2 p}  (\frac{\beta }{r})^{2 p}  \Gamma \left(p+\frac{s}{2}-1\right)}{(\tilde m \beta)^{2p+s}\Gamma (p+1) \Gamma \left(\frac{s}{2}\right)}
			\sum_{n=1}^\infty \frac{1}{n^{s+1}}\nonumber,\\
			&\equiv \sum_{p=0}^\infty \log Z_2^{[p]}.
		\end{align}
		First let us consider only the term for $p=0$ from the summand above and let us carry out the integral
		\begin{align}
			\log Z_2^{[0]}=\frac{1}{2\pi i} \int_{-i\infty+a}^{i\infty+a} ds 
			\sum_{n=1}^\infty	\frac{r^2 \tilde m^{2-s} n^{-s-1} \beta ^{-s} \Gamma \left(\frac{s}{2}-1\right) \Gamma (s)}{\Gamma \left(\frac{s}{2}\right)}.
		\end{align}
		To evaluate this integral, we sum over the residues at the poles  of the integrand on the $s$-plane. \\
		
		The residue at $s=2$
		\begin{align}\label{s=2 res}
			\sum_{n=1}^\infty	\frac{r^2 \tilde m^{2-s} n^{-s-1} \beta ^{-s} \Gamma \left(\frac{s}{2}-1\right) \Gamma (s)}{\Gamma \left(\frac{s}{2}\right)}\Big|_{{\rm Res\ at\ }s=2}=\frac{2 r^2 \zeta (3)}{\beta ^2}.
		\end{align}
		
		The residue at $s=-k$ for $k$ being non-negative integers.
		\begin{align}\label{s=k res}
			\sum_{k=0}^\infty	\Big[\sum_{n=1}^{\infty}	\frac{r^2 \tilde  m^{2-s} n^{-s-1} \beta ^{-s} \Gamma \left(\frac{s}{2}-1\right) \Gamma (s)}{\Gamma \left(\frac{s}{2}\right)}&\Big|_{{\rm Res \ at\ }s=k }\Big]\nonumber\\
			&=\frac{2 \left(\beta\tilde   m r^2 \text{Li}_2\left(e^{-\tilde m \beta }\right)+r^2 \text{Li}_3\left(e^{-\tilde m \beta }\right)-r^2 \zeta (3)\right)}{\beta ^2}.
		\end{align}
		Thus adding the contributions due to the poles at the integer values of $s$ such that $s\le 2$, by combining \eqref{s=2 res} and \eqref{s=k res}, we have	the leading most term of the small $\frac{\beta}{r}$ expansion of $\log Z_2$ to be 
		\begin{align}\label{Z2[0]}
			\log Z_2^{[0]}&=\frac{1}{2\pi i} \int_{-i\infty+a}^{i\infty+a} ds	\sum_{n=1}^\infty	\frac{r^2 \tilde m^{2-s} n^{-s-1} \beta ^{-s} \Gamma \left(\frac{s}{2}-1\right) \Gamma (s)}{\Gamma \left(\frac{s}{2}\right)},\\
			&=\frac{2 r^2 \left(\beta  \tilde m \text{Li}_2\left(e^{-\tilde m \beta }\right)+\text{Li}_3\left(e^{-\tilde m \beta }\right)\right)}{\beta ^2}.
		\end{align}		
		Note that this agrees with the leading term in $\log Z_2$ given in \eqref{small beta log Z} which was obtained in \cite{David:2024pir}.\footnote{We refer to the leading most term or $p=0 $ term of the expansion given in equation (2.15) of \cite{David:2024pir}. }\\
		
		For the rest of the terms from \eqref{contour int} with $p\ge 1$, we have to evaluate  the following integral
		\begin{align}
			\log Z_2^{[p\ge1]}=\frac{1}{2\pi i} \int_{-i\infty+a}^{i\infty+a} ds {\Gamma(s)} 
			\frac{(-1)^{p+1}  \left(1-2^{1-2p}\right) (\tilde m r)^2 B_{2 p}  (\frac{\beta }{r})^{2 p}  \Gamma \left(p+\frac{s}{2}-1\right)}{(\tilde m \beta)^{2p+s}\Gamma (p+1) \Gamma \left(\frac{s}{2}\right)}
			\sum_{n=1}^\infty n^{-s-1}.
		\end{align}
		For this case of $ p\ge 1$, the above integral has contributions only from the poles of the $\Gamma(s)$ in the above expression for $s=-k$ where $k$ is a non-negative integer, as  the integrand has no other pole. Thus, we can evaluate the residue at $s=-k$, with $k$ being non-negative integers, and we obtain the following expression
		\begin{align}\label{p>0}
			\log Z_2^{[p\ge 1]}=	 
			\sum_{n=1}^\infty	\frac{(-1)^{p+1} 4^{-p} \left(4^p-2\right) (\tilde m r)^2 B_{2 p}  (\frac{\beta }{r})^{2 p}}{n \Gamma (p+1)(\tilde m\beta)^{2p}}
			\sum_{k=0}^{\infty}\frac{(-1)^k \beta ^k \tilde m^k n^k \Gamma \left(-\frac{k}{2}+p-1\right)}{k! \Gamma \left(-\frac{k}{2}\right)}.
		\end{align}		
		Now we can perform the infinite sum over $k$ present in  the above expression \eqref{p>0} in terms of modified Bessel function of second kind as given below,	
		\begin{align}\label{k sum}
			\sum_{k=0}^{\infty}\frac{(-1)^k \beta ^k \tilde m^k n^k \Gamma \left(-\frac{k}{2}+p-1\right)}{k! \Gamma \left(-\frac{k}{2}\right)}
			&=\sum_{k=0}^\infty \frac{\sqrt{\pi } 2^{-k} \beta ^k \tilde m^k n^k (-1)^{k+p-1}}{\Gamma \left(\frac{k}{2}+\frac{1}{2}\right) \Gamma \left(\frac{k}{2}-p+2\right)}\nonumber,\\
			&=
			\frac{2^{\frac{3}{2}-p} (\beta  \tilde m n)^{p-\frac{1}{2}} K_{p-\frac{3}{2}}(\tilde m n \beta )}{\sqrt{\pi }}.
		\end{align}
		The use of the truncated series formula for the modified Bessel function of the second kind with half integer order, as given below, will allow us to perform the sum over $n$ in \eqref{p>0},	
		\begin{align}\label{Bessel truncated formula}
			K_{p-\frac{3 }{2}}(\tilde mn \beta)=	\sum_{j=0}^{|p-\frac{3}{2}|-\frac{1}{2}} \frac{\sqrt{\pi }  e^{-\tilde m\beta   n} \left(-j+\left| p-\frac{3}{2}\right| +\frac{1}{2}\right)_{2 j} (\beta  \tilde m n)^{-j-\frac{1}{2}}}{2^{j+\frac{1}{2}}j!}.
		\end{align}
		Finally by combining \eqref{p>0}, \eqref{k sum} and \eqref{Bessel truncated formula}, and performing sum over $n$ resulting in the Polylogarithm functions, we obtain
		\begin{align}
			\log Z_2^{[p\ge 1]}=\big(\frac{\beta}{r}\big)^{2p-2}	\sum_{j=0}^{|p-\frac{3}{2}|-\frac{1}{2}} 
			\frac{(-1)^{p+1} \left(4^p-2\right) B_{2 p}  \left(\frac{1}{2} (-2 j+| 3-2 p| +1)\right)_{2 j}  \text{Li}_{j-p+2}\left(e^{-\tilde m \beta }\right)}{2^{j+3p-1}(\beta \tilde m)^{j+p-1}j! \Gamma (p+1)}.
		\end{align}
		It is also easy to realize that the term $\log Z_2^{[0]}$ computed in \eqref{Z2[0]}  fits in the above formula, thus we have the small $\frac{\beta}{r}$ expansion of \eqref{appen log Z2} as given by
		\begin{align}
			\log Z_2=\sum_{p=0}^\infty\big(\frac{\beta}{r}\big)^{2p-2}	\sum_{j=0}^{|p-\frac{3}{2}|-\frac{1}{2}} 
			\frac{(-1)^{p+1} \left(4^p-2\right) B_{2 p}  \left(|p-\frac{3}{2}|-j+\frac{1}{2}\right)_{2 j}  \text{Li}_{j-p+2}\left(e^{-\tilde m \beta }\right)}{2^{j+3p-1}(\beta \tilde m)^{j+p-1}j! \Gamma (p+1)}.
		\end{align}
		This precisely agrees with the expansion given in the 2nd term  of \eqref{small beta log Z} derived  in \cite{David:2024pir}. The calculation presented here generalizes the method formulated for the massless free conformal scalars on $S^1\times S^2$ by Cardy \cite{Cardy:1991kr} to the case of a massive scalar field.

		\section{The large $N$ simplification on $S^1\times S^2$ }\label{appen C}
		In this appendix we show that only the zero mode of the auxiliary field denoted by $\zeta_0$ contributes at the leading order in large $N$ in the action given in  \eqref{intro of zeta}, while the non-zero modes $\tilde \zeta$ contributes in a sub-leading order. This important simplification has been discussed in many papers on this subject, 
		see for example in \cite{Romatschke:2019ybu}. 
		But,  in all cases the model was considered on $S^1\times R^2$, in this appendix, we show that this simplification holds also for the $O(N)$ model on $S^1 \times S^2$.  Let us begin with the partition function 
		after the Hubbard-Stratonovich transformation. 
		\begin{align}\label{action append C}
			\tilde{Z}= \sqrt{ \frac{N \beta V}{8\pi \lambda} } \int_{S^1\times S^2}^{} {\cal D}\phi {\cal D}\zeta e^{-\frac{1}{2}\int_0^\beta\int d\tau d^2x [\partial_\mu\phi_i\partial_\mu\phi_i+\frac{1}{4r^2}\phi_i\phi_i+\frac{\zeta^2N}{4\lambda}+i\zeta\phi_i\phi_i]}.
		\end{align}
		Here  $V = 4\pi r^2$ and 
		we have kept track of the prefactor\footnote{This factor is not taken into account in the main text, as such factors do not contribute at the leading order in the large-$N$ expansion.} in the partition function which ensures that on integrating out 
		the auxiliary field $\zeta$, we obtain the partition function with the quartic action. 
		We recall  the definition given in equation \eqref{zero+nz}, to separate the zero and the non-zero mode of the auxiliary field $\zeta$ used to linearise the quartic interaction term in $\phi_i$ in the  action \eqref{quartic action}
		\begin{align}\label{zero+nz appen}
			\zeta(\tau,\vec x)=\zeta_0+\tilde \zeta(\tau,\vec x),
		\end{align}
		Substituting this decomposition of the auxiliary field, the partition function can be written as 
		\begin{eqnarray}
			\tilde Z = \sqrt{ \frac{N \beta V}{8\pi \lambda} } \int_{S^1 \times S^2 } 
			d\zeta_0 e^{ - \frac{ \zeta_0^2  N \beta V}{8 \lambda} }
			\int {\cal D} \phi {\cal D} \tilde \zeta e^{-S_0 - S_I}
		\end{eqnarray}
		where 
		\begin{eqnarray} \label{sosi}
			S_0 &=& \frac{1}{2} \int_0^\beta d\tau \int d^2 x \sqrt{g} \left( 
			\partial_\mu \phi_i \partial^\mu \phi_i   + \frac{1}{4 r^2} \phi_i \phi_i  + i \zeta_0 \phi_i\phi_i
			+ \frac{ \tilde \zeta^2  N }{ 4\lambda}  \right) , \\ \nonumber
			S_I &=& \frac{i}{2}  \int_0^\beta d\tau \int d^2 x \sqrt{g}  \tilde \zeta \phi_i \phi_i
		\end{eqnarray}
		Note that there is no cross term between $\tilde \zeta$ and the zero mode $\zeta_0$. The 
		integral over $S^1\times S^2$ kills such a term due to momentum conservation on $S^1$  and angular momentum conservation on $S^2$. 
		
		The crucial point is that at the leading order in large  $N$, the interaction term $S_I$ can be neglected, 
		as the  contribution  to the partition function obtained 
		by integrating $\tilde \zeta$ and $\phi_i$ from these terms is sub-leading in $N$. 
		If this is indeed the case then the  path integral over $\tilde \zeta$ just produces a normalization factor 
		leaving behind the integrals 
		\begin{eqnarray}
			\lim_{N\rightarrow \infty} \tilde{Z} &=& \int d\zeta_0 {\cal D} 
			\phi e^{ - \frac{ \zeta_0^2  N \beta V}{8 \lambda}  -\hat S_0 } , \\ \nonumber
			\hat S_0 &=&  \frac{1}{2} \int_0^\beta d\tau \int d^2 x \sqrt{g} \left( 
			\partial_\mu \phi_i \partial^\mu \phi_i   + \frac{1}{4 r^2} \phi_i \phi_i  + i \zeta_0 \phi_i\phi_i \right) 
		\end{eqnarray}
		Now the path integral over the scalar $\phi_i$ is that of  $N$ massive scalars and  is of the form 
		$ \exp \big[  N \log Z(\tilde m^2 = i \zeta_0) \big] $ as given in  (\ref{path int to int}). 
		After which we perform the integral over 
		$\zeta_0$  using the saddle point  results in the gap equation, which has been used in (\ref{saddle pt cond}). 
		In the rest of the appendix, we show that the contribution due to $S_I$ is indeed subleading in large $N$ and thus only the zero mode of the auxillary field contributes at the leading order in $N$. 
		
		\subsubsection*{Contribution from $S_I$ is subleading in $N$}
		
		Let us now justify that the contribution from $S_I$ is subleading in $N$. 
		For this it is suitable to expand both $\phi_i$ and $\tilde \zeta$ in terms of the Matsubara fequencies 
		and spherical harmonics. 
		\begin{align}\label{phi fourier}
			\phi_i(\tau,\vec x)=\sum_{n,l=0}^{\infty}\sum_{m=-l}^l e^{\frac{2\pi i n\tau}{\beta}} Y_{l,m}(\theta,\varphi) \tilde \phi_{i;n,l,m}
		\end{align}
		We recall that $i=1,\cdots , N.$
		Note that, $\tilde \phi_{i;n,l,m}(r) $ are the Fourier modes on $S^1\times S^2$, $Y_{l,m}$'s denote spherical harmonics, $\beta$ being the inverse temperature. Similarly for the non-zero mode of the auxiliary field $\tilde\zeta$, we have
		\begin{align}\label{zeta fourier}
			\tilde{\zeta}(\tau,\vec x) &=\sum_{n,l=0}^{\infty}\sum_{m=-l}^l e^{\frac{2\pi i n\tau}{\beta}} Y_{l,m}(\theta,\varphi) \tilde \zeta'_{n,l,m}, 
		\end{align}
		Again $\tilde\zeta'_{n,l,m}$ denotes the Fourier modes of $\tilde \zeta(\tau,\vec x) $ on $S^1\times S^2$, Note that as  the non-zero mode $\tilde \zeta$ contains no constant mode since we have separated  the constant modes $\zeta_0$, we must have 
		\begin{align}
			\tilde\zeta'_{0,0,0}=0 \label{zeta' is 0}
		\end{align}

		Let us perform the path integral over $\tilde \zeta$ and $\phi_i$ first. 
		For this it is convenient to expand 
		the exponential $e^{\frac{i}{2}{\int d\tau d^2x} \tilde\zeta \phi_i\phi_i}$ in the action \eqref{action append C}  and consider the following path integral
		\begin{align}
			\hat{Z}&= \sqrt{ \frac{N \beta V}{8\pi \lambda} }  \int_{S^1\times S^2}^{} {\cal D}\phi{\cal D}\tilde\zeta  e^{-\frac{1}{2}\int_0^\beta\int d\tau d^2x [\partial_\mu\phi_i\partial_\mu\phi_i+( \frac{1}{4r^2} + i \zeta_0) \phi_i\phi_i+\frac{\tilde\zeta^2N}{4\lambda}]}
			\sum_{k=0}^\infty\frac{1}{k!}\Big(\frac{i}{2}{\int d\tau d^2x} \tilde\zeta \phi_i\phi_i\Big)^k \nonumber \\ 
			&\equiv \sqrt{ \frac{N \beta V}{8\pi \lambda} } \int {\cal D}\tilde\zeta e^{-\frac{1}{2}\int d\tau d^2x \frac{\tilde\zeta^2N}{4\lambda}}\sum_{k=0}^\infty \hat  Z^{(k)}(\tilde\zeta)\equiv \sum_{k=0}^\infty {\cal Z}^{(k)} \label{def def}
		\end{align}
		This path integral involves the integral over $\phi_i$ and $\tilde \zeta$ which does not contain the 
		zero mode. 
		
		The $k=0$ term is when we do not consider the interaction $S_I$. To show that contributions 
		from $k>0$ are subleading in $N$ we first show in the expansion, any self contraction within $\phi_i\phi_i$  which may result in an $O(N)$ term vanishes due to angular momentum conservation on $S^2$ and momentum conservation on $S_1$. 
		We will demonstrates  the term with $k=1$ and $k=2$ in the above equation.
		Before we proceed we would need the propagator of $\phi_i$ for each Matsubara mode and angular momentum mode.
		By substituting the expansions  (\ref{phi fourier}), (\ref{zeta fourier})  in the action, this propagator is given by 
		\begin{align}\label{2 pt func1}
			\langle\tilde \phi_{i;n,l,m}\tilde \phi_{j;n',l',m'}\rangle	&
			=\frac{(-1)^m}{\beta r^2}\frac{\delta_{i,j} \delta_{n+n',0} \delta_{l,l'}\delta_{m+m',0}}{\big(\frac{2\pi n}{\beta}\big)^2+\frac{(l+\frac{1}{2})^2}{r^2}  +  i \zeta_0 }
		\end{align}
		The $(-1)^m$ arises, due the relation between spherical harmonics $Y_{l, m  }$ and $Y_{l, -m}$.   
		Now expanding the fields in the interaction term for $k=1$  using 
		\eqref{phi fourier} and \eqref{zeta fourier} we have
		\begin{align}\label{ylm int}
			\hat Z^{(1)}[\zeta] &=\frac{i}{2}\beta r^2
			\hat Z^{(0)}
			\sum_{\substack{n,l,m\\n',l',m'\\n'',l'',m''}}\Big[ \int d\tau d\Omega e^{\frac{2\pi i \tau}{\beta}(n+n'+n'')} Y_{l,m}(\theta,\varphi) Y_{l',m'}(\theta,\varphi)Y_{l'',m''}(\theta,\varphi)\nonumber\\
			& \qquad \qquad\qquad\qquad\qquad \times \zeta'_{n,l,m} \times \langle\phi_{i;n,l,m}\phi_{i;n',l',m'}\rangle \Big]
		\end{align}
		where 
		$
		d\Omega=\sin \theta d\theta d\phi 
		$. Note that the $\hat Z^{(0)}$ the partition function without any interaction 
		occurs as a prefactor, since the two point function 
		in (\ref{2 pt func1}) is the normalized correlator. 
		Using the fact that 
		\begin{align}
			\int_0^{\beta } e^{\frac{2 \pi  i \tau  (n+n'+n'')}{\beta }} \, d\tau=\beta \delta_{n+n'+n'',0}
		\end{align}	
		and using equation \eqref{2 pt func1} in \eqref{ylm int}, we obtain
		\begin{align}
			\hat{Z}^{(1)}[\zeta]=\frac{i}{2}N  \hat Z^{(0)} \sum_{\substack{l,m\\l',m'}}\frac{\zeta'_{0,l,m}}{\frac{(l+\frac{1}{2})^2}{r^2} +i \zeta_0 } (-1)^{m'} \int d\Omega Y_{l,m} Y_{l',m'} Y_{l',-m'}
		\end{align}
		Now we have the following identity for the integral for three spherical harmonics occurred in the above equation
		\begin{align}
			\int d\Omega Y_{l',m'} Y_{l',-m'} Y_{l,m}=\sqrt{\frac{(2l+1)(2l'+1)^2}{4\pi}}  \Big( 
			\begin{array}{ccc}
				l' & l' & l \\
				0 & 0 & 0 \\
			\end{array}
			\Big)  \Big( 
			\begin{array}{ccc}
				l' & l' & l \\
				m' & -m' & m \\
			\end{array}
			\Big)
		\end{align}
		where $ \Big( 
		\begin{array}{ccc}
			l & l' & l'' \\
			m & m' & m'' \\
		\end{array}
		\Big)$ denotes the Wigner's $3j$-symbol, this  vanishes unless $m+m'+m''=0 $, which forces $m=0$ in the above equation.
		\begin{align}
			\hat  Z^{(1)}[\zeta]=\frac{i}{2}N  \hat  Z^{(0)} \sum_{l,l'}\frac{\zeta'_{0,l,0}}{\frac{(l+\frac{1}{2})^2}{r^2} + i \zeta_0 } \sqrt{\frac{(2l+1)(2l'+1)^2}{4\pi}} \Big( 
			\begin{array}{ccc}
				l' & l' & l \\
				0 & 0 & 0 \\
			\end{array}
			\Big)
			\sum_{m'=-l}^l (-1)^{m'} \Big( 
			\begin{array}{ccc}
				l' & l' & l \\
				m' & -m' & 0 \\
			\end{array}
			\Big)
		\end{align}
		Now we use the following relation to perform the sum over $m'$ in the above equation
		\begin{align}
			\sum_{m'=-l}^{l} (-1)^m \Big( 
			\begin{array}{ccc}
				l' & l' & l \\
				m' & -m' & 0 \\
			\end{array}
			\Big)=(-1)^{l'} \sqrt{2l'+1} \delta_{l,0}
		\end{align}
		in the above equation to obtain, 
		\begin{align}
			\hat Z^{(1)}[\zeta]&=\frac{i}{4\sqrt{\pi}}N \hat Z^{(0)}\sum_{l'} \frac{\zeta'_{0,0,0}}{\frac{1}{4r^2}+i\zeta_0} (-1)^{l'}(2l'+1)^{3/2} \Big( 
			\begin{array}{ccc}
				l' & l' & 0 \\
				0 & 0 & 0 \\
			\end{array}
			\Big)\\
			&=\frac{i}{4\sqrt{\pi}}N \hat Z^{(0)}\sum_{l'} \frac{\zeta'_{0,0,0}}{\frac{1}{4r^2}+i\zeta_0} (2l'+1)=0
		\end{align}
		We observe that only the zero mode contributes in the expression above.
		We know that the zero mode of $\tilde\zeta$ is zero, i.e., $\tilde \zeta'_{0,0,0}=0$ from the definition of 
		$\tilde \zeta$  as in \eqref{zeta' is 0}, thus the above term vanishes.
		The calculation shows that putative which may contribute at $O(N)$ vanishes due to momentum conservation. 
		It is important to note that we have arrived at this result before the $\tilde \zeta $ integration. 
		
		Similarly, one can evaluate the contribution of the self contractions in the  $k=2$ term. 
		This   also vanishes  due to  $\tilde \zeta'_{0,0,0}=0$
		\begin{align}
			\hat Z^{(2)}(\zeta)|_{\text{self contraction}} =-\frac{1}{8} N^2 \Big[\frac{\zeta'_{0,0,0}}{\frac{1}{4r^2}+i\zeta_0}\frac{1}{2\sqrt{\pi}}(2l+1)\Big]^2=0
		\end{align}
		It is easy to carry this analysis forward to show that the contribution to  $\hat {Z}^{(k)}$ for all $k$ , any contraction which involves self contraction, that is contraction among $\phi_i\phi_i$ which occurs in $S_I$ vanishes. 
		The reason this vanishes is that the contraction picks  up the zero mode of $\tilde \zeta'$, which itself is zero.

		We are therefore led to  consider only contractions between $\phi_i$ and $\phi_j$ which occur among  2 different $S_I$ in the expansion (\ref{def def}).  This can be shown to be subleading in $N$, at the most they are of $O(N^0)$. 
		For example for the term ${{\cal Z}^{(2)}}$ with non-self interaction\footnote{For odd $n$, ${\cal Z}^{(n)}=0$ as the path integral in $\tilde \zeta$ vanishes in presence of odd number of $\tilde \zeta$ fields.}
		\begin{align}
			{\cal Z}^{(2)}|_{\rm no\;self\; contractions}
			=	\langle \tilde\zeta\tilde\zeta \phi_i \phi_i \phi_j\phi_j\rangle\sim O(N^0)
		\end{align}
		This is because when contract between $\phi_i$ and $\phi_j$(with different indices) and it is proportional to $\delta_{ij}\delta_{ij}\sim N$. This factor of 
		$N$ cancels  the $\frac{1}{N} $ which arises from the contraction between $\langle\tilde \zeta\tilde \zeta\rangle$
		which is given by 
		\begin{eqnarray}
			\langle \tilde \zeta \tilde \zeta \rangle \sim   \frac{1}{N}
		\end{eqnarray}
		This fact can be seen from the normalization of the quadratic term in (\ref{def def}). 
		Note that we need the normalized expectation value, since we have the overall  prefactor in (\ref{def def}). 
		
		Similarly, one can consider ${\cal Z}^{(4)}$, and similarly, counting the power of $N$ leads to the fact that the highest order of $N$ it can produce is $O(N^0)$
		\begin{align}
			{\cal Z}^{(4)} |_{\rm no\; self\; contractions} \sim O(N^0 )
		\end{align}
		Similarly for ${\cal Z}^{(k)}$ with any value $k$, the non-self contracted contributions will appear in order $O(1)$ terms or even more suppressed in large $N$.

		Thus, it is clear from the above discussion that the non-trivial contribution to the partition function from $S_I$  in (\ref{sosi}) occurs at 
		$O(N^0) $ in large $N$ and therefore can be neglected in comparison with $S_0$ which contributes at the at the leading order.   $S_0$ contains the zero mode $\tilde \zeta_0$ which behaves as the  mass.

			\bibliographystyle{JHEP}
			\bibliography{references} 
		\end{document}